\documentclass[paperpaper,twocolumn,english,superscriptaddress,aps,prb,floatfix,amsfonts,amssymb]{revtex4-1}
\usepackage{float}
\usepackage{epsfig}
\usepackage{graphicx}
\usepackage{dcolumn}
\usepackage{bm}
\usepackage{appendix}
\usepackage{subfigure}
\usepackage{color}
\usepackage{natbib}

\begin{document}  
\title{Zigzag materials: selective interchain couplings control the coexistence of one-dimensional physics and
deviations from it}


\author{J. M. P. Carmelo}
\affiliation{Center of Physics of University of Minho and University of Porto, P-4169-007 Oporto, Portugal}
\affiliation{CeFEMA, Instituto Superior T\'ecnico, Universidade de Lisboa, Av. Rovisco Pais, P-1049-001 Lisboa, Portugal}
\affiliation{Boston University, Department of Physics, 590 Commonwealth Ave, Boston, MA 02215, USA}
\author{P. D. Sacramento}
\affiliation{CeFEMA, Instituto Superior T\'ecnico, Universidade de Lisboa, Av. Rovisco Pais, P-1049-001 Lisboa, Portugal}
\author{T. Stauber}
\affiliation{Instituto de Ciencia de Materiales de Madrid, CSIC, E-28049 Madrid, Spain}
\author{D. K. Campbell}
\affiliation{Boston University, Department of Physics, 590 Commonwealth Ave, Boston, MA 02215, USA}

\begin{abstract}
The coexistence in the low-temperature spin-conducting phases 
of the zigzag materials BaCo$_2$V$_2$O$_8$ and SrCo$_2$V$_2$O$_8$ of  
one-dimensional (1D) physics with important deviations from it is not well understood. 
The studies of this paper account for an important selection rule that follows from 
interchain spin states being coupled more strongly within the spin dynamical structure factor of such zigzag materials 
whenever they are connected by a specific symmetry operation of the underlying lattice. 
In the case of excited states, this symmetry operation is only a symmetry in spin-space if {\it no} electronic spin flip is performed
within the generation of such states.
The corresponding selective interchain couplings both protect the 1D physics 
and are behind important deviations from it concerning the enhancement
of the spectral-weight intensity of $S^{zz} (k,\omega)$. 
Strong evidence is provided that this justifies, beyond interchain mean-field theory and in contrast to 1D physics, the 
experimental low-energy dominance in both zigzag materials of the longitudinal nuclear-magnetic-resonance relaxation rate 
term $1/T_1^{\parallel}$ for the whole magnetic-field interval of the spin-conducting phases.
To further understand the role of selective interchain couplings concerning their contradictory effects in
protecting the 1D physics and controlling deviations from it, the physical-spins scattering processes
behind the experimentally observed sharp peaks in the dynamic structure factor components
are investigated. Indeed, the experimentally observed Bethe strings in $S^{+-} (k,\omega)$ cannot be expressed 
in terms of configurations of usual spinons. We find that the line shape at
and near the sharp peaks of the spin dynamic structure factor experimentally observed 
in BaCo$_2$V$_2$O$_8$ and SrCo$_2$V$_2$O$_8$ is fully controlled by unbound-unbound and 
unbound-bound scattering of singlet pairs of physical spins $1/2$. 
Our results on both the role of selective interchain couplings in protecting the 1D physics and being behind deviations from it
and on the dynamical properties being controlled by scattering of singlet pairs of physical spins $1/2$ 
open the door to a key advance in the understanding of the physics of the spin 
chains in BaCo$_2$V$_2$O$_8$ and SrCo$_2$V$_2$O$_8$. 
\end{abstract}
\maketitle

\section{Introduction}
\label{SECI}

The spin chains in the zigzag materials BaCo$_2$V$_2$O$_8$ 
and SrCo$_2$V$_2$O$_8$ are systems of considerable scientific interest and intense study
\cite{He_05,He_06,Kimura_07,Okunishi_07,Kimura_08,Suga_08,Canevet_13,Kimura_13,Okutani_15,Klanjsek_15,Grenier_15A,Grenier_15,Wang_18,Shen_19,Wang_19,Horvatic_20,Bera_20,Han_21,Cui_22}.
However, the coexistence in their low-temperature spin-conducting phases of 
one-dimensional (1D) physics with important 
deviations from it is not well understood. 

For instance, magnetization experimental results for these materials are explained well in terms of a 
1D spin-$1/2$ Heisenberg-Ising chain in longitudinal magnetic fields with
anisotropy $\Delta\approx 2$ \cite{Kimura_07,Okunishi_07,Kimura_08,Han_21}. In addition,
for their low-temperature spin-conducting phases,
the magnetic-field dependencies of the energies of the sharp peaks both in the transverse 
components of the spin dynamic structure factor 
observed by optical experiments \cite{Wang_18,Wang_19}
and in the longitudinal component $S^{-+} (k,\omega)$ observed by neutron
scattering \cite{Bera_20} have been quantitatively described by that purely 
1D chain. Such spin-conducting phases occur for longitudinal magnetic fields $h_{c1}<h<h_{c2}$,
where $h_{c1}\approx 3.8$\,T, $h_{c2}\approx 22.9$\,T for BaCo$_2$V$_2$O$_8$ and
$h_{c1}\approx 3.8$\,T, $h_{c2}\approx 28.7$\,T for SrCo$_2$V$_2$O$_8$.
The 1D physics of these zigzag materials also includes the experimental identification of finite-energy sharp peaks in
the transverse component $S^{+-} (k,\omega)$ associated with excited states containing
exotic complex Bethe strings of length two and three \cite{Wang_18,Wang_19,Bera_20} described by the 
exact Bethe-ansatz solution \cite{Gaudin_71,Gaudin_14,Takahashi_99,Yang_19,Carmelo_22}
of the spin-$1/2$ $XXZ$ chain. 

Interchain mean-field theory \cite{Okunishi_07} provides interesting qualitative information on the
physics of BaCo$_2$V$_2$O$_8$ and SrCo$_2$V$_2$O$_8$. However, some of the experimental observations 
\cite{Grenier_15,Shen_19} highlight the complex magnetic properties in these zigzag materials and evidence 
the inadequacy of that theory. This is in part due to their complicated structure of individual nearest-neighbor (NN) 
and next-nearest-neighbor (NNN) interchain couplings \cite{Klanjsek_15}.

The zigzag materials BaCo$_2$V$_2$O$_8$ and SrCo$_2$V$_2$O$_8$ have similar 
chain structures along the $c$-axis, being almost iso-structural. In this paper we use symmetries 
that follow from the one-particle potential transforming according to the underlying lattice symmetries
to clarify issues concerning the coexistence in their low-temperature spin-conducting phases 
of 1D physics with important deviations from it. This is achieveded by accounting for an 
important selection rule. It results from interchain spin states being coupled more strongly 
within the spin dynamical structure factor whenever they are connected by a specific symmetry 
operation of the underlying lattice: In the case of excited states, this symmetry operation is 
only a symmetry in spin-space if {\it no} electronic spin flip is performed within the generation of such states.

The corresponding selective interchain couplings protect the 1D physics 
of the components $S^{+-} (k,\omega)$ and $S^{-+} (k,\omega)$, which are associated with excited states 
that involve an electronic spin flip. For such states interchain coupling should tend to zero or be very small.
On the other hand, such selective interchain couplings are found to be behind deviations from
the 1D physics associated with an enhancement of the spectral-weight intensity of 
$S^{zz} (k,\omega)$ whose excitations do not involve electronic spin flips.

The latter enhancement is then found to be behind the experimental low-energy dominance of the 
longitudinal nuclear-magnetic-resonance (NMR) relaxation rate 
term $1/T_1^{\parallel} \propto \sum_k\vert A_{\parallel} (k)\vert^2 S^{zz} (k,\omega_0)$
for the whole magnetic-field interval  $h\in [h_{c1},h_{c2}]$ of the spin-conducting phases\cite{Klanjsek_15,Cui_22}.
For magnetic fields $h\in [h_*,h_{c2}]$ where the transverse
term ${1\over T_1^{\perp}} \propto  \sum_k\vert A_{\perp} (k)\vert^2 (S^{+-} (k,\omega_0)+S^{-+} (k,\omega_0))$
is supposed to dominate, this contradicts the 1D physics \cite{Dupont_16}. 
The experimental values of $h_*$ for BaCo$_2$V$_2$O$_8$ and 
SrCo$_2$V$_2$O$_8$ suggested by neutron scattering read 
$h_* \approx 8.5$\,T and $h_* \approx 7.0$\,T, respectively \cite{Grenier_15,Shen_19}.

To further understand the role of selective interchain couplings concerning their contradictory effects in
protecting the 1D physics and controlling deviations from it, the physical-spins scattering processes
behind the experimentally observed sharp peaks in the dynamic structure factor components
are investigated. Our results clarify the microscopic processes 
in terms of scattering of physical spins $1/2$ configurations that control and determine the line shape at and near the 
experimentally observed sharp peaks of the spin dynamical structure factor \cite{Wang_18,Wang_19,Bera_20}. 
To describe such scattering processes, we use 
an exact representation of the spin-$1/2$ $XXZ$ chain in a longitudinal magnetic field
$h\in [h_{c1},h_{c2}]$ in terms of both singlet pairs of physical spins $1/2$
and unpaired physical spins $1/2$ that is valid for the whole Hilbert space \cite{Carmelo_22}. 

That physical-spins representation is a generalization for 
anisotropy $\Delta >1$ of that used for the $\Delta =1$ isotropic point of the
spin-$1/2$ Heisenberg chain \cite{Carmelo_15,Carmelo_17}. For anisotropy $\Delta >1$, the spin projection $S^z$ remains a good 
quantum number whereas spin $S$ is not. It is replaced by the $q$-spin $S_q$ in the eigenvalue 
of the Casimir generator of the continuous $SU_q(2)$ symmetry \cite{Pasquier_90}. 
Concerning that symmetry, the issue that matters for our present study is that 
$q$-spin $S_q$ has exactly the same values for anisotropy $\Delta >1$ as spin $S$ for $\Delta =1$. 
This includes their relation to the values of $S^z$. Hence singlet and multiplet refer in this paper to 
physical spins configurations with zero and finite $S_q$, respectively.

One of the reasons for our use of the physical-spins representation is that
the $S^{+-} (k,\omega)$'s Bethe strings of lengths two and three experimentally identified and realized in 
SrCo$_2$V$_2$O$_8$ and BaCo$_2$V$_2$O$_8$ \cite{Wang_18,Wang_19} cannot be expressed 
in terms of configurations of usual spinons. On the other hand, within the physical-spins representation
the unbound elementary magnetic configurations described by $n=1$ single real Bethe rapidities
and the $n=2,3,...$ bound elementary magnetic configurations described by
Bethe $n$-strings are singlet $S^z = S_q =0$ pairs of physical spins $1/2$. 

By the use of a dynamical theory that accounts for the scattering processes 
of unbound-unbound pairs and unbound-bound pairs of physical spins $1/2$ \cite{Carmelo_22} 
(see Appendix \ref{C} for a summary of that theory), 
we derive expressions for the line shape near the sharp peaks that are experimentally observed in the
spin dynamic structure factor for BaCo$_2$V$_2$O$_8$ and SrCo$_2$V$_2$O$_8$ \cite{Wang_18,Wang_19,Bera_20}.

That dynamical theory is similar to that used for the isotropic point $\Delta =1$ \cite{Carmelo_20}.
The theory belongs to the same general class as that introduced in Ref. \onlinecite{Carmelo_05} for
another integrable model. The latter is a generalization to the whole interaction range 
of an approach used for the infinite interaction limit \cite{Karlo_97}. 
For integrable problems, such a class of dynamical theories is 
equivalent to the mobile quantum impurity model scheme \cite{Imambekov_09,Imambekov_12}, 
accounting for exactly the same microscopic elementary excitation processes \cite{Carmelo_18}.
In the low-energy limit, that dynamical theory recovers the corresponding operator description \cite{Carmelo_94}.
Momentum-dependent exponents in the expressions
of dynamical correlation functions have also been obtained by other methods \cite{Sorella_96,Sorella_98}.

Thus the motivation and main results of this paper are: 1) The physical origin in terms
of selective interchain couplings of the coexistence in BaCo$_2$V$_2$O$_8$ and SrCo$_2$V$_2$O$_8$ of 1D 
physics with important deviations from it; 2) The further understanding of the dynamical properties BaCo$_2$V$_2$O$_8$ and 
SrCo$_2$V$_2$O$_8$ in the low-temperature spin-conducting phases by clarifying the 
role in them of scattering of both unbound and bound singlet pairs of physical spins $1/2$.

It is convenient to start by comparing the experimental data on the dynamical properties of
BaCo$_2$V$_2$O$_8$ and SrCo$_2$V$_2$O$_8$ with their theoretical descriptions 
involving physical-spins scattering to clarify which properties refer to 1D physics and 
deviates from it, respectively. To reach this goal, we use the above mentioned 
suitable physical-spins representation. After handling such issues, we then address that of the
role of selective interchain couplings in the physics of the zigzag materials under study.

The paper is organized as follows. The physical-spins representation used
in our studies is introduced Sec. \ref{SECII}. In Sec. \ref{SECIII} the scattering processes in terms of physical spin $1/2$ configurations
that control the line shapes at and near the sharp peaks in the spin dynamical structure
factor experimentally observed in the zigzag materials for fields $h_{c1}<h<h_{c2}$
are studied. The effects of selective interchain couplings concerning both the protection 
of the 1D physics of BaCo$_2$V$_2$O$_8$ and SrCo$_2$V$_2$O$_8$ and important experimental 
deviations from it is the issue addressed in Sec. \ref{SECIV}.
The concluding remarks are presented in Sec. \ref{SECV}. In addition, in Appendix \ref{A} some basic quantities needed
for the studies of this paper are provided, in Appendix \ref{B} the applicability of the physical-spins representation
to the whole Hilbert space is discussed,
and a summary of the dynamical theory used in our studies is presented in Appendix \ref{C}.

\section{The physical-spins representation}
\label{SECII}

\subsection{The 1D quantum problem and its representation}
\label{SECIIA}

We start by describing the superexchange interactions between the magnetic moments in the spin chains of 
BaCo$_2$V$_2$O$_8$ and SrCo$_2$V$_2$O$_8$ by the Hamiltonian 
of the spin-$1/2$ Heisenberg-Ising chain \cite{Wang_18,Yang_19,Bera_20}. It describes 
$N=\sum_{\sigma =\uparrow,\downarrow}N_{\sigma}$ physical spins $1/2$ of projection $\sigma =\uparrow,\downarrow$. 
For the anisotropy parameter range $\Delta=\cosh\eta\geq 1$ and thus $\eta\geq 0$,
spin densities $m = 2m^z = (N_{\uparrow}-N_{\downarrow})/N \in [0,1]$, exchange integral $J$, and length $L\rightarrow\infty$
for $N/L$ finite, that Hamiltonian in a longitudinal magnetic field $h$ becomes,
\begin{eqnarray}
\hat{H}_{\parallel} & = & \hat{H}_{\Delta} + g\mu_B\,h\sum_{j=1}^{N} \hat{S}_j^z
\hspace{0.40cm}{\rm where}
\nonumber \\
\hat{H}_{\Delta} & = & J\sum_{j=1}^{N}\left({\hat{S}}_j^x{\hat{S}}_{j+1}^x + {\hat{S}}_j^y{\hat{S}}_{j+1}^y + 
\Delta\,{\hat{S}}_j^z{\hat{S}}_{j+1}^z\right) \, .
\label{Hphi}
\end{eqnarray}

Here $\hat{\vec{S}}_{j}$ is the spin-$1/2$ operator at site $j=1,...,N$ with components $\hat{S}_j^{x,y,z}$
and $\mu_B$ is the Bohr magneton. 
For $\Delta >1$, spin-insulating, spin-conducting, and fully-polarized ferromagnetic quantum phases
occur for spin density $m=0$ and magnetic fields $0\leq h<h_{c1}$, spin
densities $0<m<1$ and fields $h_{c1}< h<h_{c2}$, and spin density $m=1$ and fields $h> h_{c2}$, respectively.
The critical fields $h_{c1}$ and $h_{c2}$ have known Bethe-ansatz expressions \cite{Takahashi_99} given
in Eq. (\ref{hc1c2}) of Appendix \ref{A}. 

In this paper the $h\rightarrow h_{c1}$ and $h\rightarrow h_{c2}$ limits are from $h> h_{c1}$
and $h< h_{c2}$ values, respectively, and we use natural units in which the lattice spacing and the 
Planck constant are equal to one. 

By using the $SU_q(2)$ symmetry algebra, we find that each energy eigenstate with 
$q$-spin in the range $0\leq S_q\leq N/2$ is populated by physical spins $1/2$ in two types of configurations \cite{Carmelo_22}:
A set of $M=2S_q$ physical spins $1/2$ that participate in a multiplet configuration, and a complementary 
set of even number $2\Pi=N-2S_q$ of physical
spins $1/2$ that participate in singlet configurations. This holds for {\it all} $2^N$ energy eigenstates.
The {\it unpaired spins $1/2$} and {\it paired spins $1/2$} are the members of such two sets of $M = 2S_q$ and 
$2\Pi = N-2S_q$ physical spins $1/2$, respectively. 

Within the corresponding representation in terms of the $N$ physical spins $1/2$ described by the
Hamiltonian, Eq. (\ref{Hphi}), the designation {\it $n$-pairs} refers both to
{\it $1$-pairs} and {\it $n$-string-pairs} for $n>1$: 

- The internal degrees of freedom of a $1$-pair correspond to one unbound singlet
pair of physical spins $1/2$. It is described by a $n=1$ single real Bethe rapidity. Its translational 
degrees of freedom refer to the $1$-band momentum $q_j \in [q_1^-,q_1^+]$ where $j = 1,...,L_1$ 
carried by each such a pair.

- The internal degrees of freedom of a 
$n$-string-pair refer to a number $n>1$ of singlet pairs of physical spins $1/2$. They are
bound within a configuration described by a corresponding complex Bethe $n$-string. Its translational degrees of freedom 
refer to the $n>1$ $n$-band momentum $q_j \in [q_n^-,q_n^+]$ where $j = 1,...,L_n$ carried by each such a $n$-pair.

For each $n$-band, the $q_j$'s have for both $n=1$ and $n>1$ discrete values $q_j \in [q_n^-,q_n^+]$ with 
separation $q_{j+1}-q_j = {2\pi\over L}$. Here $j=1,...,L_n$ and $L_n = N_n + N^h_{n}$ where $N_n$ is the number of occupied
$q_j$'s and thus of $n$-pairs and $N^h_{n} = 2S_q +\sum_{n'=n+1}^{\infty}2(n'-n)N_{n'}$ that of unoccupied $q_j$'s 
and thus of $n$-holes. The present results refer to the thermodynamic limit. In that limit, the 
the number $2\Pi = N-2S_q$ of paired physical spins $1/2$ of an energy eigenstate
can be exactly expressed as \cite{Carmelo_22}, $2\Pi = \sum_{n=1}^{\infty}2n\,N_n$.

The Bethe-ansatz quantum numbers \cite{Gaudin_71} $I_j^n$ are actually
the $n$-band momentum values $q_j = {2\pi\over L}I_j^n$
in units of ${2\pi\over L}$. They are given by $I_j^n = 0,\pm 1,...,\pm {L_n -1\over 2}$ for $L_n$ odd
and $I_j^n =\pm 1/2,\pm 3/2,...,\pm {L_n -1\over 2}$ for $L_n$ even. 
Such numbers and thus the set $\{q_j\}$ of $n$-band discrete momentum values have Pauli-like 
occupancies: The corresponding momentum distributions read $N_n (q_j) = 1$ and 
$N_n (q_j) = 0$ for occupied and unoccupied $q_j$'s, respectively.

The energy eigenvalues are specified by the set of $n=1,...,\infty$ 
distributions $\{N_n (q_j)\}$ and described by a corresponding set of rapidity functions $\{\varphi_{n} (q_{j})\}$
defined by Bethe-ansatz equations \cite{Gaudin_71,Carmelo_22}.
Such functions are the real part of corresponding $n=1$ real and
$n>1$ complex rapidities \cite{Gaudin_71,Carmelo_22}. 

The $q_j$'s of ground states and excited states that contribute to the dynamical properties
can in the thermodynamic limit be described by continuous variables $q \in [q_n^-,q_n^+]$. Here
$q_1^{\pm} = \pm k_{F\uparrow}$ and $q_n^{\pm} = \pm (k_{F\uparrow}-k_{F\downarrow})$ for $n>1$
where $k_{F\uparrow} = {\pi\over 2}(1+m)$ and $k_{F\downarrow} = {\pi\over 2}(1-m)$.
Ground states refer to a $1$-band Fermi sea $q \in [-k_{F\downarrow},k_{F\downarrow}]$ 
with $1$-holes for $\vert q\vert\in [k_{F\downarrow},k_{F\uparrow}]$ and empty $n$-bands for $n>1$
with $n$-holes for $q' \in [-(k_{F\uparrow}-k_{F\downarrow}),(k_{F\uparrow}-k_{F\downarrow})]$.
In real space, a ground-state $1$-band momentum $q$ occupied by one 
unbound singlet pair of physical spins $1/2$ 
refers to a superposition of local configurations with the 
weight decreasing with increasing lattice distance between the two paired physical spins.

In addition to the $2\Pi=N-2S_q$ paired physical spins $1/2$ in the $\Pi=N/2-S_q$ $n$-pairs singlet configurations,
the representation accounts for the remaining $M=2S_q$ unpaired physical spins $1/2$ 
of an energy eigenstate: The question is where in the Bethe-ansatz solution are the $M = 2S_q$ 
unpaired physical spins $1/2$? The clarification of this issue involves a squeezed space 
construction \cite{Carmelo_22,Kruis_04}. 

This issue involves the description of the translational degrees of freedom
and spin internal degrees of freedom of the $M=2S_q$ unpaired physical spins $1/2$,
which is addressed in Appendix \ref{B}. That the physical-spins representation
accounts for the latter internal degrees of freedom in shown in that Appendix 
to ensure it applies to the whole Hilbert space.

Indeed, the Bethe ansatz refers only to subspaces spanned either by the 
highest weight states (HWSs) or the lowest weight states (LWSs) of the $SU_q(2)$ symmetry \cite{Gaudin_71,Carmelo_22}.	
For such states, all the $M=2S_q$ unpaired physical spins $1/2$ 
have the same $\uparrow$ or $\downarrow$, respectively,
spin projection. This implies that $S^z = S_q$ and $S^z = -S_q$, respectively.
In this paper we use a HWS Bethe ansatz. 

Finally, concerning representations of spin chains other than the physical spins representation used in this paper,
the most often used are in terms of spinons \cite{Caux_08} at vanishing spin density $m=0$ 
and psinons and antipsinons for finite spin density $0<m<1$ \cite{Karbach_02}.
In the thermodynamic limit they are well defined in subspaces with no $n$-strings 
or with a vanishing density of $n$-strings. 

Spinons are $1$-holes within excited energy eigenstates of the $m=0$ ground state. Psinons and antipsinons 
are $1$-holes that emerge or are moved to inside the $1$-band Fermi sea and 
$1$-pairs that emerge or are moved to outside that sea, respectively. They occur 
in excited energy eigenstates of ground states 
corresponding to spin-conducting quantum phases for $h_{c1}<h<h_{c2}$. 

However, such representations do not describe the spin configurations of Bethe strings
and the dynamical properties of the present quantum system are naturally 
and directly described by physical-spins $n$-pairs scattering.

\subsection{The $n$-pair energy dispersions}
\label{SECIIB}

Important quantities of the physical spins representation 
are the energy dispersions $\varepsilon_{n} (q)$ of the $n$-pairs
given in Eqs. (\ref{equA4n})-(\ref{equA6}) of Appendix \ref{A}. The expressions of the spectra 
of the spin dynamic structure factor components considered below in Sec. \ref{SECIII}
are expressed in terms of such energy dispersions for $n=1,2,3$. 
Indeed, only Bethe strings of length two and three contribute to that factor.
The corresponding $n=2$ and $n=3$ $n$-string-pair energy dispersions $\varepsilon_{n} (q')$ are plotted 
in units of $J$ in Figs. \ref{figure12PR} and \ref{figure13PR} of Appendix \ref{A}, respectively, as a function of $q'/\pi$ for
$n$-band momentum $q'\in [-(k_{F\uparrow} - k_{F\downarrow}),(k_{F\uparrow} - k_{F\downarrow})]$, 
spin densities $m=0.2$, $m=0.5$, $m=0.8$, and several anisotropy values.

The energy dispersion $\varepsilon_{1} (q)$ is plotted in Fig. 1 of Ref. \onlinecite{Carmelo_22}.
For $n=1$, that dispersion $\varepsilon_{1} (q)>0$ and minus it $-\varepsilon_{1} (q)>0$ are
for $\vert q\vert \in [k_{F\downarrow},k_{F\uparrow}]$ and $q \in [-k_{F\downarrow},k_{F\downarrow}]$
the energy required to create in a ground state for fields $h_{c1}<h<h_{c2}$
one $1$-pair and one $1$-hole, respectively. As mentioned above, ground states are not populated by $n>1$ $n$-string pairs.
Their energy dispersion $\varepsilon_{n} (q')>0$ is the energy required to create in a ground state
one $n$-pair of $n$-band momentum $q' \in [-(k_{F\uparrow}-k_{F\downarrow}),(k_{F\uparrow}-k_{F\downarrow})]$. 

For $n\geq 1$ the zero-energy level of the dispersions $\varepsilon_{n} (q)$
refers to that of the ground state corresponding to a given fixed value of the longitudinal magnetic field. 
The related $n\geq 1$ energy dispersions $\varepsilon_{n}^0 (q)$ differ from $\varepsilon_{n} (q)$
in the zero-energy level: it corresponds to that of the $h=0$ absolute ground state.
However, relative to a ground state for a given fixed field value 
$h_{c1}<h<h_{c2}$ of the spin-conducting phases, the energy $-E_{1}^{\uparrow\downarrow} (h)>0$ where,
\begin{equation}
E_{1}^{\uparrow\downarrow} (h) = \varepsilon_{1}^0 (k_{F\downarrow}) = - g\mu_B\,h
\hspace{0.20cm}{\rm for}\hspace{0.20cm}h_{c1}<h<h_{c2}  \, ,
\label{Eminm}
\end{equation}
is the excitation energy for the annihilation of one $1$-pair giving rise to two physical spins of {\it opposite} projection, 
whereas $-\varepsilon_{1} (k_{F\downarrow})=0$, where $\varepsilon_{1} (q) = \varepsilon_{1}^0 (q) + g\mu_B\,h$
for $h_{c1}<h<h_{c2}$, is the vanishing energy for the annihilation of one $1$-pair leading 
to two physical spins with the {\it same} $\uparrow$ projection. 

On the other hand, the energy dispersions of $n$-string-pairs can for spin densities $0\leq m\leq 1$ be written as,
\begin{eqnarray}
\varepsilon_{n} (q') & = & \varepsilon_{n}^0 (q') + n\,g\mu_B\,h\hspace{0.20cm}{\rm where}
\nonumber \\
\varepsilon_{n}^0 (q') & = & E_{{\rm bind},n} + T_{n} (q') 
\nonumber \\
E_{{\rm bind},n} & = & \varepsilon_{n}^0 (0) < 0 \hspace{0.20cm}{\rm and}
\nonumber \\
T_{n} (q') & = & \varepsilon_{n}^0 (q') - \varepsilon_{n}^0 (0) = \varepsilon_{n} (q') - \varepsilon_{n} (0) \geq 0 \, .
\label{EbnTn}
\end{eqnarray}
Here the binding energy $E_{{\rm bind},n}$ and the energy $T_{n} (q')$ refer to the  internal and translation degrees of freedom, 
respectively, of a $n$-string pair. Each of the $n>1$ energy terms $g\mu_B\,h$ of the additional magnetic energy $n\,g\mu_B\,h$ 
is associated with creation of one physical spin pair. It
can either result from the energy $-E_{1}^{\uparrow\downarrow} (h)=g\mu_B\,h$ associated with
the annihilation of one $1$-pair giving rise to two physical spins of opposite projection or to
the energy $g\mu_B\,h$ needed to flip one ground-state unpaired physical spin $1/2$, Eq. (\ref{EEGS}) for $n_z=1$,
which pairs with another ground-state unflipped unpaired physical spin $1/2$.

In the case of creation of one $2$-pair and one $3$-pair to generate the $2$-string and $3$-string excited states, respectively,
considered below in Sec. \ref{SECIII} that contribute to $S^{+-} (k,\omega)$, one unpaired physical spin $1/2$ is flipped and one $1$-pair and
two $1$-pairs, respectively, are annihilated. In the case of creation of one $2$-pair to generate the $2$-string 
excited states also considered in that section that contribute to $S^{zz} (k,\omega)$, two $1$-pairs are annihilated and no unpaired 
physical spin $1/2$ is flipped.

Analytical expressions valid in the two limiting cases (i) $h \in [0,h_{c1}]$ and $m=0$ and (ii) for $h\rightarrow h_{c2}$ and $m\rightarrow 1$,
respectively, of the energy dispersions $\varepsilon_{n} (q)$ and $\varepsilon_{n}^0 (q)$ for $n\geq 1$,
binding energy $E_{{\rm bind},n}$ for $n>1$, and energy $T_{n} (q')$ for $n>1$ are given in 
Eqs. (\ref{vareband1m0m1})-(\ref{BnTnexpresshc2}) of Appendix \ref{A}.
\begin{figure}
\begin{center}
\centerline{\includegraphics[width=8.5cm]{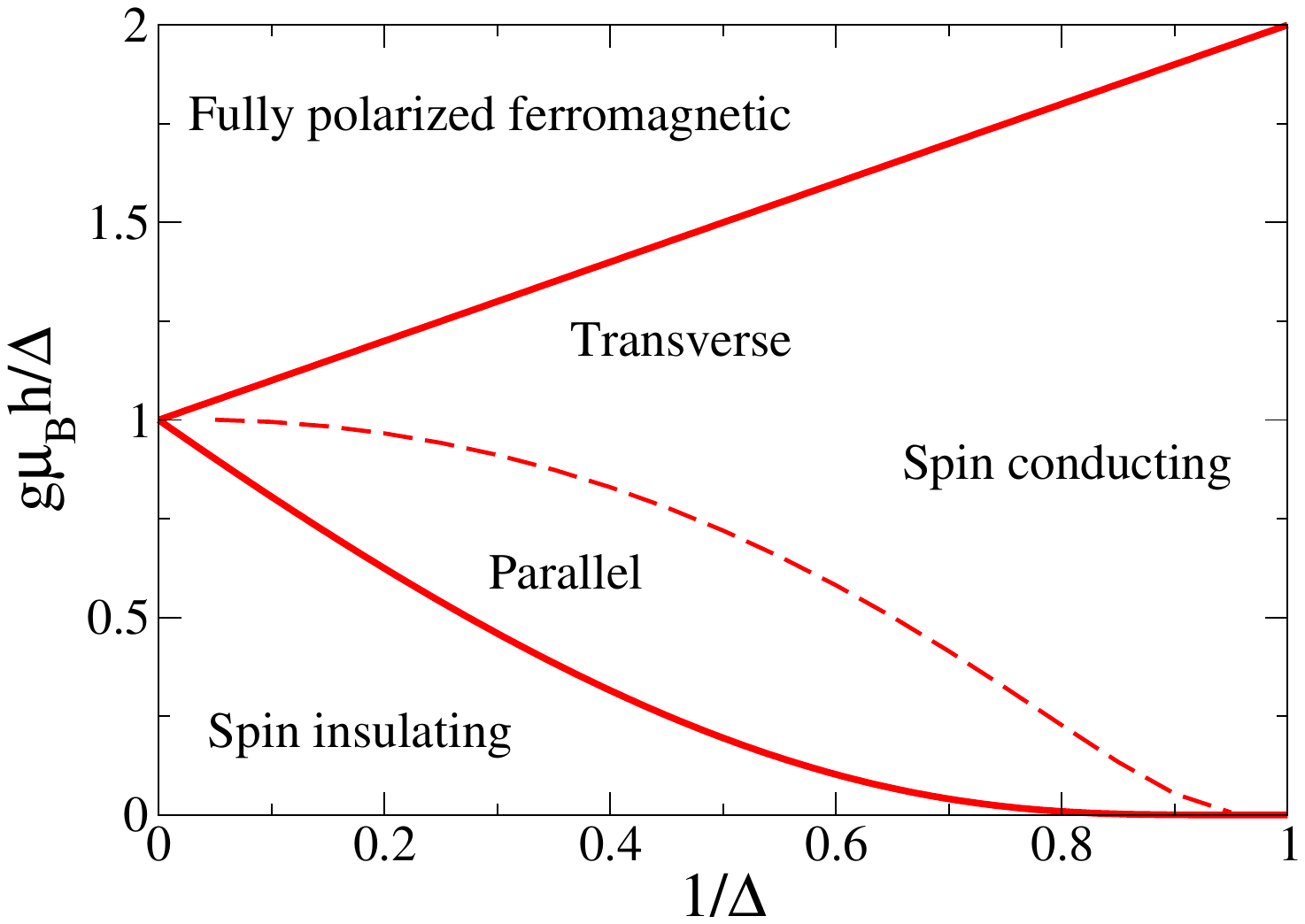}}
\caption{The spin-$1/2$ $XXZ$ chain phase diagram of the magnetic energy over anisotropy, $g\mu_B h/\Delta$, 
in units of $J$ versus inverse anisotropy $\epsilon = 1/\Delta\in [0,1]$.
The energy absolute value over anisotropy lines (a) 
$\vert E_{1}^{\uparrow\downarrow} (h_{c1}^+)\vert/\Delta =\lim_{h\rightarrow h_{c1}}\vert E_{1}^{\uparrow\downarrow}(h)\vert/\Delta$,
(b) $\vert E_{1}^{\uparrow\downarrow} (h_*)\vert/\Delta$, and (c) 
$\vert E_{1}^{\uparrow\downarrow} (h_{c2}^-)\vert/\Delta = \lim_{h\rightarrow h_{c2}}\vert E_{1}^{\uparrow\downarrow}(h)\vert/\Delta$ 
separate (a) the spin-insulating phase from the spin-conducting phase with dominant longitudinal
relaxation-rate fluctuations, (b) the latter from the spin-conducting phase with dominant transverse
relaxation-rate fluctuations, and (c) the latter from the fully-polarized ferromagnetic phase, respectively.}
\label{figure1PR}
\end{center}
\end{figure}

It follows from the relation, Eq. (\ref{Eminm}), that the critical fields $h_{c1}$ and $h_{c2}$
are given by $h_{c1}= \lim_{m\rightarrow 0}\vert E_{1}^{\uparrow\downarrow}\vert/g\mu_B$
and $h_{c2}= \lim_{m\rightarrow 1}\vert E_{1}^{\uparrow\downarrow}\vert/g\mu_B$, respectively.
As illustrated in Fig. \ref{figure1PR}, such two limits of $\vert E_{1}^{\uparrow\downarrow}\vert/g\mu_B$
fully control the spin-$1/2$ $XXZ$ chain phase diagram of the magnetic energy 
over anisotropy, $g\mu_B h/\Delta$, versus $\epsilon = 1/\Delta\in [0,1]$. 
The middle dashed line in that diagram refers to $g\mu_B h_*/\Delta$ where
the magnetic field $h_{*}=\vert E_{1}^{\uparrow\downarrow}\vert_{m_*}/g\mu_B$
and the corresponding spin density $m_*$ are those
at which for the purely 1D spin-$1/2$ $XXZ$ chain
the parameter $\xi$ in Eq. (\ref{x-aa}) of Appendix \ref{A} reads $\xi =1/\sqrt{2}$.
As discussed below in Sec. \ref{SECIV} and illustrated in that figure, $h_*$ 
separates the field regions $h_{c1}<h<h_*$ and $h_{*}<h<h_{c2}$ where the 
longitudinal and and transverse term of the NMR relaxation rate $1/T_1$ dominates, respectively. 
Another reference magnetic field of interest is $h_{1/2}=\vert E_{1}^{\uparrow\downarrow}\vert_{m=1/2}/g\mu_B$.
It refers to spin density $m=1/2$ and defines the field intervals $h\in [h_{c1},h_{1/2}]$ 
and $h\in [h_{1/2},h_{c2}]$ for which some of the sharp peaks studied below in Sec. \ref{SECIII} exist. 

As in other 1D spin systems \cite{Sacramento_95}, the
magnetic energy $g\mu_B h_{c1} = \lim_{m\rightarrow 0}\vert E_{1}^{\uparrow\downarrow}\vert$ 
equals a minimum $h=0$ spin energy gap, in the present case that 
of the transverse spin dynamic structure factor \cite{Takahashi_99,Caux_08}. 
The parameter sets $\Delta =2.17$, $J=2.60$ meV, and $g =6.2$ for BaCo$_2$V$_2$O$_8$ for BaCo$_2$V$_2$O$_8$ and
$\Delta =2.00$, $J=3.55$ meV, and $g =6.2$ for SrCo$_2$V$_2$O$_8$  \cite{Kimura_07,Suga_08,Canevet_13,Wang_18}
have been chosen so that $h (m) = \vert E_{1}^{\uparrow\downarrow}\vert/g\mu_B$ gives for $m\rightarrow 0$,
$m=1/2$, and $m\rightarrow 1$ the experimental values for $h_{c1}$, $h_{1/2}$,
and $h_{c2}$, respectively. Indeed, $E_{1}^{\uparrow\downarrow} (h) = \varepsilon_{1}^0 (k_{F\downarrow})$, Eq. (\ref{Eminm}),
can be expressed in terms of known Bethe-ansatz quantities \cite{Gaudin_71,Gaudin_14,Takahashi_99}: See 
Eqs. (\ref{equA4n}) and (\ref{equA4n10})-(\ref{equA6}) of Appendix \ref{A} for $n=1$.

\section{The dynamical properties of the two zigzag materials for fields $h_{c1}<h<h_{c2}$}
\label{SECIII}

\subsection{Sharp peaks in the $(k,\omega)$-plane}
\label{SECIIIA}

Electronic spin resonance measurements can detect the spin dynamic structure
factor components of SrCo$_2$V$_2$O$_8$ and BaCo$_2$V$_2$O$_8$ 
only at specific momentum values $k=0$, $k=\pi/2$, $k=\pi$,
and $k=3\pi/2$ \cite{Wang_18,Wang_19}. Due to inversion symmetry, the 
momentum values $k=\pi/2$ and $k=3\pi/2$ are equivalent. In addition,
the excitations that are allowed in such optical experiments obey the selection 
rules $\delta S^z = \pm 1$, which limits the corresponding studies to sharp peaks 
in the transverse components $S^{+-} (k,\omega)$ and $S^{-+} (k,\omega)$.
On the other hand, sharp peaks of $S^{zz} (k,\omega)$ have been studied by neutron 
scattering in SrCo$_2$V$_2$O$_8$ \cite{Bera_20}.

In Refs. \onlinecite{Wang_18} and \onlinecite{Wang_19} it was shown that 
for the parameter sets suitable to SrCo$_2$V$_2$O$_8$ and BaCo$_2$V$_2$O$_8$,
respectively, the frequencies/energies of the sharp peaks experimentally observed in $S^{+-} (k,\omega)$ and $S^{-+} (k,\omega)$
by high-resolution terahertz spectroscopy agree with those predicted for the spin-conducting 
phases of the spin-$1/2$ $XXZ$ in a longitudinal magnetic field. The same applies to
the sharp peaks observed in $S^{zz} (k,\omega)$ by neutron scattering \cite{Bera_20}.
However, no analytical expressions for the line shapes at and near the sharp peaks were given in 
previous studies for finite-size systems, only the energies of such peaks \cite{Wang_18,Wang_19,Bera_20,Yang_19}.
\begin{figure*}
\includegraphics[width=0.45\textwidth]{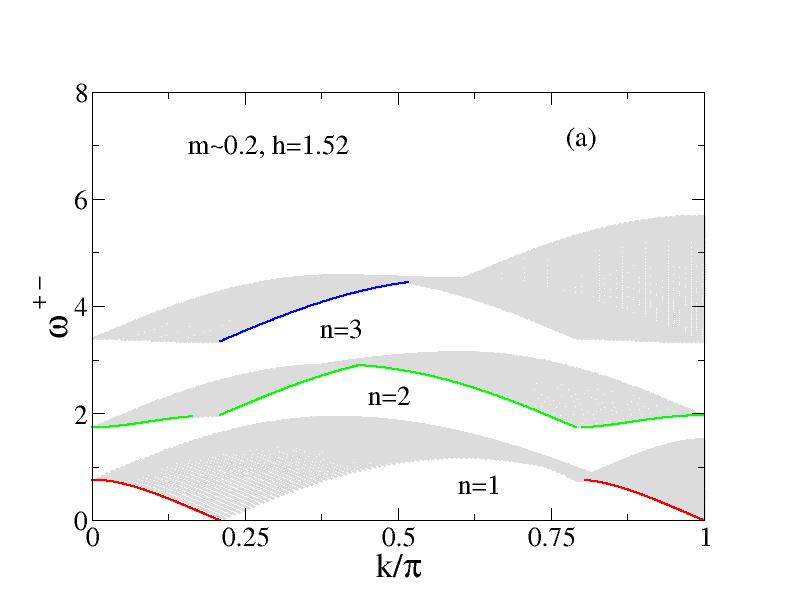}
\includegraphics[width=0.45\textwidth]{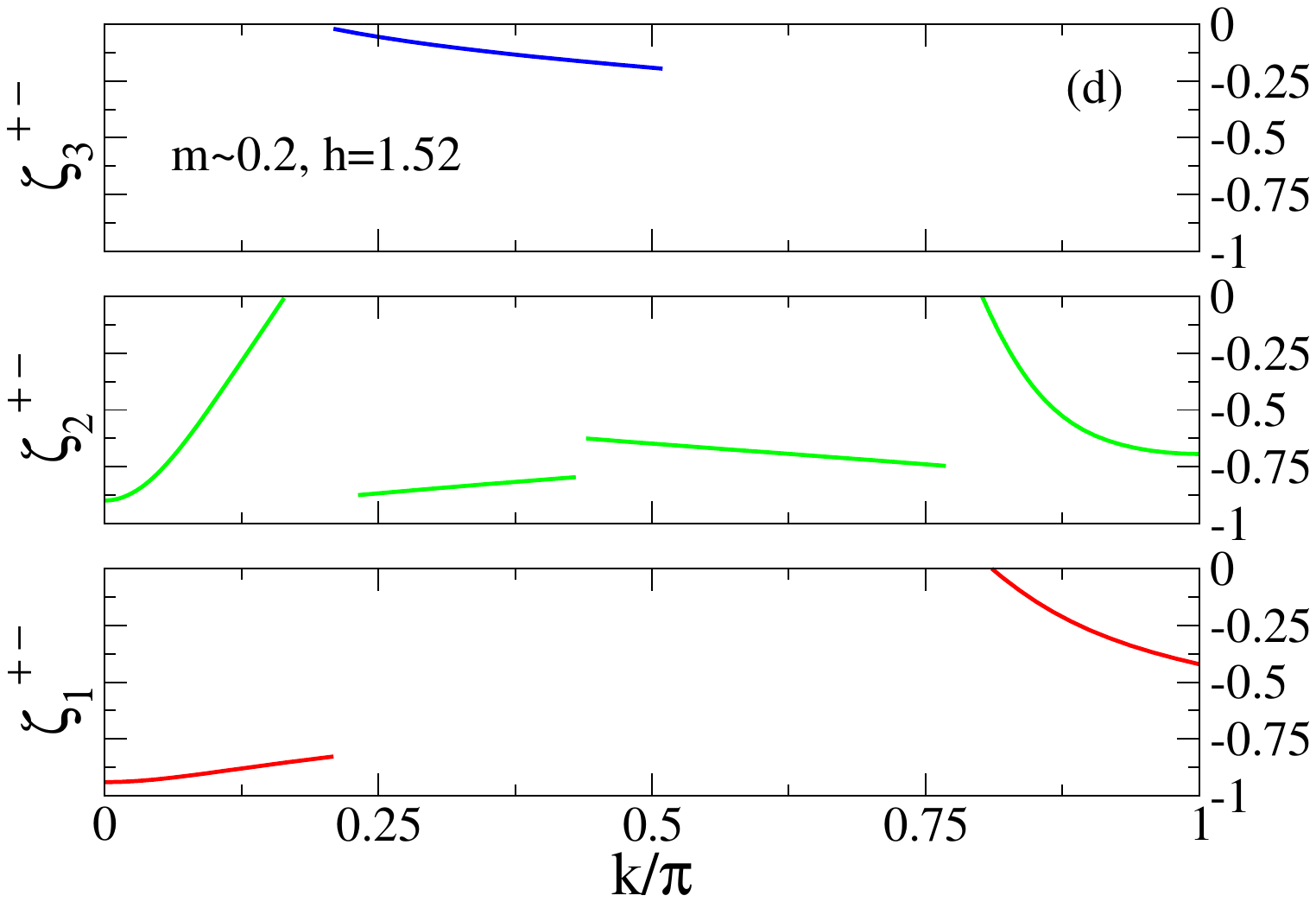}
\includegraphics[width=0.45\textwidth]{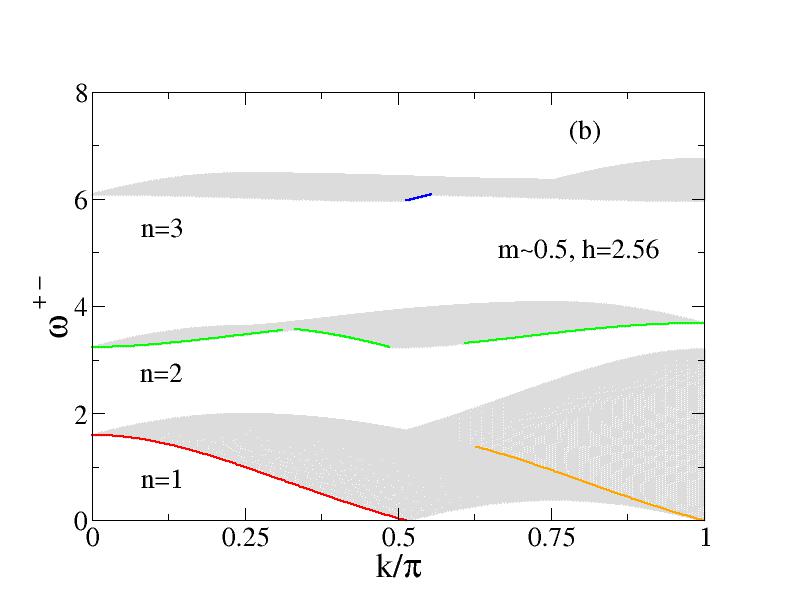}
\includegraphics[width=0.45\textwidth]{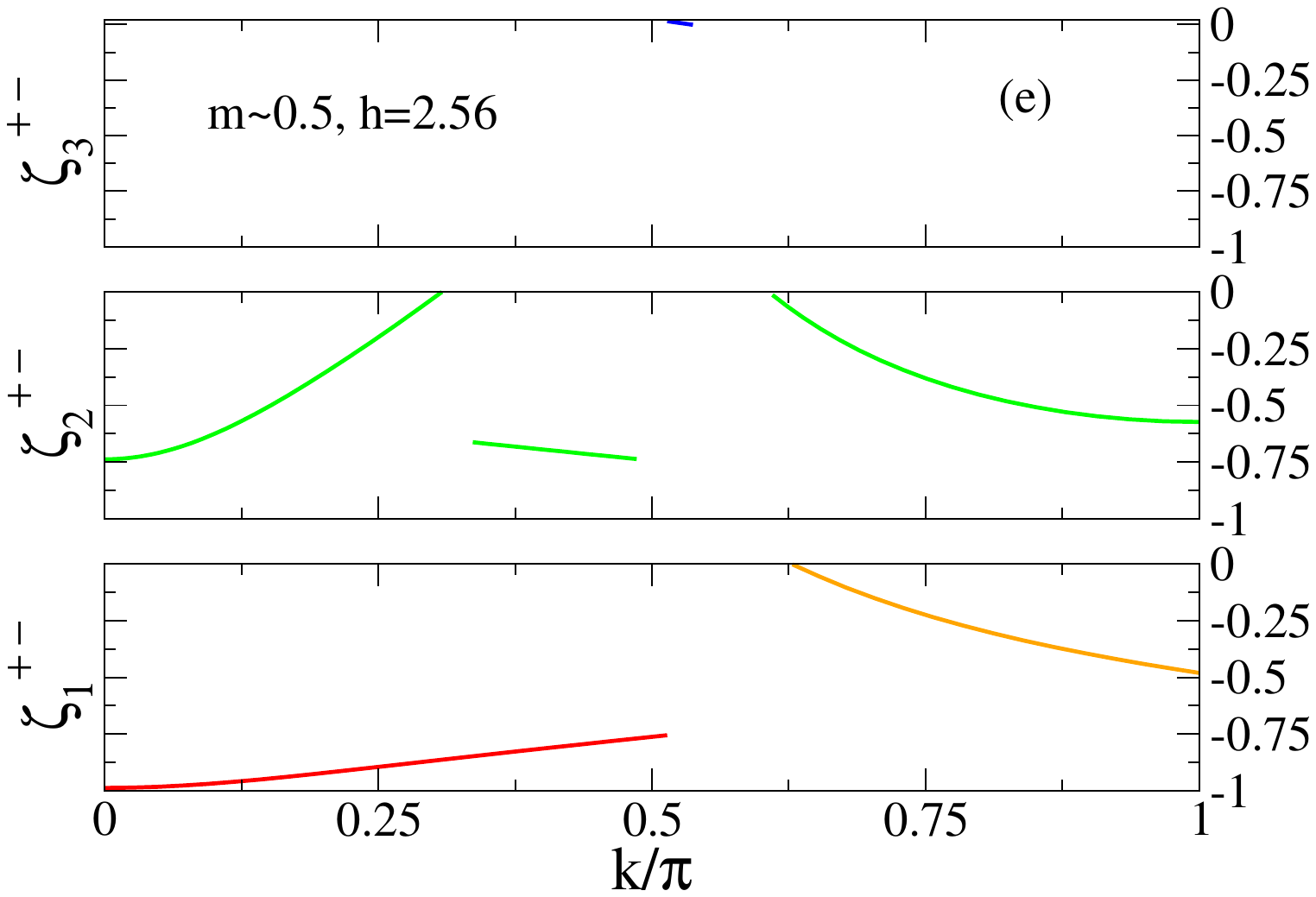}
\includegraphics[width=0.45\textwidth]{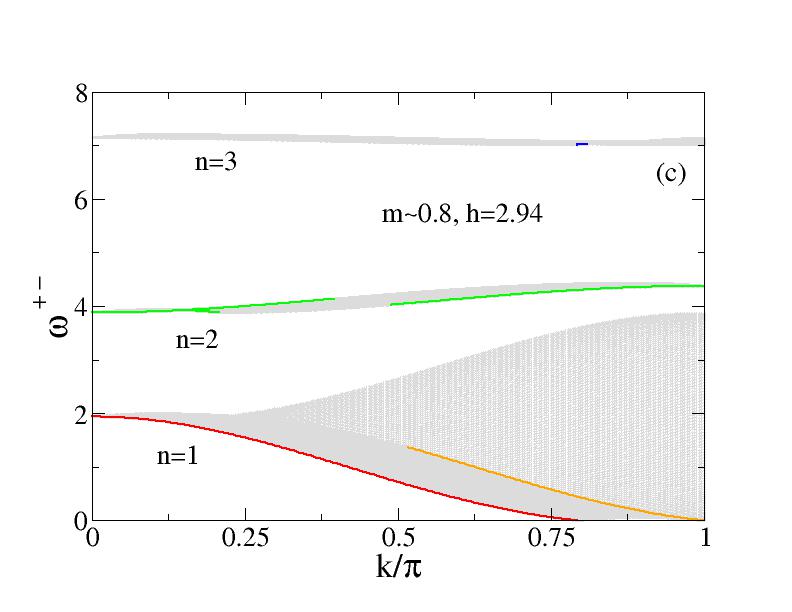}
\includegraphics[width=0.45\textwidth]{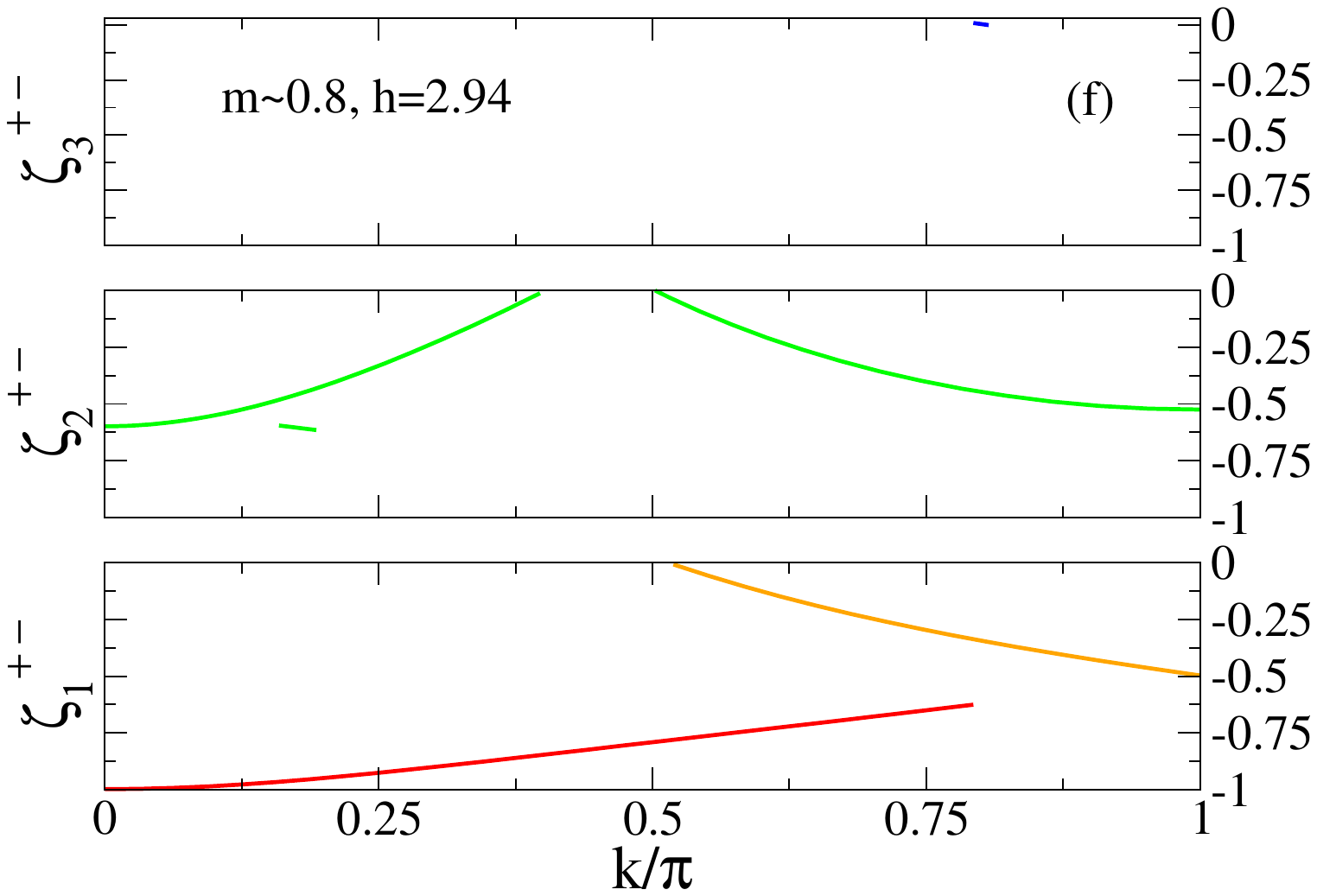}
\caption{\label{figure2PR}
The $(k,\omega)$-plane (a) $n=1$, (b) $n=2$, and (c) $n=3$ $n$-continua where in the thermodynamic limit there is significant
spectral weight in $S^{+-} (k,\omega)$ for the spin-$1/2$ Heisenberg-Ising chain with anisotropy $\Delta =2$ in
a longitudinal magnetic field. Very similar spectra are obtained for anisotropy $\Delta =2.17$.
The corresponding negative $k$ dependent exponents that
control the line shape $S^{+-} (k,\omega)\propto (\omega - E^{+-}_n (k))^{\zeta^{+-}_n (k)}$ 
in the $k$ intervals near the lower thresholds of such continua (d)-(f). The spin densities 
in (a),(b), and (c) are $m=0.209\approx 0.2$, $m=0.514\approx 0.5$, and $m=0.793\approx 0.8$, respectively. The corresponding
$h$ values are given in units of $J/(g\mu_B)$. The exponents are negative
in the $k$ intervals of the $n$-continua lower thresholds marked in the spectra (a)-(c)
and near the branch line running through the $1$-continuum in (b) and (c). On the marked lines in the 
$(k,\omega)$-plane $S^{+-} (k,\omega)$ displays sharp peaks.}
\end{figure*} 
\begin{figure*}
\includegraphics[width=0.45\textwidth]{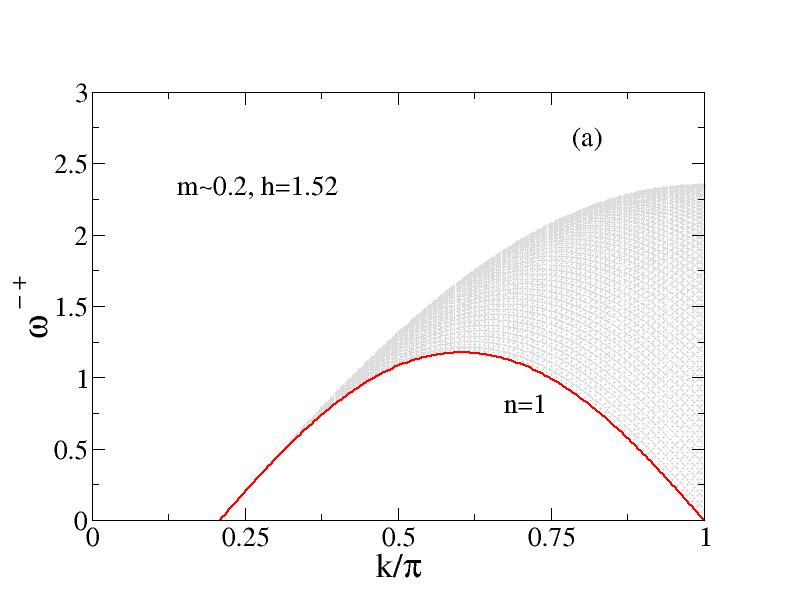}
\includegraphics[width=0.45\textwidth]{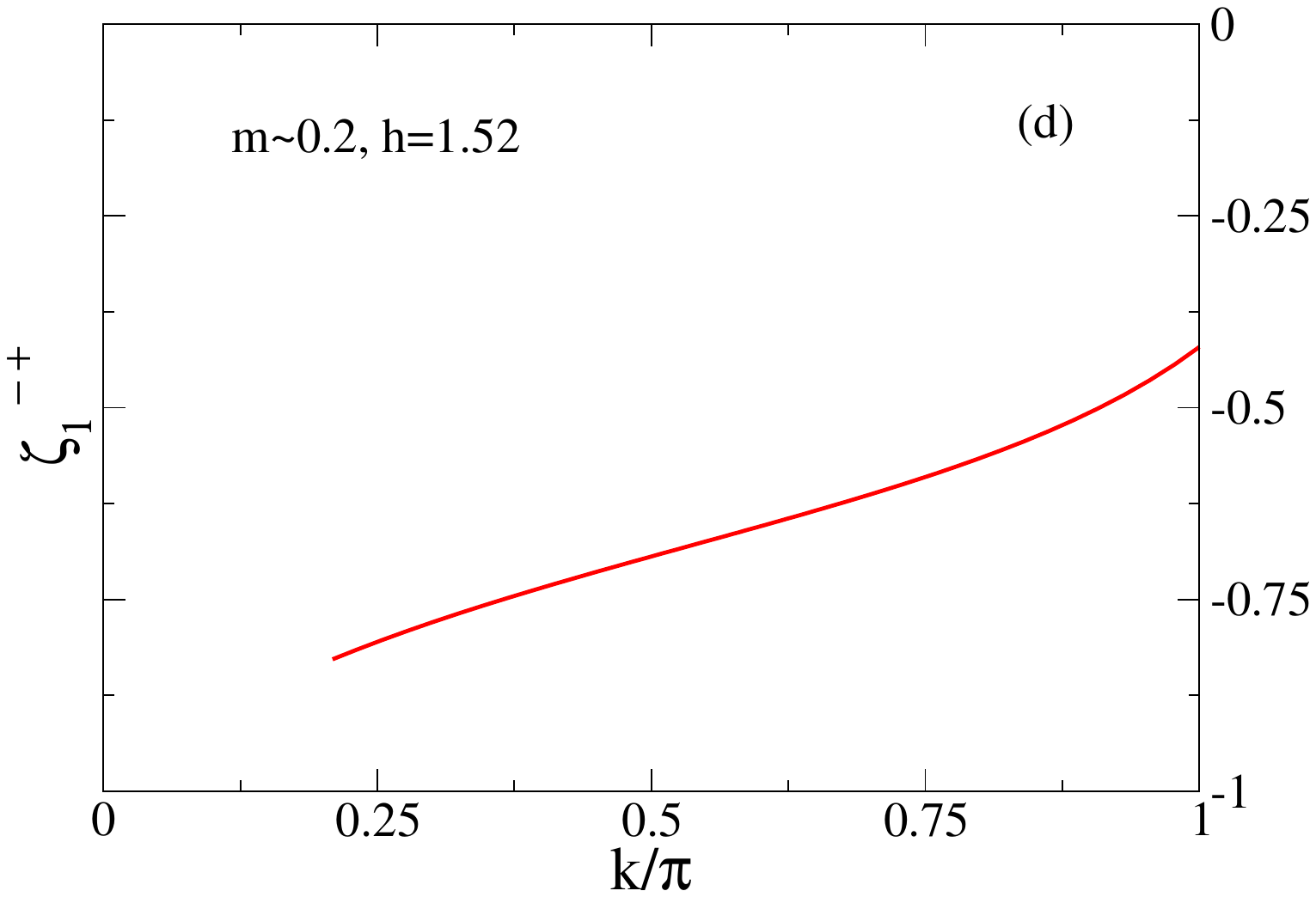}
\includegraphics[width=0.45\textwidth]{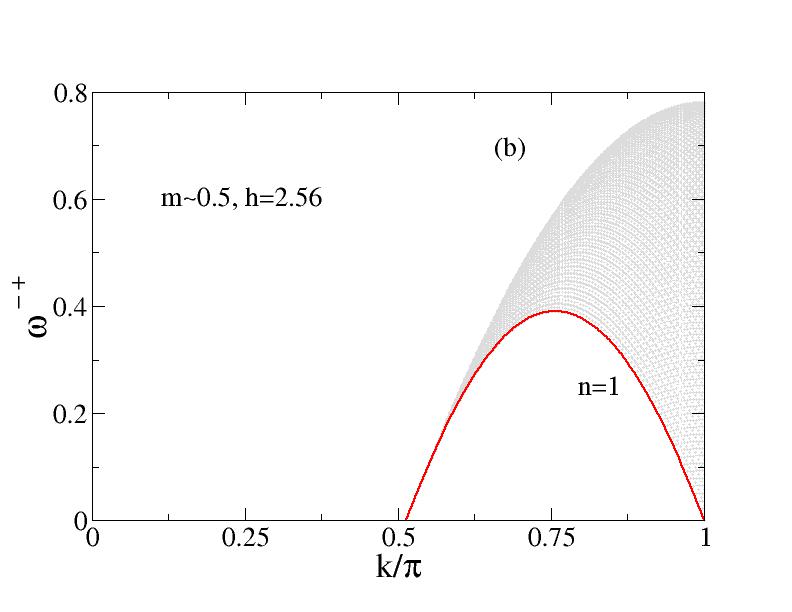}
\includegraphics[width=0.45\textwidth]{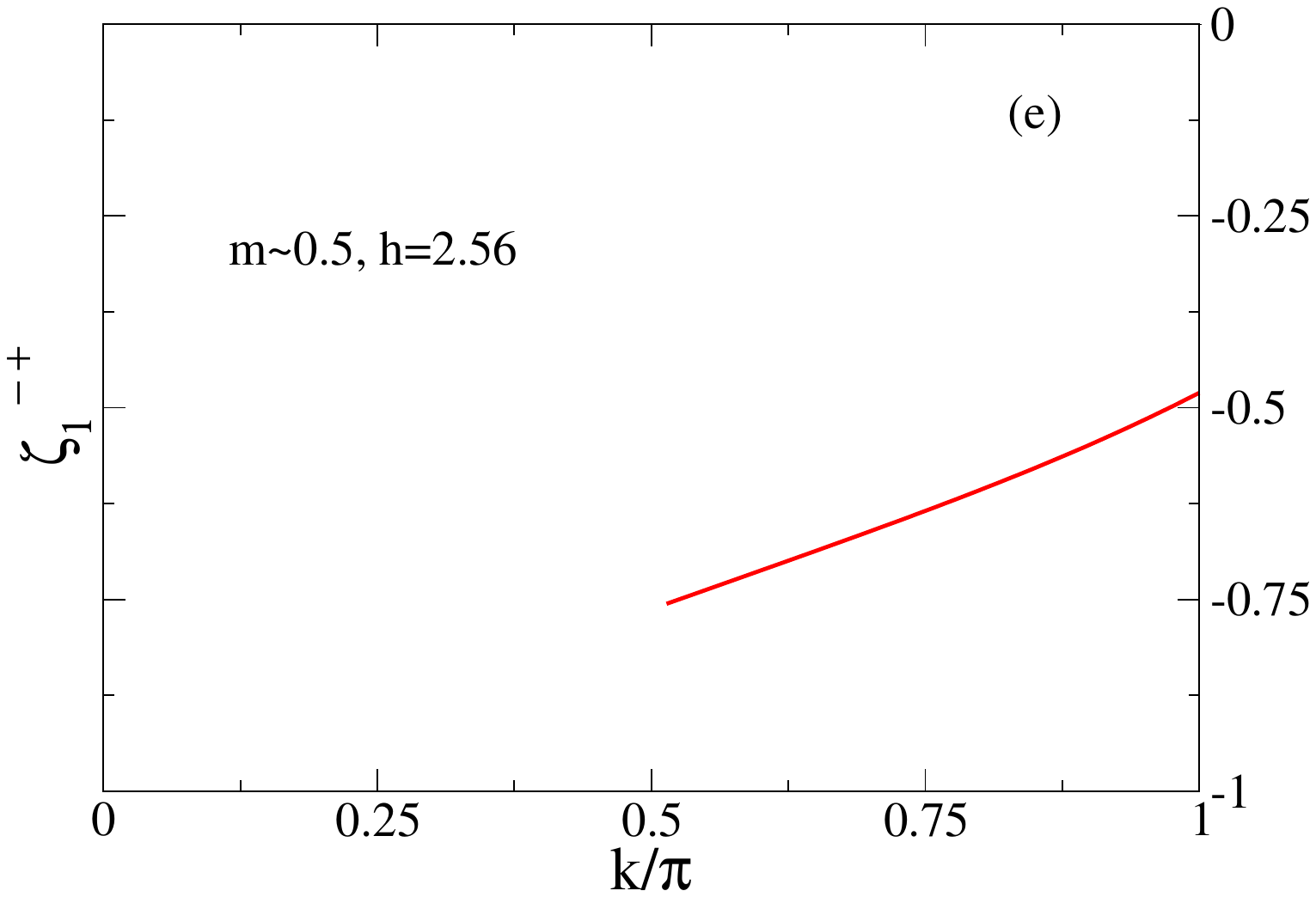}
\includegraphics[width=0.45\textwidth]{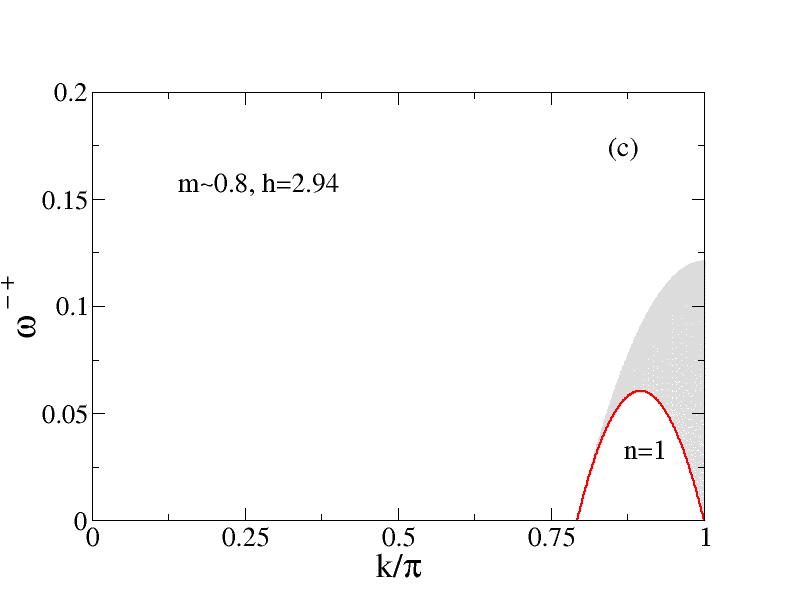}
\includegraphics[width=0.45\textwidth]{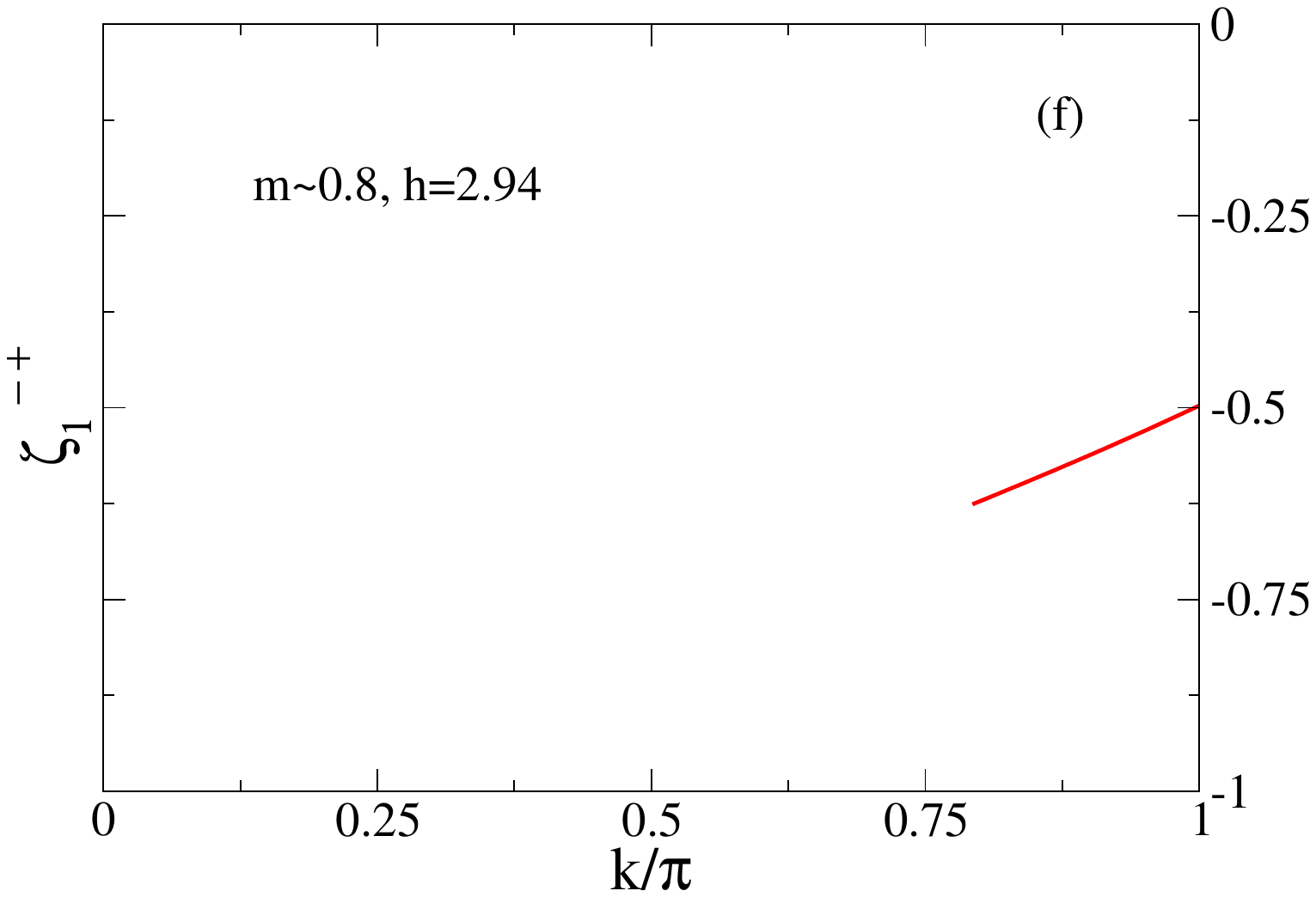}
\caption{\label{figure3PR}
The $(k,\omega)$-plane $1$-continuum where in the thermodynamic limit there is significant spectral weight in 
$S^{-+} (k,\omega)$ for the spin-$1/2$ Heisenberg-Ising chain with anisotropy $\Delta =2$ in
a longitudinal magnetic field (a)-(c). As in the case of Fig. \ref{figure2PR}, very similar spectra are obtained 
for anisotropy $\Delta =2.17$. The corresponding negative $k$-dependent exponent that
controls the line shape $S^{-+} (k,\omega)\propto (\omega - E^{-+}_1 (k))^{\zeta^{-+}_1 (k)}$ 
at and near the lower threshold of such $1$-continuum for its whole $k$ interval (d)-(f).
The spin densities in (a), (b), and (c) are $m=0.209\approx 0.2$, $m=0.514\approx 0.5$, and $m=0.793\approx 0.8$, respectively.
The corresponding $h$ values are given in units of $J/(g\mu_B)$. On this $1$-continuum lower threshold 
$S^{-+} (k,\omega)$ displays sharp peaks.}
\end{figure*} 
\begin{figure*}
\includegraphics[width=0.45\textwidth]{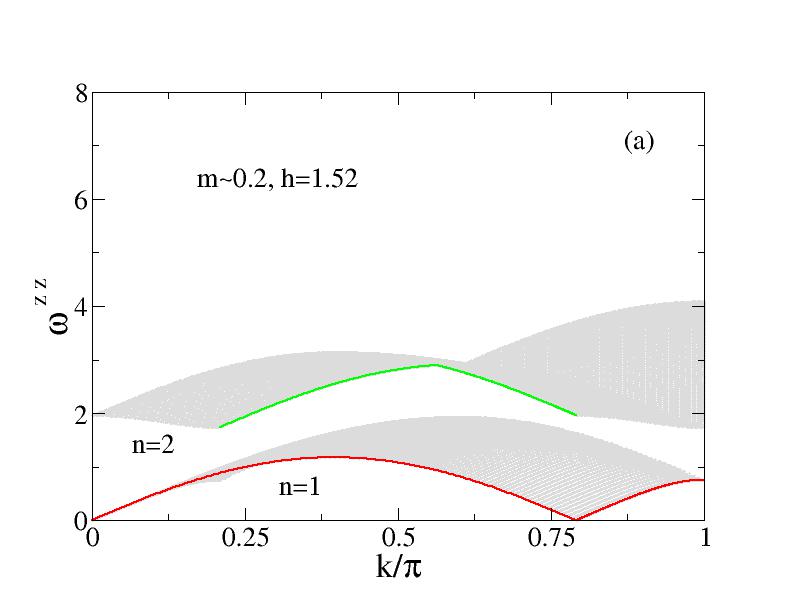}
\includegraphics[width=0.45\textwidth]{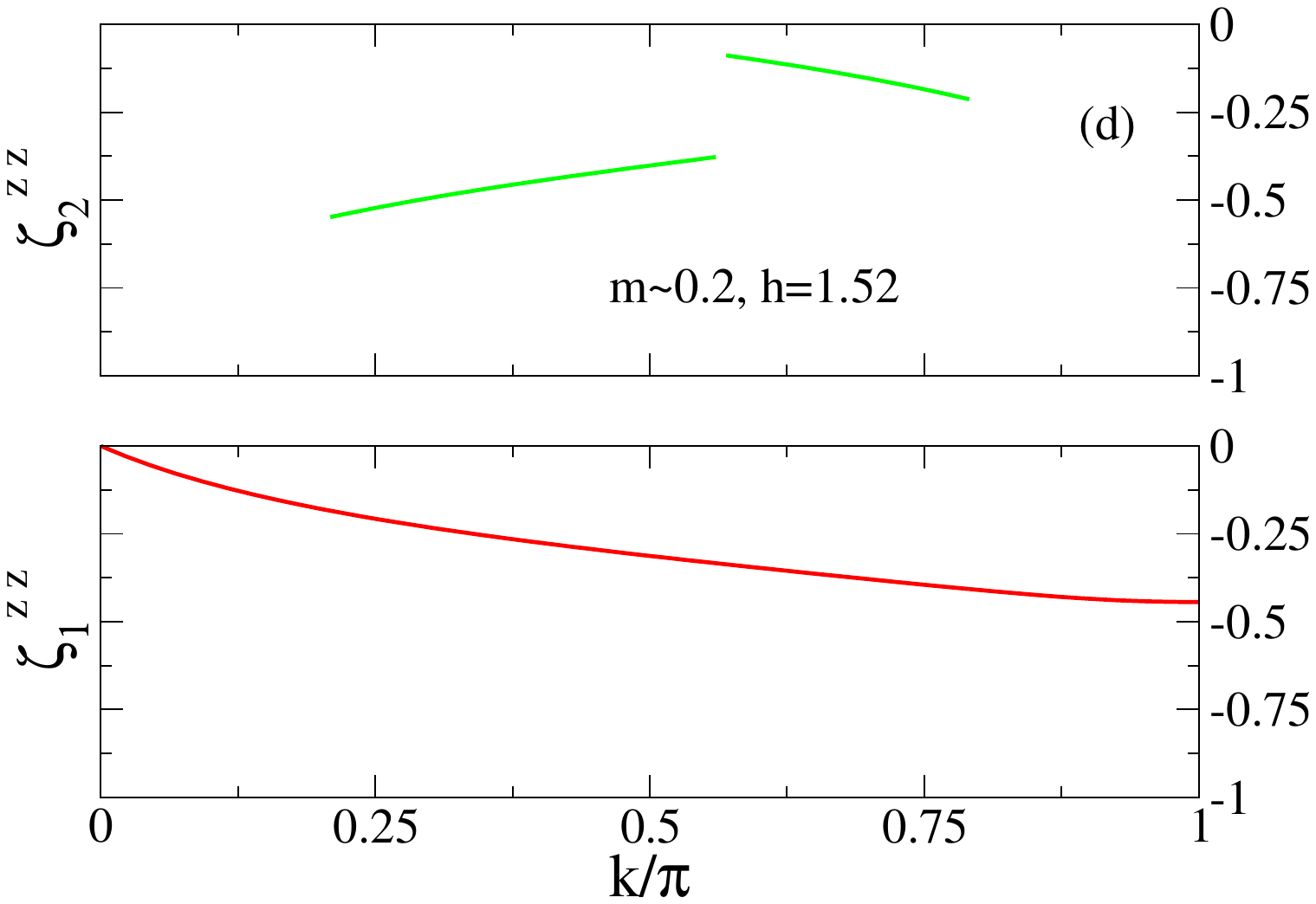}
\includegraphics[width=0.45\textwidth]{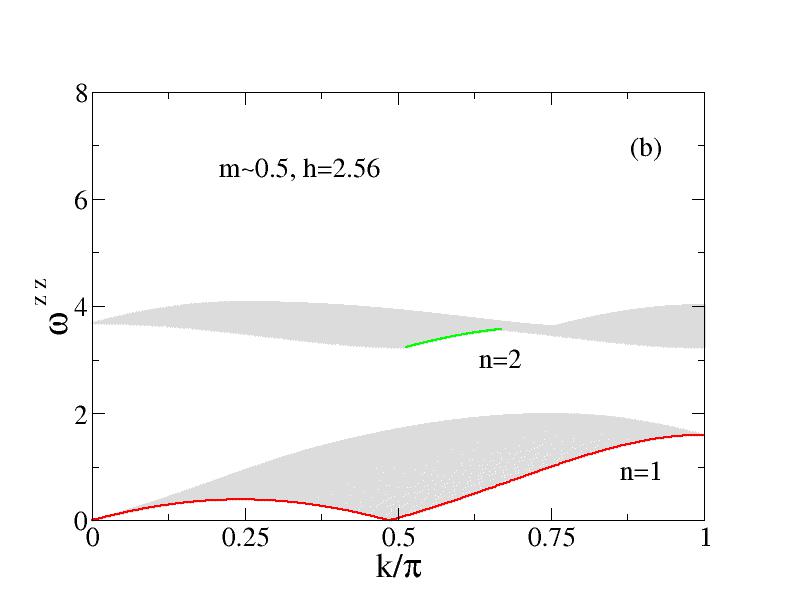}
\includegraphics[width=0.45\textwidth]{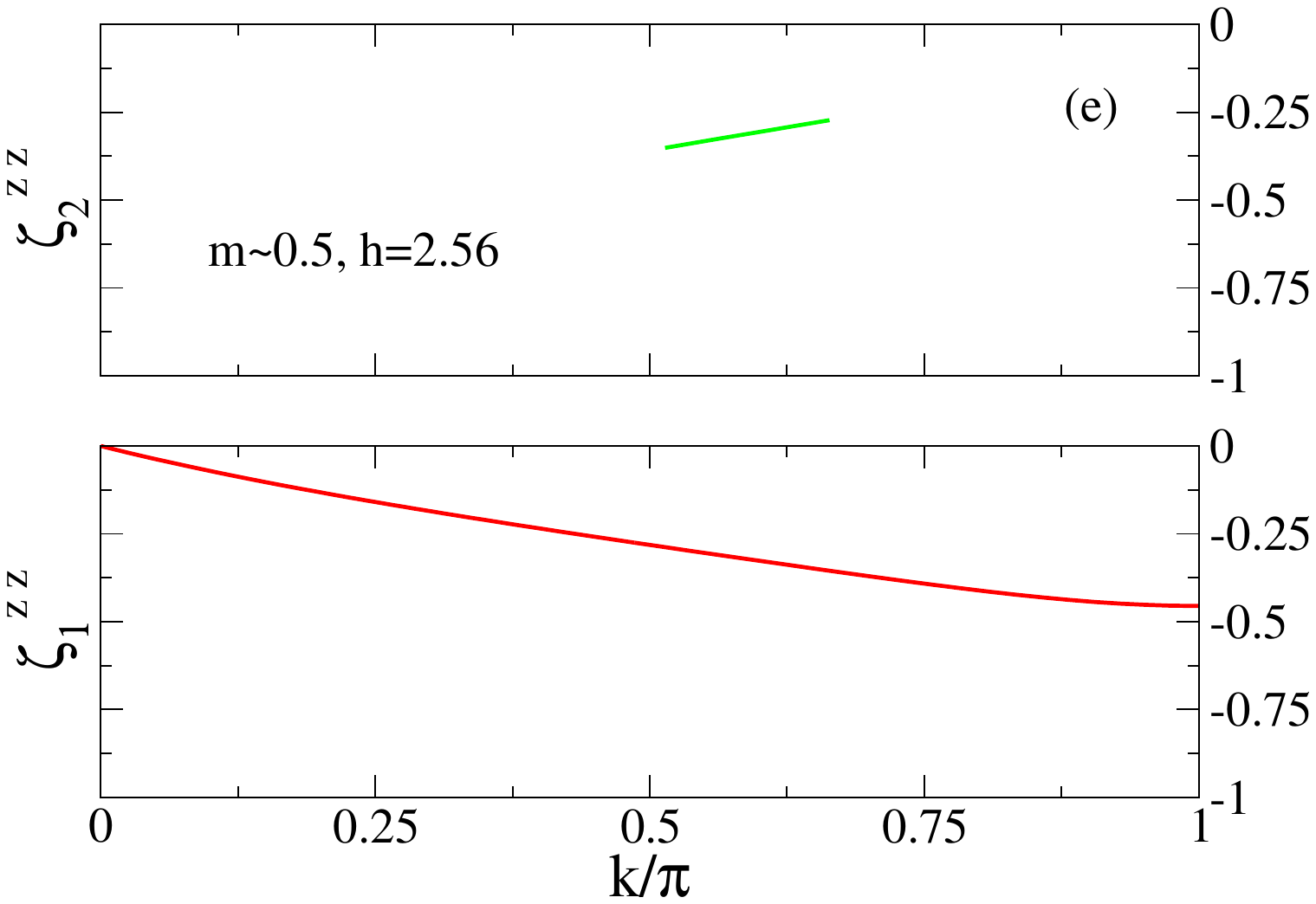}
\includegraphics[width=0.45\textwidth]{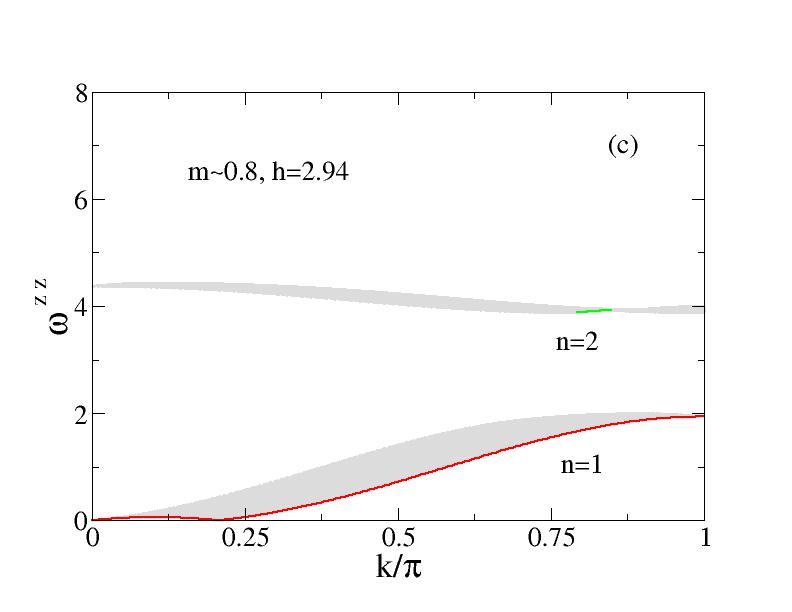}
\includegraphics[width=0.45\textwidth]{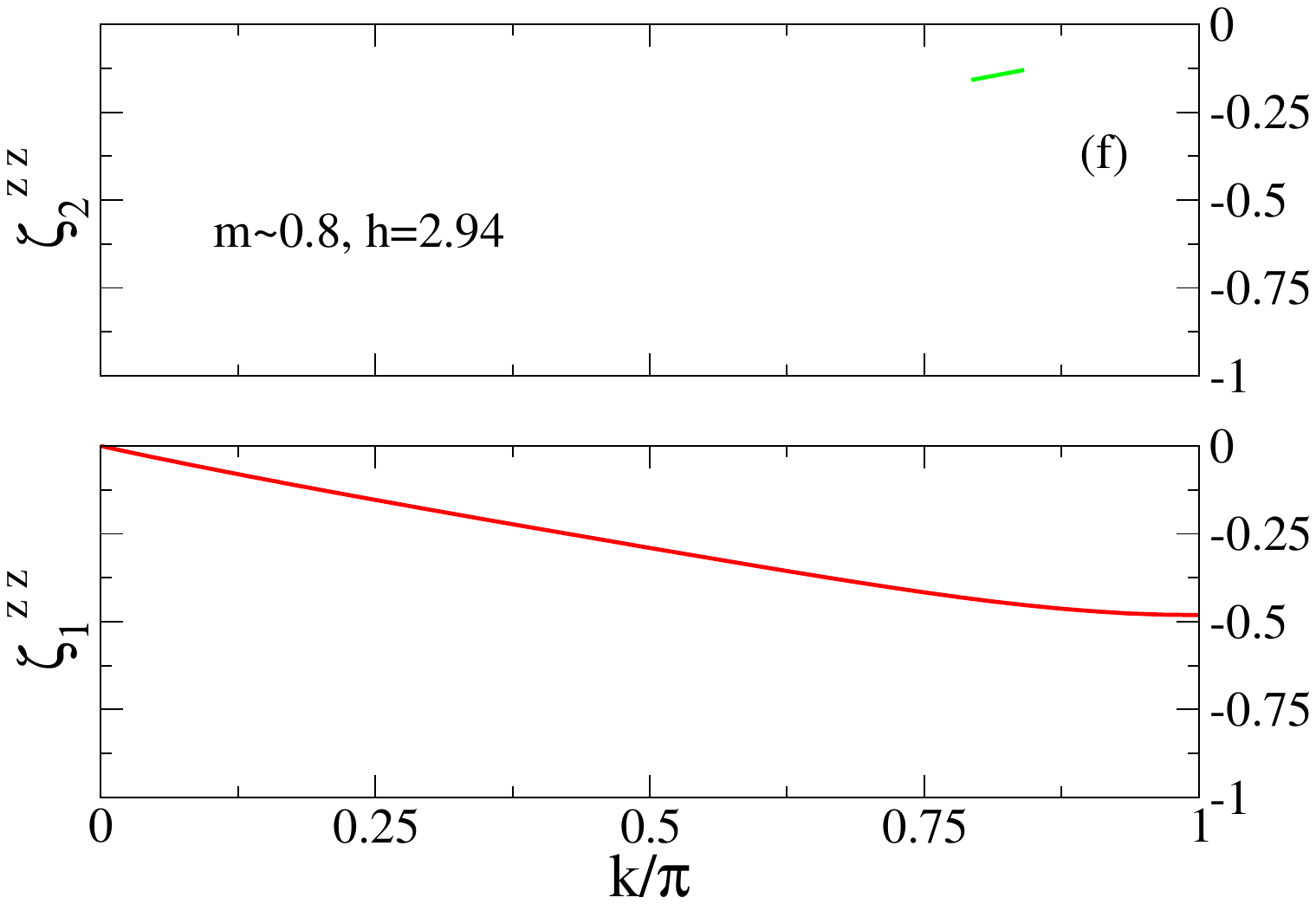}
\caption{\label{figure4PR}
The $(k,\omega)$-plane $n=1$ and $n=2$ $n$-continua where in the thermodynamic limit there is significant
spectral weight in $S^{zz} (k,\omega)$ for the spin-$1/2$ Heisenberg-Ising chain with anisotropy $\Delta =2$ in
a longitudinal magnetic field (a)-(c). As in the case of Fig. \ref{figure2PR}, very similar spectra are obtained 
for anisotropy $\Delta =2.17$. The corresponding negative $k$ dependent exponents that
control the line shape $S^{zz} (k,\omega)\propto (\omega - E^{zz}_n (k))^{\zeta^{zz}_n (k)}$ 
in the $k$ intervals at and near the lower thresholds of such continua (d)-(f).
The spin densities in (a), (b), and (c) are $m=0.209\approx 0.2$, $m=0.514\approx 0.5$, and $m=0.793\approx 0.8$,
respectively. The corresponding $h$ values are given in units of $J/(g\mu_B)$. The exponents are negative
in the $k$ intervals of these lower thresholds marked in the spectra. On the marked lines in the $(k,\omega)$-plane
$S^{zz} (k,\omega)$ displays sharp peaks.}
\end{figure*} 

In Figs. \ref{figure2PR} (a)-(c), \ref{figure3PR} (a)-(c), and \ref{figure4PR} (a)-(c)
we show the regions in the $(k,\omega)$-plane where there is significant spectral weight in 
$S^{+-} (k,\omega)$, $S^{-+} (k,\omega)$, and $S^{zz} (k,\omega)$, respectively, for
anisotropy $\Delta =2$. Very similar spectra are obtained for anisotropy $\Delta =2.17$.
The panels (a),(b),(c) of these figures refer to spin densities $m=0.209\approx 0.2$, $m=0.514\approx 0.5$, 
and $m=0.793\approx 0.8$, respectively. The field $h$ values corresponding to the above spin densities given in
these figures are in units of $J/(g\mu_B)$. In these units the critical fields
and the intermediate field $h_{1/2}$ that refers to spin density $m=1/2$
read $h_{c1} = 0.39$, $h_{1/2} = 2.53$, and $h_{c2} = 3.00$ for $\Delta =2$
and $h_{c1} = 0.52$, $h_{1/2} = 2.69$, and $h_{c2} = 3.17$ for $\Delta =2.17$.

The $(k,\omega)$-plane continua in such figures are classified as $n$-continua
where $n=1$, $n=2$, and $n=3$, respectively. This is according to the corresponding excited states having no $n>1$ Bethe $n$-strings, a 
single $2$-string, and a single $3$-string, respectively. In terms of singlet pairs of physical spins $1/2$,
this corresponds to such states having no $n$-string-pairs, a 
single $2$-string-pair, and a single $3$-string-pair, respectively.
The $2$-continuum and the $3$-continuum are gapped.

In the following we show that the $1$-pair phase shifts resulting from 
physical-spins $1$-pair - $1$-pair and $1$-pair - $n$-pair scattering whose scattering centers are $2n$-physical-spins $n$-pairs
for $n=1,2,3$ and $1$-holes control the line shape at and near the experimentally observed 
sharp peaks in $S^{+-} (k,\omega)$, $S^{-+} (k,\omega)$, and $S^{zz} (k,\omega)$ that
are located in the $n$-continua lower thresholds. This applies to the two
zigzag materials under study. We calculate and plot the negative momentum
dependent exponents that control such a line shape for the parameter sets suitable to
both such materials. 

The main aim of Figs. \ref{figure2PR} (a)-(c), \ref{figure3PR} (a)-(c), and \ref{figure4PR} (a)-(c)
is to provide the location in the $(k,\omega)$-plane  
of the marked $n$-continua lower thresholds $k$ intervals where there are sharp peaks. The experimentally observed sharp peaks
refer to specific momentum and energy values in these lower thresholds $k$ intervals.
However, the figures do not provide detailed information on the relative 
intensities of the spectral-weight distribution over the $n$-continua. The shapes of these continua are to be compared with 
those in the following figures of Ref. \onlinecite{Yang_19}:
Figs. 3 (a1)-(c1) for $S^{-+} (k,\omega)$, Figs. 3 (a2)-(c2) for $S^{+-} (k,\omega)$,
and Fig. 8 (d)-(f) for $S^{zz} (k,\omega)$ for a finite-size system, which also
provide this information.

Within the dynamical theory used in our studies, the line shapes of 
$S^{+-} (k,\omega)$, $S^{-+} (k,\omega)$, and $S^{zz} (k,\omega)$ 
have for extended $k$ intervals the general power-law form given in
Eqs. (\ref{MPSs}) and (\ref{zetaabk})-(\ref{functional}) of Appendix \ref{C}.
The general expression, Eq. (\ref{MPSs}) of that Appendix, applies at and just above the $(k,\omega)$-plane 
$n$-continua lower thresholds $k$ intervals for $n=1,2,3$ where there are sharp peaks.  

That line shape is controlled by exponents $\zeta_{n}^{ab} (k)$ 
whose general expression is given in Eqs. (\ref{zetaabk}) and (\ref{functional}) of Appendix \ref{C}.
They are negative in the lower thresholds $k$ intervals marked in Figs. \ref{figure2PR} (a)-(c),
\ref{figure3PR} (a)-(c), and \ref{figure4PR} (a)-(c). These figures
refer to lines of sharp peaks located in $k$ intervals much beyond the few momentum
values in these lines of the sharp modes experimentally observed \cite{Wang_18,Wang_19,Bera_20}.
The latter were considered in studies of finite-size systems \cite{Yang_19}.

The $k$ dependence of the corresponding negative exponents is shown 
in Figs. \ref{figure2PR} (d)-(f), \ref{figure3PR} (d)-(f), and \ref{figure4PR} (d)-(f)
for the components $S^{+-} (k,\omega)$, $S^{-+} (k,\omega)$, and $S^{zz} (k,\omega)$, respectively. 
The exponent values plotted in these figures refer to anisotropy $\Delta = 2$.
Very similar exponent values are obtained for anisotropy $\Delta =2.17$.

Within the physical-spins $1$-pair - $1$-pair and $1$-pair - $n$-pair scattering that controls the line shape at and near the 
sharp peaks located in the $(k,\omega)$-plane $n$-continua lower thresholds, 
the $1$-pairs at the $1$-band Fermi points $q=\pm k_{F\downarrow}$ are the scatterers and
the $1$-holes and $1$-pairs created under the transitions to excited states
at $1$-band momenta $q \in [-k_{F\downarrow},k_{F\downarrow}]$
and $\vert q\vert \in [k_{F\downarrow},k_{F\uparrow}]$, respectively, and the $n$-string-pairs 
created under such transitions at $n$-band momenta $q' \in [-(k_{F\uparrow}-k_{F\downarrow}),(k_{F\uparrow}-k_{F\downarrow})]$
are the scattering centers.

Important $n$-pair scattering quantities that control the momentum dependent exponents of the
spin dynamic structure factor components are the corresponding phase shifts acquired by a $1$-pair at the $1$-band Fermi 
momentum $\iota k_{F\downarrow} =\pm k_{F\downarrow}$ (i) $2\pi\Phi_{1,n}(\iota k_{F\downarrow},q)$ where $n=1,2,3$
and (ii) $-2\pi\Phi_{1,1}(\iota k_{F\downarrow},q)$. Those are due
to creation (i) of one $n$-pair at $n$-band momentum $q$ and (ii) of one $1$-hole at $1$-band momentum $q$,
respectively, under a transition to an excited state.
(See Eq. (\ref{Phi-barPhi}) of Appendix \ref{A} with $q = \iota k_{F\downarrow}$
and Eq. (\ref{x-aa}) of that Appendix.) 

\subsection{Selected sharp peaks at fixed momenta $k=0,\pi/2,\pi$ in the $(h,\omega)$-plane}
\label{SECIIIB}

Besides momentum dependencies, our study includes extracting the longitudinal
magnetic field $h$ dependencies in the thermodynamic limit of the negative exponents that control the line shape at and near the 
sharp peaks in $S^{+-} (k,\omega)$, $S^{-+} (k,\omega)$, and $S^{zz} (k,\omega)$
at the momentum values $k=0$, $k=\pi/2$, and $k=\pi$ 
at which they were experimentally observed \cite{Wang_18,Wang_19,Bera_20}. 
This is carried out by using Eqs. (\ref{MPSs}) and (\ref{zetaabk})-(\ref{functional}) of Appendix \ref{C}.
In order to provide information on the frequency/energy $\omega$
values of the sharp peaks under study, we also plot their energies, which are to be compared
with those obtained by finite-size methods \cite{Yang_19} used in previous studies \cite{Wang_18,Wang_19,Bera_20}.

The momentum values $k=0$, $k=\pi/2$, and $k=\pi$ 
of the sharp peaks observed experimentally \cite{Wang_18,Wang_19,Bera_20}
belong to the marked $k$ intervals of the $n$-continua lower thresholds 
shown in Figs. \ref{figure2PR} (a)-(c), \ref{figure3PR} (a)-(c), and  \ref{figure4PR} (a)-(c).
When at such momenta the corresponding lower threshold is not marked,
the exponent is not negative and there is no sharp peak.

The following thermodynamic-limit results are for the spin-$1/2$ Heisenberg-Ising chain in 
a longitudinal field $h_{c1}<h<h_{c2}$ with anisotropies $\Delta =2$ and $\Delta =2.17$ 
representative of the 1D physics of SrCo$_2$V$_2$O$_8$ and BaCo$_2$V$_2$O$_8$, respectively.
At and near the sharp peaks denoted by $R^{+-}_{0}$, $R^{+-}_{\pi/2}$, $R^{-+}_{\pi/2}$, $R^{zz}_{\pi}$, 
$\chi^{(2)}_0$, $\chi^{(2)}_{\pi/2}$, $\chi^{(2)}_{\pi}$, and 
$\chi^{(3)}_{\pi/2}$ in Refs. \onlinecite{Wang_18},\onlinecite{Wang_19}, except for $R^{zz}_{\pi}$,
which is called $R^{\rm PAP(zz)}_{\pi}$ in Fig. 5-b of Ref. \onlinecite{Bera_20}, the dynamical theory
used in our study gives for small values of the energy deviation $(\omega - E^{ab}_{n} (k,h))\geq 0$
from the $ab=-+,+-,zz$ $n$-continuum lower-threshold energy $E^{ab}_{n} (k,h)$
at momentum $k$ and field $h$ a line shape of power-law form,
\begin{eqnarray}
R_{k}^{ab} & = & S^{ab} (k,\omega) = \bar{C}^{ab}_1 (k) \,
\Bigl(\omega - E^{ab}_{1} (k,h)\Bigr)^{\zeta_{1}^{ab} (k,h)}  
\nonumber \\
\chi^{(n)}_k & = & S^{+-} (k,\omega) = \bar{C}^{+-}_n (k)
\Bigl(\omega - E^{+-}_{n} (k,h)\Bigr)^{\zeta_{n}^{+-} (k,h)} 
\nonumber \\
&& {\rm where}
\nonumber \\
\bar{C}^{ab}_ n (k) & = & {C_{ab}^n (k)\over (4\pi\,B_1^{ab}\,v_1 (k_{F\downarrow}))^{\zeta_{n}^{ab} (k,h)}} 
\hspace{0.20cm}{\rm for}\hspace{0.20cm}n = 1,2,3  \, .
\label{Rkab}
\end{eqnarray}
These line shapes refer to zero temperature. Hence
we expect that the sharp modes observed in low-temperature experiments 
\cite{Wang_18,Wang_19,Bera_20} to be a bit smeared by thermal fluctuations and coupling to phonons. 

According to the set of sharp peaks experimentally observed in SrCo$_2$V$_2$O$_8$ and BaCo$_2$V$_2$O$_8$,
the excitation momentum $k$, spin component $ab$, and $n$-pair number $n$
in Eq. (\ref{Rkab}) have the values $k = 0,\pi/2$ for $ab = +-$ and $n=1$, $k = \pi/2$ for $ab = -+$ and $n=1$,
$k = \pi$ for $ab = zz$ and $n=1$, $k = 0,\pi/2,\pi$ for $+-$ and $n=2$, and $k = \pi/2$ for $+-$ and $n=3$. 
In that equation, $v_1 (k_{F\downarrow})$ is the $1$-band group velocity 
$v_1 (q) = \partial\varepsilon_1 (q)/ \partial q$ at $q=k_{F\downarrow}$, the $\eta$ and $m$ dependent parameter
$B_1^{ab}$ has values in the range $0<B_1^{ab}\leq 1$, and $C_{ab}^n (k)$ is given in Eq. (\ref{Cabn}) of Appendix \ref{C}.
\begin{figure}
\begin{center}
\subfigure{\includegraphics[width=8.50cm]{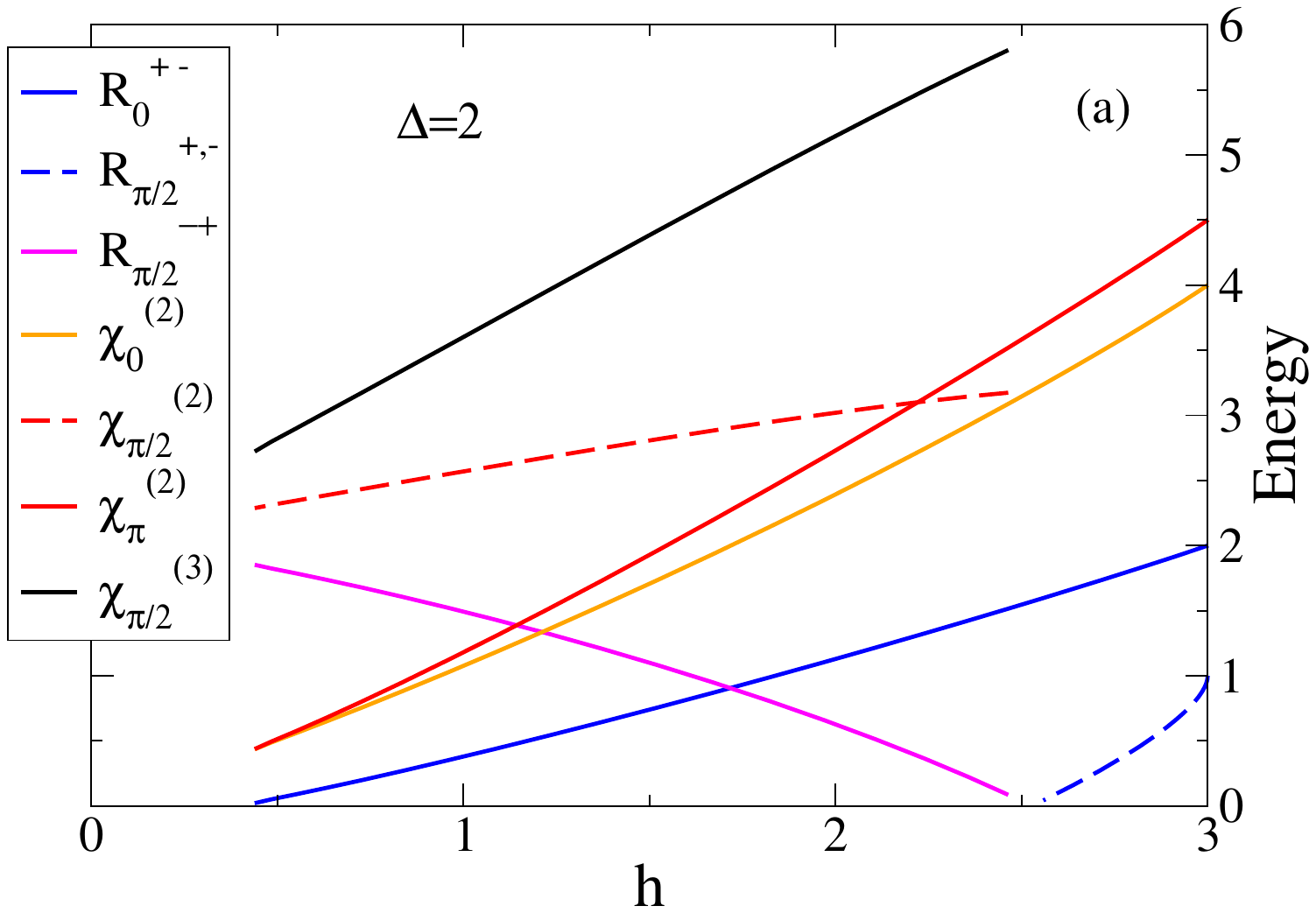}}
\hspace{0.50cm}
\subfigure{\includegraphics[width=8.50cm]{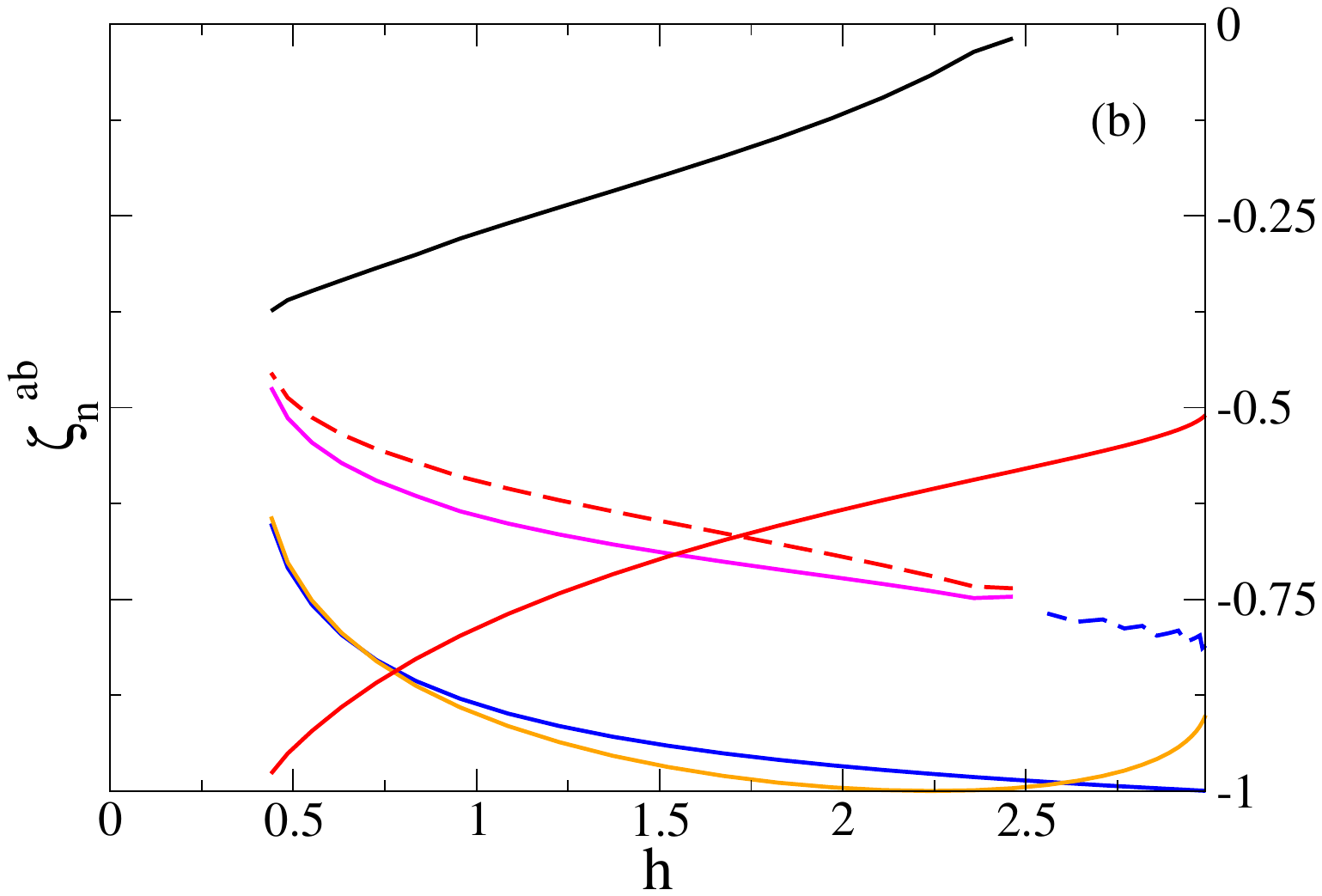}}
\caption{The energies in units of $J$ of the sharp peaks $R^{+-}_{0}$, $R^{+-}_{\pi/2}$, $R^{-+}_{\pi/2}$, $\chi^{(2)}_0$,
$\chi^{(2)}_{\pi/2}$, $\chi^{(2)}_{\pi}$, and  $\chi^{(3)}_{\pi/2}$ in the transverse components $S^{+-} (k,\omega)$ and $S^{-+} (k,\omega)$
versus the magnetic field $h$ for $h\in [h_{c1},h_{c2}]$ in units of $J/(g\mu_B)$ (a);
The corresponding magnetic field $h$ dependencies of the negative exponents 
that control the line shape near such sharp peaks (b). 
The expressions of these energies and exponents are given in Eqs. 
(\ref{EPM0})-(\ref{EPM3PI2}) of Appendix \ref{C}. The energy curves plotted here are to be compared 
with those shown in Fig. 5 of Ref. \onlinecite{Yang_19} for a finite-size system.}
\label{figure5PR}
\end{center}
\end{figure}

The $n=1,2,3$ lower threshold energies $E^{+-}_{n} (k,h)$, $n=1$ lower threshold energy $E^{-+}_{1} (k,h)$,
$n=1$ lower threshold energy $E^{zz}_{1} (k,h)$, and exponents $\zeta_{n}^{ab} (k,h)$, Eqs. (\ref{zetaabk}) 
and (\ref{functional}) of Appendix \ref{C}, appearing in the expressions, Eq. (\ref{Rkab}), of the line shape 
at and near the sharp peaks at anisotropies $\Delta =2$ and $\Delta =2.17$ representative
of SrCo$_2$V$_2$O$_8$ and BaCo$_2$V$_2$O$_8$, respectively, are given in Eqs.
(\ref{EPM0})-(\ref{EPM3PI2}) of Appendix \ref{C}.

The sharp peaks $R^{+-}_{\pi/2}$ and $R^{zz}_{\pi}$ whose energy interval in Eqs. (\ref{EZZPI}) and (\ref{EPM2PI2}) of Appendix \ref{C},
respectively, was not given for anisotropy $\Delta =2.17$ have not been experimentally studied for BaCo$_2$V$_2$O$_8$.
The same applies to the sharp peak $\chi^{(2)}_0$. However, as it is associated with $2$-string states, its energy interval
was given for $\Delta =2.17$ in Eq. (\ref{EPM20}) of Appendix \ref{C}. 

For simplicity, we do not discuss here a spectral feature denoted by $R^{+-,b}_{\pi/2}$ within 
finite-size studies \cite{Yang_19}: It is not
among the sharp modes experimentally observed that are
displayed in Fig. 4 of Ref. \onlinecite{Wang_18} for SrCo$_2$V$_2$O$_8$
and in Fig. 4 (b) of Ref. \onlinecite{Wang_19} for BaCo$_2$V$_2$O$_8$.
\begin{figure}
\begin{center}
\subfigure{\includegraphics[width=8.50cm]{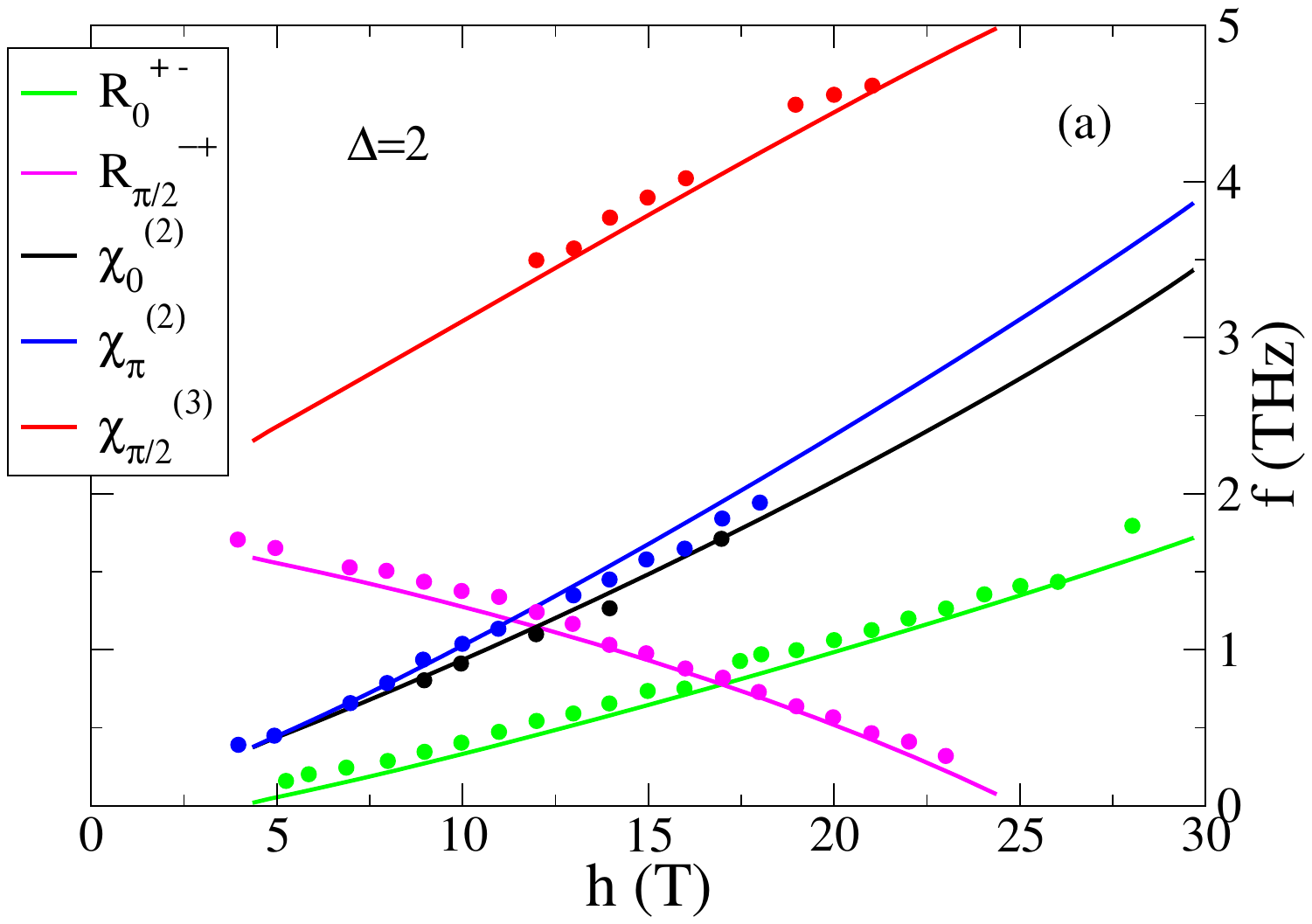}}
\hspace{0.50cm}
\subfigure{\includegraphics[width=8.50cm]{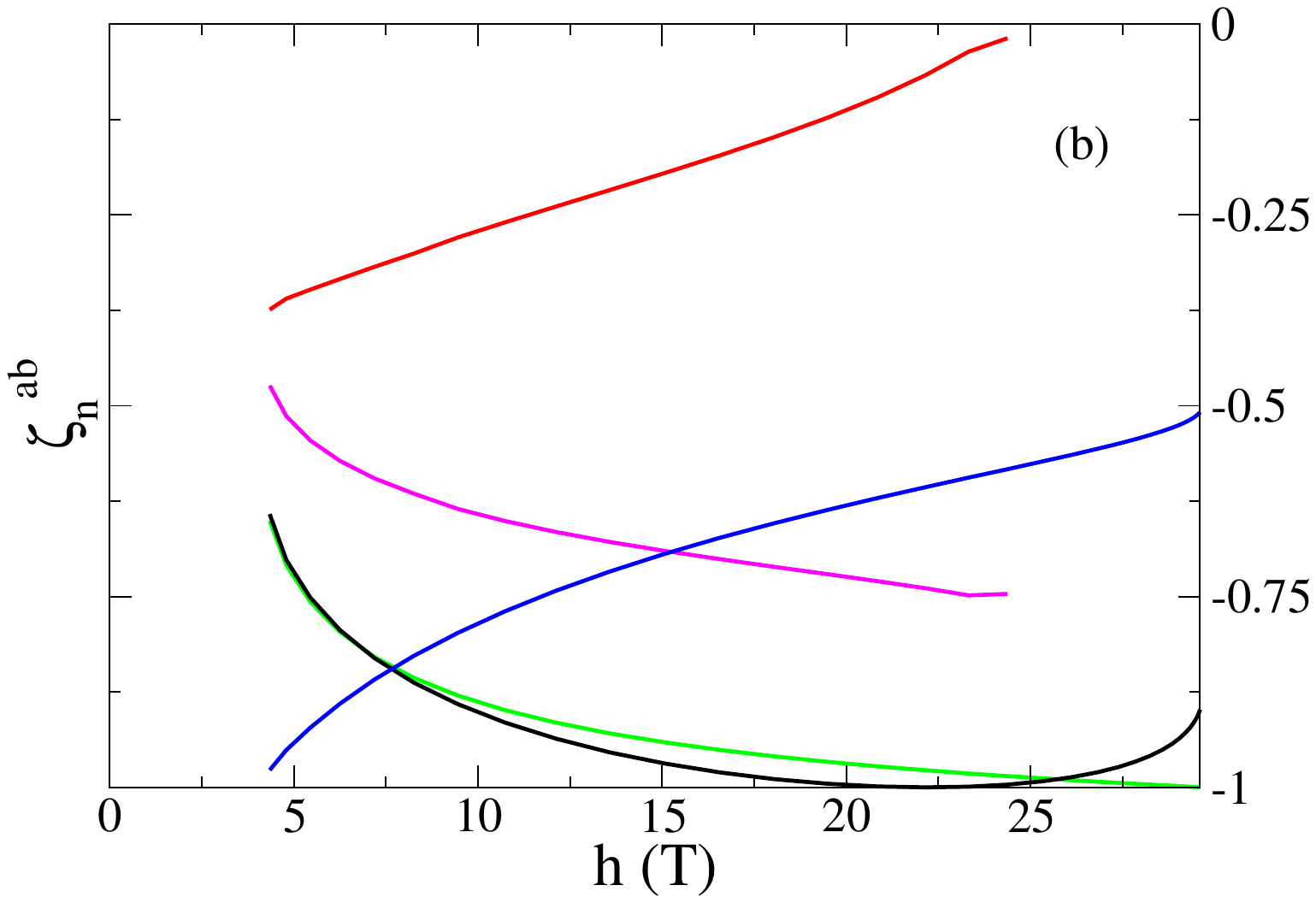}}
\caption{The dependencies on the magnetic field $h\in [h_{c1},h_{c2}]$ 
in Tesla of the frequencies in units of THz associated with the 
energies of the transverse sharp peaks $R_{0}^{+-}$, $R_{\pi/2}^{-+}$, 
$\chi^{(2)}_{0}$, $\chi^{(2)}_{\pi}$, and $\chi^{(3)}_{\pi/2}$, respectively,
experimentally observed in SrCo$_2$V$_2$O$_8$ (a); The corresponding negative exponents (b). 
Expressions both of the energies corresponding to these frequencies and of the latter 
exponents are given in Eqs. (\ref{EPM0}), (\ref{EMPPI2}), (\ref{EPM20}), (\ref{EPM2pi}), and 
(\ref{EPM3PI2}) of Appendix \ref{C}. Such theoretical frequency dependencies on $h\in [h_{c1},h_{c2}]$ 
are to be compared with those of the corresponding sharp peaks points experimentally 
observed in SrCo$_2$V$_2$O$_8$ also shown in (a), which are
those displayed in Fig. 4 of Ref. \onlinecite{Wang_18} with $h_{c1}=B_c$ and $h_{c2}=B_s$.}
\label{figure6PR}
\end{center}
\end{figure}

All above sharp peaks are located in $n$-continua lower thresholds. 
On the other hand, the momentum $k=\pi/2$ sharp peak $R_{\pi/2}^{zz}$ called $R^{\rm PAP(zz)}_{\pi/2}$ in Fig. 5-a 
of Ref. \onlinecite{Bera_20} is located in the $1$-continuum {\it upper threshold} of $S^{zz} (\pi,\omega)$.
The line shape at and near it is for small values of the energy deviation $(\omega - E^{zz}_1 (\pi/2,h))\geq 0$
provided in Eq. (\ref{EZZPI2}) of Appendix \ref{C}. A discussion of the processes behind that sharp peak
is given in a text below that equation.

As given in Eqs. (\ref{EPM0})-(\ref{EZZPI2}) of Appendix \ref{C}, 
depending on which specific sharp peaks, they occur for four
ranges of magnetic fields: $h\in [h_{c1},h_{c2}]$, $h\in [h_{c1},h_{1/2}]$, $h\in [h_{1/2},h_{c2}]$,
and $h\in [h_{c1},h_{\diamond}]$. The theoretical dependencies on the magnetic field $h$ in units of $J/(g\mu_B)$
of the energies in units of $J$ and of the corresponding exponents given in Eqs. (\ref{EPM0})-(\ref{EPM3PI2}) of Appendix \ref{C}
of the transverse sharp peaks $R^{+-}_{0}$, $R^{+-}_{\pi/2}$, $R^{-+}_{\pi/2}$, $\chi^{(2)}_0$,
$\chi^{(2)}_{\pi/2}$, $\chi^{(2)}_{\pi}$, and  $\chi^{(3)}_{\pi/2}$
are plotted in Figs. \ref{figure5PR} (a) and (b), respectively, for anisotropy $\Delta =2$.
Corresponding results for anisotropy $\Delta =2.17$ are very similar.
The specific energy lines $h$ ranges in these figures are those for which in the thermodynamic limit the 
corresponding exponents are negative. Only for such ranges there are sharp peaks.

While the field dependencies of the longitudinal sharp peaks $R_{\pi}^{zz}$ and $R_{\pi/2}^{zz}$ are discussed 
below in Sec. \ref{SECIIIC}, the energy of the peak $R_{\pi}^{zz}$ obeys the equality $E^{zz}_1 (\pi,h) = E^{+-}_1 (0,h)$
for the whole magnetic field interval $h \in [h_{c1},h_{c2}]$, so that it is also plotted in Fig. \ref{figure5PR} (a). 

The overall behavior of the $(h,\omega)$-plane energy versus field lines of the sharp peaks plotted 
in Fig. \ref{figure5PR} (a) for $\Delta = 2$ are to be compared 
with those shown in Fig. 5 of Ref. \onlinecite{Yang_19} for a finite-size system with $N=200$ spins 
and anisotropy $\Delta =2.00$. There is agreement concerning the general trends of the $h$ dependencies
of the lines associated with the sharp peaks common to the two figures. In the present case, each point of the solid lines refers to an existing sharp peak. 

Other sharp peaks included in Fig. 5 of Ref. \onlinecite{Yang_19} refer to specific $(k,\omega)$-plane points
that correspond to the momenta $k=0,\pi/2,\pi$ on the lines marked in 
Figs. \ref{figure2PR} (a)-(c) and \ref{figure3PR} (a)-(c). The line shape at and near such other 
sharp peaks is also of the form given in Eq. (\ref{MPSs}) of Appendix \ref{C}.
\begin{figure}
\begin{center}
\subfigure{\includegraphics[width=8.50cm]{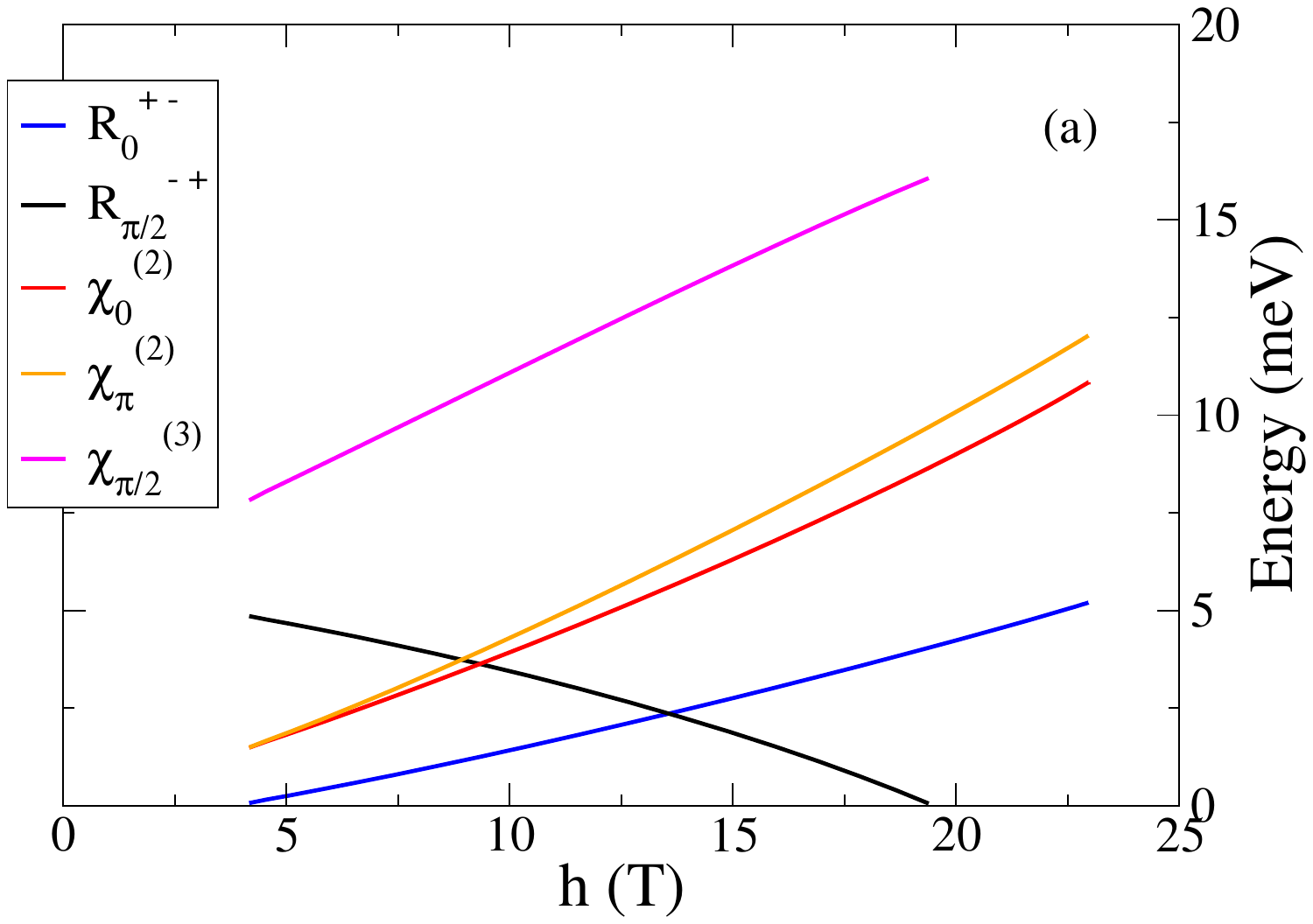}}
\hspace{0.50cm}
\subfigure{\includegraphics[width=8.50cm]{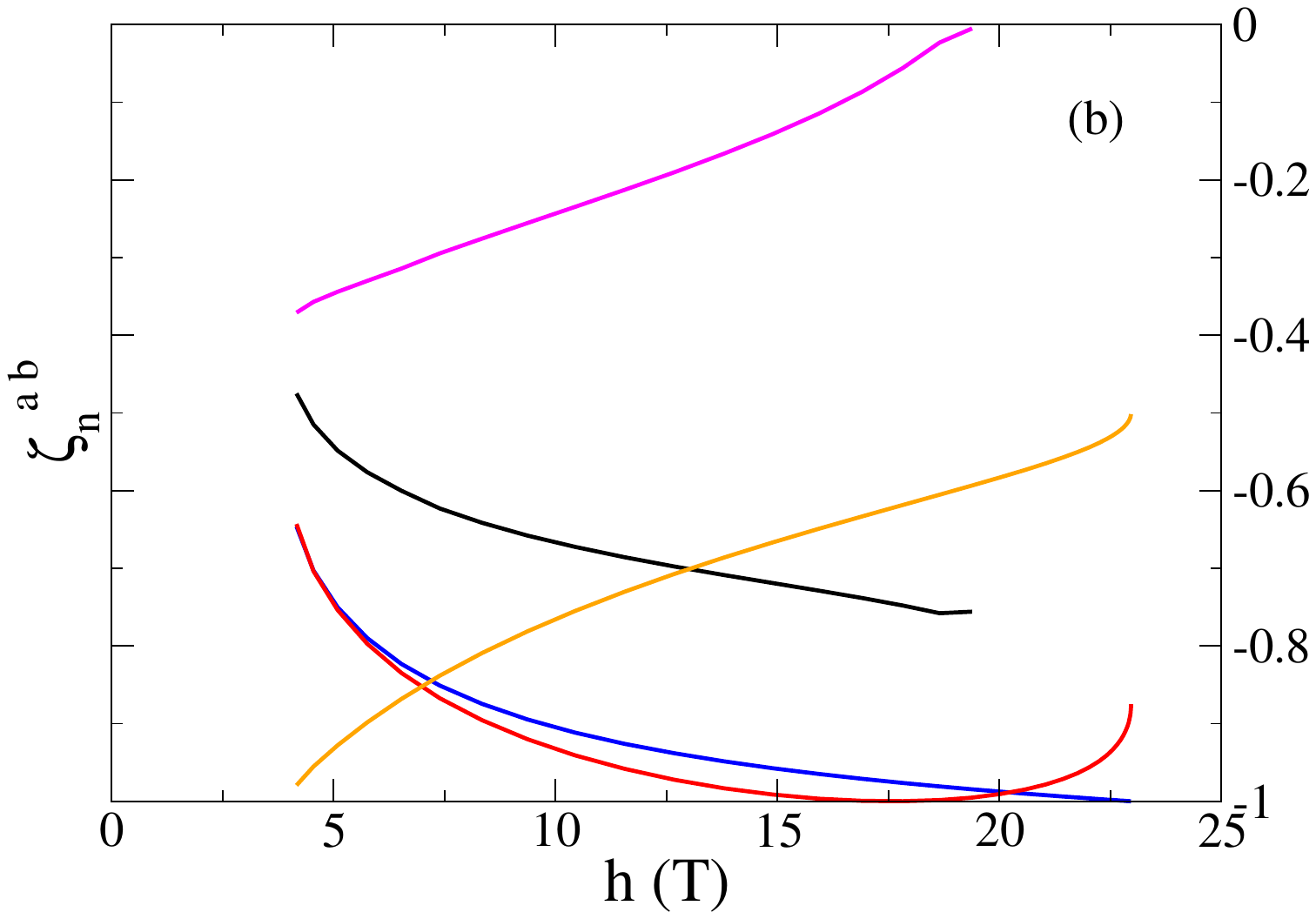}}
\caption{The $(h,\omega)$-plane lines of energy versus magnetic field 
$h\in [h_{c1},h_{c2}]=[3.76\,$T$,22.97\,$T$]$ in meV and Tesla, respectively, 
of the transverse sharp peaks $R_{0}^{+-}$, $R_{\pi/2}^{-+}$, $\chi^{(2)}_{\pi}$, and
$\chi^{(3)}_{\pi/2}$, Eqs. (\ref{EPM0}), (\ref{EMPPI2}), (\ref{EPM2pi}), and (\ref{EPM3PI2}) of Appendix \ref{C},
respectively, experimentally observed in BaCo$_2$V$_2$O$_8$ plus those of the
sharp peak $\chi^{(2)}_{0}$, Eq. (\ref{EPM20}) of Appendix \ref{C}, (a); The corresponding negative exponents (b).}
\label{figure7PR}
\end{center}
\end{figure}

\subsection{The sharp peaks experimentally observed in SrCo$_2$V$_2$O$_8$ and BaCo$_2$V$_2$O$_8$}
\label{SECIIIC}

Here the parameter sets $\Delta =2.00$, $J=3.55$ meV, and $g =6.2$ suitable to 
SrCo$_2$V$_2$O$_8$ and $\Delta =2.17$, $J=2.60$ meV, and $g =6.2$ suitable to BaCo$_2$V$_2$O$_8$
are again used. Our results concerning the sharp peaks experimentally observed in 
SrCo$_2$V$_2$O$_8$ and BaCo$_2$V$_2$O$_8$
refers to the line shape at and near them and to scattering processes that control it, 
in the following we also confirm that our thermodynamic-limit results for their energies 
agree with those experimentally observed in SrCo$_2$V$_2$O$_8$ and BaCo$_2$V$_2$O$_8$, as
already reported in Refs. \onlinecite{Wang_18},\onlinecite{Wang_19},\onlinecite{Bera_20} by use
of finite-size algebraic Bethe-ansatz theoretical results.

Our thermodynamic-limit theoretical dependencies on the magnetic 
field $h$ in Tesla for the ranges of the frequencies in THz corresponding 
to the lower-threshold energies given in Eqs. (\ref{EPM0})-(\ref{EPM3PI2}) of Appendix \ref{C} 
of the subset of sharp peaks $R^{+-}_{0}$, $R^{-+}_{\pi/2}$, $\chi^{(2)}_0$, $\chi^{(2)}_{\pi}$, and 
$\chi^{(3)}_{\pi/2}$ experimentally observed in SrCo$_2$V$_2$O$_8$ by optical experiments 
are plotted in Fig. \ref{figure6PR} (a). The corresponding experimental points in the $(h,\omega)$ plane
that describe the $h$ dependencies of the frequencies displayed in Fig. 4 of Ref. \onlinecite{Wang_18} for SrCo$_2$V$_2$O$_8$
are also shown in Fig. \ref{figure6PR} (a). The negative exponents that control the line shape 
near such peaks that have not been previously studied by other authors and
whose expressions are given in Eqs. (\ref{EPM0})-(\ref{EPM3PI2}) of Appendix \ref{C}
are plotted as a function of the magnetic field $h$ in Fig. \ref{figure6PR} (b). 

Comparison with the experimental dependence on $h\in [h_{c1},h_{c2}]$ of the frequencies of the
sharp peaks displayed in Fig. 4 of Ref. \onlinecite{Wang_18} for
SrCo$_2$V$_2$O$_8$ with those plotted in Fig. \ref{figure6PR} (a) for
the spin-$1/2$ chain with $\Delta =2$ and $J=3.55$\,meV confirms the excellent quantitative agreement
previously reported in Ref. \onlinecite{Wang_18}.

The $(h,\omega)$-plane lines of the energy in meV versus magnetic field in Tesla of the
sharp peaks $R^{+-}_{0}$, $R^{-+}_{\pi/2}$, $\chi^{(2)}_0$, $\chi^{(2)}_{\pi}$, and 
$\chi^{(3)}_{\pi/2}$ that, except for $\chi^{(2)}_0$, have been
experimentally observed in BaCo$_2$V$_2$O$_8$ by optical experiments \cite{Wang_19}
are plotted for the parameter set suitable to that material
in Fig. \ref{figure7PR} (a) for $h\in [h_{c1},h_{c2}]=[3.76\,$T$,22.97\,$T$]$.
The corresponding field dependencies of the $ab= +-,-+$, $n=1,2,3$, and $k=0,\pi/2,\pi$ negative exponents
$\zeta_{n}^{ab} (k)$, Eq. (\ref{MPSs}) of Appendix \ref{C}, that heve not been previously studied
by other authors are plotted in Fig. \ref{figure7PR} (b). 

\begin{figure}
\begin{center}
\centerline{\includegraphics[width=8.5cm]{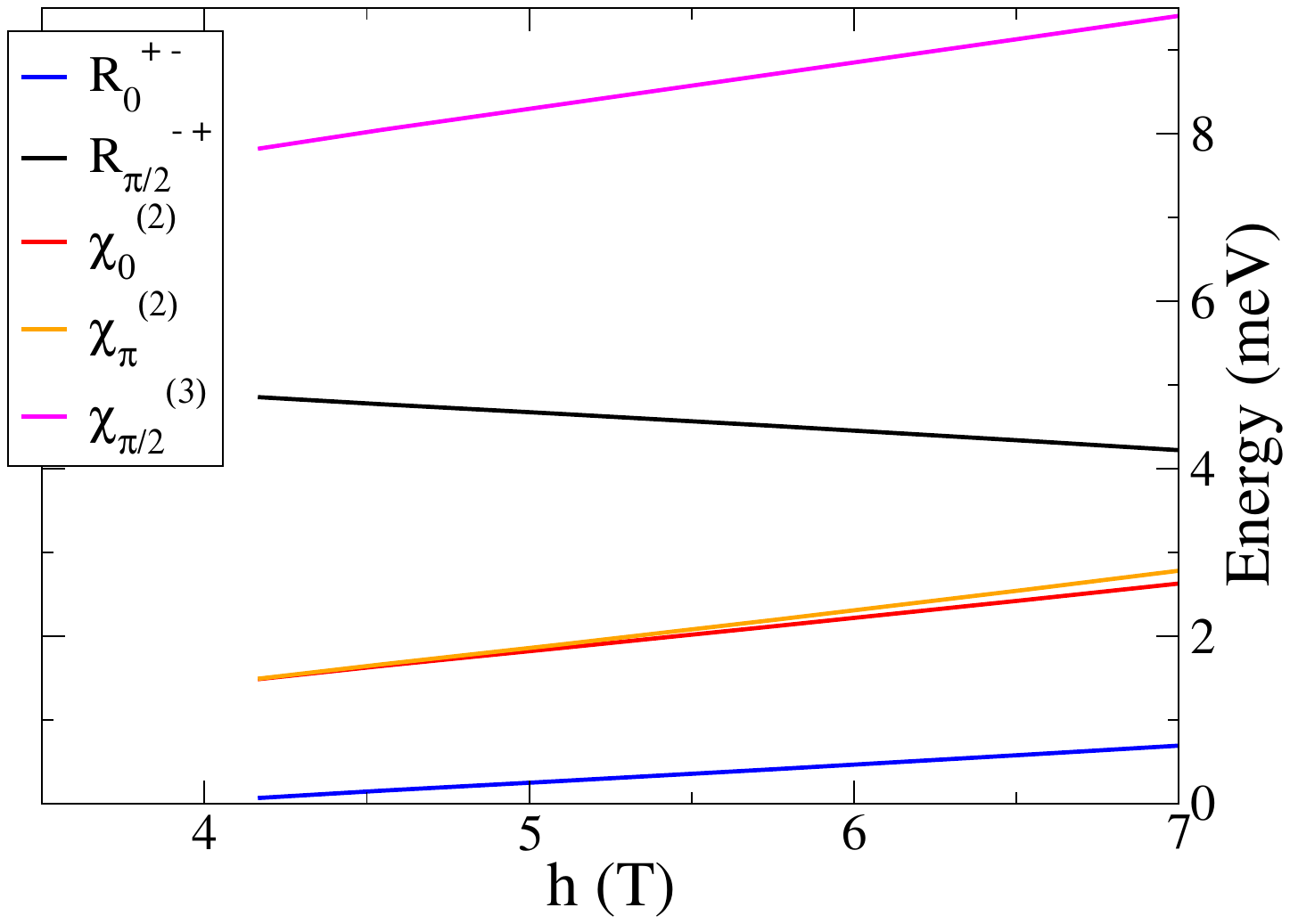}}
\caption{The same sharp-peaks $(h,\omega)$-plane lines of energy versus field as in Fig. \ref{figure7PR} for the 
smaller magnetic field interval $h\in [h_{c1},7\,$T$] = [3.76\,$T$,7\,$T$]$. 
After shifting upward the energies of the sharp peaks $R_{0}^{+-}$ and $R_{\pi/2}^{-+}$
by $\delta E_1 = 0.30$\,meV and those of the sharp peaks
$\chi^{(2)}_{\pi}$ and $\chi^{(3)}_{\pi/2}$ by $\delta E_2 = \delta E_3 = 0.50$\,meV,
their obtained energy versus field lines quantitatively agree with those
experimentally observed in BaCo$_2$V$_2$O$_8$ for $h\in [5\,$T$,7\,$T$]$ 
displayed in Fig. 4 (b) of Ref.  \onlinecite{Wang_19}.}
\label{figure8PR}
\end{center}
\end{figure}

The experimental studies of Ref. \onlinecite{Wang_19} have only considered $(h,\omega)$-plane points for
magnetic fields up to $7$\,T in the spin-conducting phase subinterval $h\in [5\,$T$,7\,$T$]$.
For the sake of comparison with corresponding experimental results
for BaCo$_2$V$_2$O$_8$, our theoretical $(h,\omega)$-plane sharp-peak energy versus field lines
are also plotted up to $7$\,T in Fig. \ref{figure8PR}, for the field subinterval $h\in [h_{c1},7\,$T$]=[3.76\,$T$,7\,$T$]$. 
The corresponding negative exponents $h$ dependencies refer for that field subinterval to
those plotted in Fig. \ref{figure7PR} (b) for $h\in [h_{c1},h_{c2}]$.

To reach agreement with the experimental values of the sharp 
peak energies, the corresponding theoretical values as obtained by the finite-size algebraic method of
Ref. \onlinecite{Yang_19} were in Ref. \onlinecite{Wang_19} shifted upward by the energy $\delta E = 0.50$\,meV, which
is smaller than the lower-energy limit of the spectroscopy of that reference. 
After shifting upward the energies of the lines plotted in Fig. \ref{figure8PR}
of the sharp peaks $R_{0}^{+-}$ and $R_{\pi/2}^{-+}$
by $\delta E_1 = 0.30$\,meV and those of the sharp peaks $\chi^{(2)}_{\pi}$ and
$\chi^{(3)}_{\pi/2}$ by $\delta E_3 = 0.50$\,meV,
their obtained energy versus field lines indeed quantitatively agree with those
experimentally observed in BaCo$_2$V$_2$O$_8$ for $h\in [5\,$T$,7\,$T$]$ 
displayed in Fig. 4 (b) of Ref.  \onlinecite{Wang_19}.

\begin{figure}
\begin{center}
\centerline{\includegraphics[width=8.5cm]{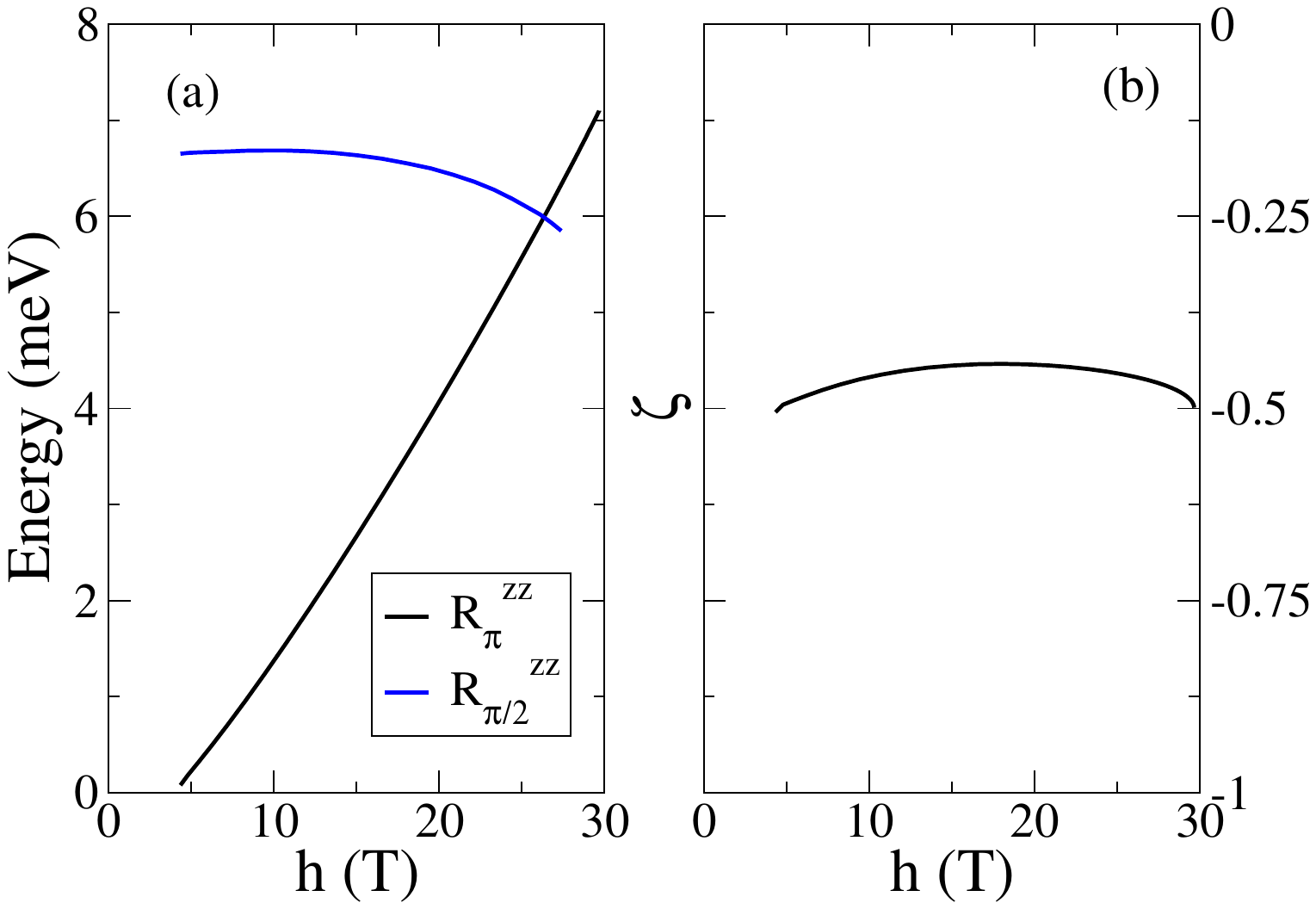}}
\caption{The dependencies on the magnetic field $h\in [h_{c1},h_{c2}]$ 
in Tesla of the energies in meV of the sharp peak $R_{\pi}^{zz}$ and 
of the sharp peak $R_{\pi/2}^{zz}$ for the fields $h\in [h_{c1},h_{\diamond}]$ 
for which it exists, both such peaks being experimentally observed in SrCo$_2$V$_2$O$_8$ 
by neutron scattering (a); The corresponding negative exponent
of $R_{\pi}^{zz}$ given in Eq. (\ref{EZZPI}) of Appendix \ref{C} (b). 
Expressions of such peaks energies are provided in Eqs. (\ref{EZZPI}) and 
(\ref{EZZPI2}) of Appendix \ref{C}, respectively.}
\label{figure9PR}
\end{center}
\end{figure}

Finally, the $(h,\omega)$-plane lines of the energy in meV versus field in Tesla of the
sharp peaks $R_{\pi}^{zz}$ and $R_{\pi/2}^{zz}$ experimentally observed in SrCo$_2$V$_2$O$_8$ by
neutron scattering are plotted in Fig. \ref{figure9PR} (a). The negative exponent that controls the line shape
near the former sharp peak is plotted as a function of the field $h$ in Fig. \ref{figure9PR} (b).
As reported in Appendix \ref{C}, the sharp peak $R_{\pi/2}^{zz}$ exists for 
spin densities $m\in [0,m_{\diamond}]$ and magnetic fields $h\in [h_{c1},h_{\diamond}]$ where 
for anisotropy $\Delta =2$ one has that $m_{\diamond} = 0.627$ and $h_{\diamond} = 2.76$
in units of $J/(g\mu_B)$ that for $J=3.55$\,meV corresponds to $h_{\diamond} = 27.30$\,T.

The experimental studies of Ref. \onlinecite{Bera_20} have considered $(h,\omega)$-plane lines for
magnetic fields up to $15$\,T in the spin-conducting phase subinterval $h\in [3.8\,$T$,15.0\,$T$]$.
Comparison with the experimental dependence on the magnetic field $h$ of the energies of the
sharp peaks $R^{\rm PAP(zz)}_{\pi/2}$ and $R^{\rm PAP(zz)}_{\pi}$ displayed in Figs. 5-a and 5-b, respectively,
with those plotted in Fig. \ref{figure9PR} (a) for the spin-$1/2$ chain with $\Delta =2$ and $J=3.55$\,meV confirms again the quantitative agreement
previously reported in Ref. \onlinecite{Bera_20}. Note that in the larger field interval $h\in [3.8\,$T$,27.3\,$T$]$ 
of Fig. \ref{figure9PR} (a) for which the sharp peak $R_{\pi/2}^{zz}$ exists 
its energy is not independent of the magnetic field $h$, as suggested from its dependence up to $15$\,T 
shown in Fig. 5-a of that reference. Indeed and as shown in Fig. \ref{figure9PR} (a) for anisotropy $\Delta =2$ and 
$J=3.55$\,meV, upon increasing the magnetic field $h$ within that interval, the theoretical energy of the sharp peak 
$R_{\pi/2}^{zz}$ decreases from $6.66$\,meV at $h=3.8$\,T to 5.79\,meV at $h=27.3$\,T.

Importantly, the experimental intensity of the longitudinal sharp peak $R^{\rm PAP(zz)}_{\pi/2}$ and
particularly of the longitudinal sharp peak $R^{\rm PAP(zz)}_{\pi}$ shown in Fig. 5-b of Ref. \onlinecite{Bera_20}
is larger than those of the transverse sharp peaks. This is an issue discussed in the ensuing section.\\

\section{Effects of selective interchain couplings}
\label{SECIV}

Here we clarify issues concerning the coexistence in BaCo$_2$V$_2$O$_8$ and SrCo$_2$V$_2$O$_8$'s
low-temperature spin-conducting phases of 1D physics with important deviations from it invoking the symmetry 
space group of their crystal structure. Both such zigzag materials have similar chain structures along the $c$-axis,
being almost iso-structural: BaCo$_2$V$_2$O$_8$ has a centro-symmetric crystal structure ($I 4_{1}/acd$, nonpolar),
while SrCo$_2$V$_2$O$_8$ has a non-centro-symmetric crystal structure ($I 4_{1}/cd$, polar) \cite{Okutani_15}.

Hopping-matrix elements associated with interchain couplings are obtained by the overlap between the 
wave functions of the excited states and the one-particle potential that transforms according to the underlying lattice symmetries. 
The overlap is largest and spin states are coupled more strongly whenever they are connected by a symmetry operation of the 
underlying lattice. The four-fold rotation with additional translation of 1/4th of the unit cell of these
zigzag materials allows for a coupling between different chains and antiferromagnetic intrachain 
coupling naturally leads to antiferromagnetic NN and NNN interchain couplings. 

The additional translation takes care of the change of chirality between adjacent chains and for an anti-ferromagnetic spin order, 
only states with the same spin-projection yield a finite overlap. On the contrary, for excitations that involve a spin-flip the 
interchain coupling should tend to zero. In the case of excited states, the symmetry operation 
involving the four-fold rotation with additional translation of 1/4th of the unit cell
is thus only a symmetry in spin-space if {\it no} electronic spin flip is performed within the generation of such states. 

We provide strong evidence that this explains why interchain couplings can be neglected concerning the spin 
dynamical structure factor transverse components $S^{+-} (k,\omega)$ and $S^{-+} (k,\omega)$: The 
transverse excitations contributing to them involve an electronic spin flip.
This though does not apply to the longitudinal component $S^{zz} (k,\omega)$ whose longitudinal 
excitations do not involve such a spin flip. 

This selection rule is thus expected to be behind selective interchain couplings that both protect the 1D physics
of BaCo$_2$V$_2$O$_8$ and SrCo$_2$V$_2$O$_8$ and lead to deviations from it, mainly associated with the 
enhancement of the spectral-weight intensity of $S^{zz} (k,\omega)$. 

\subsection{1D physics preserved by selective interchain couplings}
\label{SECIVA}

We start by discussing which low-temperature 1D physics 
is preserved and protected by selective interchain couplings based on available experimental data on the two
zigzag materials. It is well known that the 1D physics of quasi-1D spin-chain compounds occurs for the 
spin-conducting phases at low temperatures above a very small critical temperature $T_c (h)$ below 
which interchain couplings lead to three-dimensional (3D) ordered phases \cite{Okunishi_07}. 

Magnetization experimental results for BaCo$_2$V$_2$O$_8$ and SrCo$_2$V$_2$O$_8$ are 
explained well in terms of a 1D spin-$1/2$ Heisenberg-Ising chain in longitudinal magnetic fields with
anisotropy $\Delta\approx 2$ \cite{Kimura_07,Okunishi_07,Kimura_08,Han_21}.
The same applies to the magnetic field dependence of the sharp peaks 
energies experimentally observed in the dynamic structure factor \cite{Wang_18,Wang_19,Bera_20},
as we have discussed above in Sec. \ref{SECIII}. 

Other experimental studies refer for instance to the NMR relaxation rate. For both 
BaCo$_2$V$_2$O$_8$ \cite{Klanjsek_15} and SrCo$_2$V$_2$O$_8$ \cite{Cui_22} they 
have been performed on ${}^{51}{\rm V}$ nuclei. The NMR relaxation rate can be expressed as \cite{Dupont_16},
\begin{eqnarray}
{1\over T_1} & = & {1\over T_1^{\parallel}} + {1\over T_1^{\perp}}\hspace{0.20cm}{\rm where}
\nonumber \\
{1\over T_1^{\parallel}} & = & {\gamma^2\over 2}\sum_k\vert A_{\parallel} (k)\vert^2 
S^{zz} (k,\omega_0)
\hspace{0.20cm}{\rm and} 
\nonumber \\
{1\over T_1^{\perp}}  & = & {\gamma^2 \over 2}\sum_k\vert A_{\perp} (k)\vert^2 (S^{+-} (k,\omega_0)+S^{-+} (k,\omega_0)) \, .
\nonumber \\
\label{T1-1}
\end{eqnarray}
Here $\omega_0$ is the NMR frequency, $\gamma$ is the gyromagnetic ratio, and $A_{\parallel} (k)$ and $A_{\perp} (k)$
are the longitudinal and transverse hyperfine form factors, respectively.
In the case of the zigzag materials under study, these two hyperfine form factors
are peaked at  $k=2k_{F\downarrow}$ and $k=\pi$, respectively \cite{Klanjsek_15,Cui_22}.
\begin{figure}
\begin{center}
\subfigure{\includegraphics[width=4.25cm]{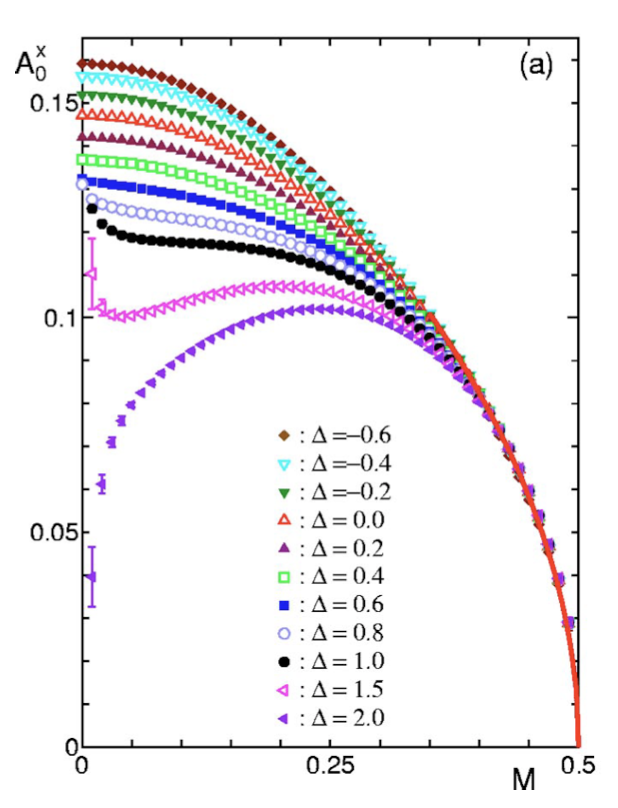}}
\subfigure{\includegraphics[width=4.25cm]{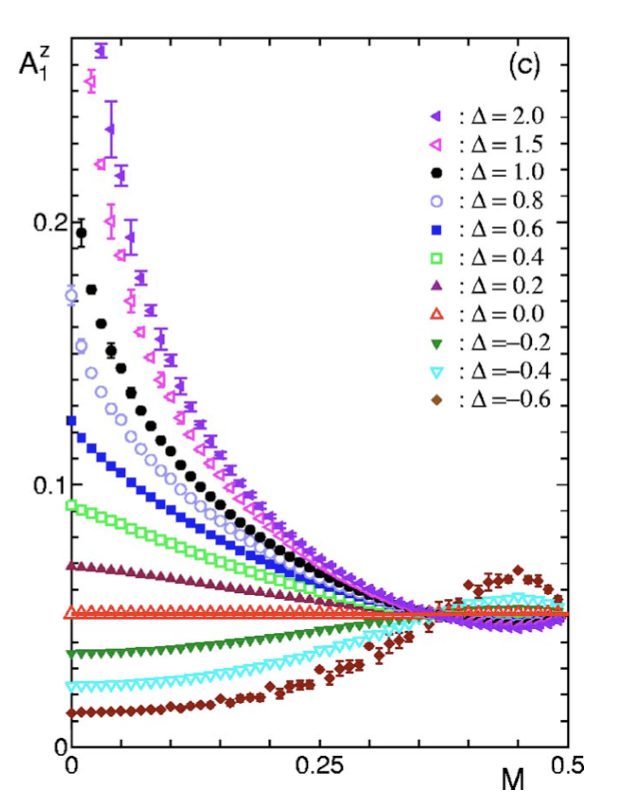}}
\caption{The dependence of the pre-factors $A_x^0$ (a) and $A_z^1$ (b) of the static spin correlation functions 
on $M = m^z = m/2$ for spin density $m\in [0,1]$ at different anisotropy 
values $\Delta$ for the spin-$1/2$ $XXZ$ chain in a longitudinal magnetic field. 
The lines of importance for this study refer to anisotropy $\Delta = 2$.
From Ref. \onlinecite{Okunishi_07}.}
\label{figure10PR}
\end{center}
\end{figure}

In case that for magnetic fields $h_{c1}<h<h_{c2}$ and small temperatures just above the very small critical temperature 
$T_c (h)$ the zigzag materials BaCo$_2$V$_2$O$_8$ and SrCo$_2$V$_2$O$_8$ were fully described by 
the 1D spin-$1/2$ $XXZ$ chain in a longitudinal magnetic field, the longitudinal and transverse terms in Eq. (\ref{T1-1}) 
of their NMR relaxation rate would have for low-energy $\omega/(k_B T)\ll 1$ the following expression \cite{Dupont_16},
\begin{eqnarray}
{1\over T_1^{\parallel}} & = & {\gamma^2\,\vert A_{\parallel} (2k_{F\downarrow})\vert^2\over 2}\,{A_1^z\cos (\pi \xi^2)\over v_1}
B (\xi^2, 1 - 2\xi^2)
\nonumber \\
& \times & \left({2\pi\,T\over v_1}\right)^{\zeta_{\parallel}}\hspace{0.20cm}{\rm and} 
\nonumber \\
{1\over T_1^{\perp}}  & = & {\gamma^2\,\vert A_{\perp} (\pi)\vert^2\over 2}\,{A_{0}^x\cos\left({\pi\over 4\xi^2}\right)\over v_1}
B \left({1\over 4\xi^2}, 1 - {1\over 2\xi^2}\right)
\nonumber \\
& \times & \left({2\pi\,T\over v_1}\right)^{\zeta_{\perp}} \, .
\label{T1-1limit-1D}
\end{eqnarray}
Here $\xi$ is the phase-shift related parameter in Eq. (\ref{x-aa}) of Appendix \ref{A}
whose direct relation to the usual Tomonaga-Luttinger liquid (TLL) parameters is
discussed below, the coordination number $c_n$ reads $c_n = 4$ for $3D$, $J'$
is the effective interchain coupling, $v_1 = v_1 (k_{F\downarrow})$ is the $1$-pair group velocity 
at $q=k_{F\downarrow}$, and $B (x,y)$ is the Euler beta function that can be expressed in
terms of the gamma function as $B (x,y)=\Gamma (x)\Gamma (y)/\Gamma (x+y)$.

The non-universal TLL pre-factors $A_0^x$ and $A_1^z$ of the static spin correlation functions 
also appearing in Eq. (\ref{T1-1limit-1D}) can be numerically computed \cite{Hikihara_04}. They
are plotted in Fig. \ref{figure10PR} (a) and (b), respectively, as a function of $M = m^z = m/2$ for spin density $m\in [0,1]$ and
several $\Delta$ values. Upon increasing $m$ and the magnetic
field $h$ for anisotropy $\Delta =2$ of interest for the zigzag materials, $A_0^x$ first increases from
$A_0^x = 0$ or a very small finite value for $m\rightarrow 0$ and $h\rightarrow h_{c1}$, goes through a maximum 
$A_0^x\approx 0.1$ at around $m=1/2$ and $h=h_{1/2}$, and then continuously decreases with final limiting behavior
$A_0^x = {c_x\over 2\sqrt{2\pi}}\sqrt{1-m}$ for $(1-m)\ll 1$ and small $(h_{c2}-h)$ 
where $c_x= \pi\sqrt{e}/(2^{1/3}A^6) = 0.92418...$ and $A$ is the Glaisher's constant. Also at $\Delta =2$ 
the pre-factor $A_1^z$ diverges as $m\rightarrow 0$ and $h\rightarrow h_{c1}$.
It continuously decreases upon increasing the spin density $m$ and the magnetic field $h$,
going through a minimum $A_1^z\approx 0.045$ at $m\approx 0.875$ and then increasing
to $A_1^z={1\over 2\pi^2}\approx 0.05$ for $m\rightarrow 1$ and $h\rightarrow h_{c2}$. 

\begin{figure}
\begin{center}
\centerline{\includegraphics[width=8.5cm]{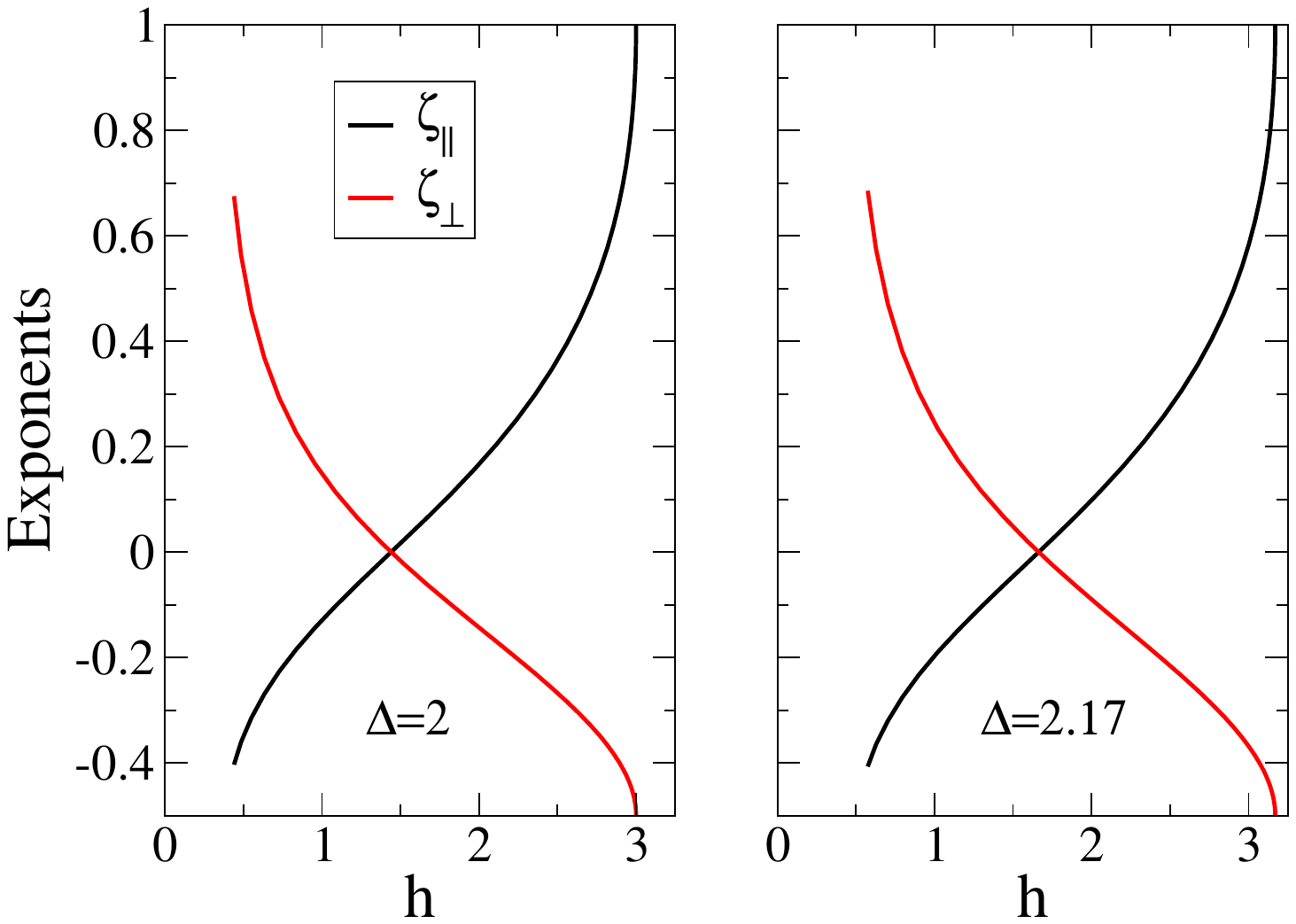}}
\caption{The exponents $\zeta_{\parallel}$ and $\zeta_{\perp}$, Eq. (\ref{exppnNMR}), 
plotted as a function of the magnetic field $h\in [h_{c1},h_{c2}]$ for anisotropies 
$\Delta =2$ and $\Delta =2.17$.}
\label{figure11PR}
\end{center}
\end{figure}

The exponents $\zeta_{\parallel}$ and $\zeta_{\perp}$ in the expressions of Eq. (\ref{T1-1limit-1D})
are given by,
\begin{equation}
\zeta_{\parallel} = 2\xi^2 - 1\hspace{0.20cm}{\rm and}\hspace{0.20cm}
\zeta_{\perp} = {1\over 2\xi^2} - 1 \, .
\label{exppnNMR}
\end{equation}
They are plotted in Fig. \ref{figure11PR} as a function of the magnetic field $h\in [h_{c1},h_{c2}]$
for anisotropies $\Delta = 2$ and $\Delta = 2.17$. Their limiting behaviors are,
\begin{eqnarray}
\zeta_{\parallel} & = & -1/2\hspace{0.20cm}{\rm and}\hspace{0.20cm}\zeta_{\perp} = 1
\hspace{0.20cm}{\rm for}\hspace{0.20cm}h\rightarrow h_{c1} 
\nonumber \\
\zeta_{\parallel} & = & \zeta_{\perp} = 0
\hspace{0.20cm}{\rm for}\hspace{0.20cm}h = h_* 
\nonumber \\
\zeta_{\parallel} & = & 1\hspace{0.20cm}{\rm and}\hspace{0.20cm}\zeta_{\perp} = -1/2
\hspace{0.20cm}{\rm for}\hspace{0.20cm}h\rightarrow h_{c2}  \, ,
\label{zetaLimits}
\end{eqnarray}
where the magnetic field $h_*$ is that where the lines for 
the exponents $\zeta_{\parallel}$ and $\zeta_{\perp}$ cross each other in Fig. \ref{figure11PR},
at which they read $\zeta_{\parallel}=\zeta_{\perp} =0$. 

The clarification above in Sec. \ref{SECIII} and in Appendix \ref{C} of the physical-spins scattering processes behind the 
1D physics's dynamical properties of the zigzag materials
is important for the discussion of which low-temperature 1D physics 
is preserved and protected by selective interchain couplings. For instance, 
the phase-shift related parameter $\xi$ and its inverse $\xi^{-1}=1/\xi$ 
appearing in Eq. (\ref{T1-1limit}) and also in the expressions of the exponents $\zeta_{\parallel}$ 
and $\zeta_{\perp}$ given in Eq. (\ref{exppnNMR}) are determined by physical-spins $1$-pair - $1$-pair 
scattering. Indeed, they are directly expressed in terms of the $1$-pair phase shift $2\pi\Phi_{1,1} (q,q')$,
Eqs. (\ref{Phi-barPhi}) and (\ref{Phis1n}) of Appendix \ref{A}, in units of $2\pi$ as follows,
\begin{equation}
\xi^{\pm 1} = 1 + \Phi_{1,1} (k_{F\downarrow},k_{F\downarrow}) \mp \Phi_{1,1}(k_{F\downarrow},-k_{F\downarrow}) \, ,
\label{x-aaPM}
\end{equation}
where in $\Phi_{1,1} (k_{F\downarrow},k_{F\downarrow})$ the two momenta differ by $2\pi/L$.

Importantly, it follows that the usual TLL parameters\cite{Horvatic_20} $K$  and $\eta = 1/(2K)$ (where
here $\eta$ {\it is not} the anisotropy parameter in $\Delta = \cosh\eta$) are determined by physical-spins $1$-pair - $1$-pair 
scattering. Indeed, they are directly related to the phase-shift parameters $\xi^{\pm 1}$, Eq. (\ref{x-aaPM}), as 
$K = \xi^2$ and $\eta = \xi^{-2}/2$, so that in terms of phase shifts in units of $2\pi$ they read
$K = (1 + \sum_{\iota = \pm1}(\iota)\Phi_{1,1} (k_{F\downarrow},\iota\,k_{F\downarrow}))^2$
and $\eta = {1\over 2}(1 + \sum_{\iota = \pm1}\Phi_{1,1} (k_{F\downarrow},\iota\,k_{F\downarrow}))^2$, respectively.

On the one hand and as discussed below in Sec. \ref{SECIVB}, an important deviation from 1D
physics is that only the longitudinal relaxation rate term $1/T_1^{\parallel}$ is experimentally
observed in both zigzag materials  \cite{Klanjsek_15,Cui_22}. On the other hand,
comparing the theoretical behavior $1/T_1=1/T_1^{\parallel}\propto (2\pi\,T/v_1)^{2\xi^2 - 1}$
of that term with the corresponding experimental data
for the whole field interval $h\in [h_{c1},h_{c2}]$, the excellent quantitative agreement for
$\eta = {1\over 2}(1 + \sum_{\iota = \pm1}\Phi_{1,1} (k_{F\downarrow},\iota\,k_{F\downarrow}))^2$
plotted in Fig. 3 (a) of Ref. \onlinecite{Klanjsek_15} for BaCo$_2$V$_2$O$_8$ and in Fig. 4 (d) of
Ref. \onlinecite{Cui_22} for SrCo$_2$V$_2$O$_8$ was reached. Also for the velocity called here
$v_1 = v_1 (k_{F\downarrow})$ there is excellent quantitative agreement between theory and experiments,
as reported in Fig. 3 (b) of Ref. \onlinecite{Klanjsek_15} for BaCo$_2$V$_2$O$_8$. 

Hence, the 1D physics phase-shift related parameters $\xi^{\pm 1} = 1 + \sum_{\iota = \pm1}(\iota)^{1\mp 1\over 2}\Phi_{1,1} (k_{F\downarrow},\iota\,k_{F\downarrow})$ and the $1$-pair group velocity $v_1 = v_1 (k_{F\downarrow})$ 
appearing in Eq. (\ref{T1-1limit-1D}) for the NMR relaxation rate 
at low-energy $\omega/(k_B T)\ll 1$ as well as that rate exponent $\zeta_{\parallel} = 2\xi^2 - 1$ 
are preserved by selective interchain couplings. 

There is overall agreement between the 1D physics distribution over the $(k,\omega)$-plane of the $k$ intervals for
which the sharp-peak exponents $\zeta_{n}^{ab} (k)$, Eqs. (\ref{zetaabk}) and (\ref{functional}) of Appendix \ref{C},
are negative and the $(k,\omega)$-plane location of the corresponding experimental observed sharp peaks.
As confirmed above in Sec \ref{SECIII}, the same applies to the distribution 
over the $(h,\omega)$-plane of the sharp peaks experimentally observed 
at the specific momenta $k=0,\pi/2,\pi$ by optical experiments in $S^{+-} (k,\omega)$ and $S^{-+} (k,\omega)$
\cite{Wang_18,Wang_19} and by neutron scattering in $S^{zz} (k,\omega)$\cite{Bera_20}. All such
agreements reveal that the sharp-peak energies and the phase shifts $2\pi\Phi_{1,n}(k_{F\downarrow},q)$ for $n=1,2,3$ 
in the expressions of the exponents that control the line shape at and near them 
are also preserved by selective interchain couplings.

\subsection{Deviations from 1D physics due to selective interchain couplings}
\label{SECIVB}

In case that for fields $h\in [h_{c1},h_{c2}]$ and low temperatures above the small critical temperature $T_c (h)$
the 1D physics fully applied to BaCo$_2$V$_2$O$_8$ and SrCo$_2$V$_2$O$_8$, the dependence 
on the magnetic field $h$ of the exponents $\zeta_{\parallel}$ and
$\zeta_{\perp}$ shown in Fig. \ref{figure11PR} would imply that the NMR 
spin-lattice relaxation rate $1/T_1 = 1/T_1^{\parallel} + 1/T_1^{\perp}$, Eq. (\ref{T1-1limit-1D}), 
was dominated by its divergent longitudinal term $1/T_1^{\parallel}$ for fields $h\in [h_{c1},h_*]$ when $\zeta_{\parallel}<0$ and 
$\zeta_{\perp} >0$ and by its divergent transverse term $1/T_1^{\perp}$ for $h\in [h_*h_{c2}]$ when $\zeta_{\parallel}>0$ 
and $\zeta_{\perp} <0$. Here $h_* = 1.441$ for $\Delta =2$ and $h_* = 1.664$ for $\Delta =2.17$ in units of 
$J/g\mu_B$, which for $J = 3.55$\,meV gives $h_* = 14.25$\,T and for $J = 2.60$\,meV 
leads to $h_* = 12.06$\,T, respectively. The corresponding magnetic energy $g\mu_B h_*$
refers to the middle dashed line in the  spin-$1/2$ $XXZ$ chain phase diagram 
of the magnetic energy over anisotropy, $g\mu_B h/\Delta$, versus $\epsilon = 1/\Delta\in [0,1]$
shown in Fig. \ref{figure1PR}.

In contrast to 1D physics, NMR experimental 
results of Ref. \onlinecite{Klanjsek_15} for BaCo$_2$V$_2$O$_8$ 
and of Ref. \onlinecite{Cui_22} for SrCo$_2$V$_2$O$_8$ though reveal that the longitudinal term
$1/T_1 = 1/T_1^{\parallel}\propto T^{\zeta_{\parallel}} = T^{2\xi^2 - 1}$ dominates for the {\it whole} 
magnetic field interval $h\in [h_{c1},h_{c2}]$ of the spin-conducting phases, including for
$h\in [h_{*},h_{c2}]$ when $1/T_1^{\perp}$ should dominate.

Note that the 1D value of $h_*$ at which 
$\zeta_{\parallel}=\zeta_{\perp} =0$ is typically larger that than that of the field $h=h_*$
at which the two $(T,h)$-plane critical-temperature $T_c^{z} (h)$ and $T_c^{x} (h)$ 
lines associated with longitudinal and transverse orders, respectively, considered below
cross each other in a system of weakly coupled chains
\cite{Okunishi_07}. The experimental values of $h_*$ for BaCo$_2$V$_2$O$_8$ and 
SrCo$_2$V$_2$O$_8$ suggested by neutron scattering are indeed lower and read 
$h_* \approx 8.5$\,T and $h_* \approx 7.0$\,T, respectively \cite{Grenier_15,Shen_19}.

One can calculate within interchain mean-field theory \cite{Okunishi_07} expressions for 
such critical temperatures $T_c^{z} (h)$ and $T_c^{x} (h)$, which read \cite{Okunishi_07},
\begin{eqnarray}
T_c^{z} (h) & = & {v_1\over 2\pi}\left(c_n\,\Delta\,J' \tilde{A}_1^z{\sin (\pi \xi^2)\over v_1}
B^2 \left({\xi^2\over 2}, 1 - \xi^2\right)\right)^{1\over 2 (1-\xi^2)}
\nonumber \\
T_c^{x} (h) & = & {v_1\over 2\pi}\left(c_n\,J' A_0^x{\sin \left({\pi\over 4\xi^2}\right)\over v_1}
B^2 \left({1\over 8\xi^2}, 1 - {1\over 4\xi^2}\right)\right)^{2\xi^2\over 4\xi^2 - 1} 
\label{TcTc}
\end{eqnarray}
Here the coordination number $c_n$ reads $c_n = 4$ for $3D$, $J'$
is the effective interchain coupling, and the other quantities are those appearing
in the relaxation rate expressions, Eq. (\ref{T1-1limit-1D}).
However, as justified below, in the expression for $T_c^{z} (h)$ given in Eq. (\ref{TcTc}), we have replaced  the TLL pre-factor $A_1^z$ plotted in 
Fig. \ref{figure10PR} (b) by a pre-factor $\tilde{A}_1^z$ which is sensitive to effects of selective interchain couplings.
The dependence on $J'$ of that pre-factor $\tilde{A}_1^z$ is beyond interchain mean-field theory. The corresponding
replacement of $A_1^z$ by $\tilde{A}_1^z$ is physically important for the following reason.  
It implies that in the expression, Eq. (\ref{T1-1limit-1D}), for the longitudinal relaxation rate term $1/T_1^{\parallel}$ the pre-factor $A_1^z$ 
is also replaced by $\tilde{A}_1^z$, so that,
\begin{eqnarray}
{1\over T_1^{\parallel}} & = & {\gamma^2\,\vert A_{\parallel} (2k_{F\downarrow})\vert^2\over 2}\,{\tilde{A}_1^z\cos (\pi \xi^2)\over v_1}
B (\xi^2, 1 - 2\xi^2)
\nonumber \\
& \times & \left({2\pi\,T\over v_1}\right)^{\zeta_{\parallel}} \, .
\label{T1-1limit}
\end{eqnarray}

Note that, in contrast to the critical temperatures shown in Eq. (\ref{TcTc}), for the purely 1D spin-$1/2$ $XXZ$ chain
the low-energy NMR relaxation rate expressions given by  Eq. (\ref{T1-1limit-1D}) do not depend explicitly on the 
effective interchain coupling $J'$. However, the component $1/T_1^{\parallel}$ as given in Eq. (\ref{T1-1limit}) 
implicitly depends on $J'$ through the pre-factor $\tilde{A}_1^z = \tilde{A}_1^z (J')$ that obeys the 
boundary condition $\tilde{A}_1^z (0) = A_1^z$. This is again beyond interchain mean-field theory.

On the other hand, it was confirmed above in Sec. \ref{SECIVA} in the basis of experimental data for the 
zigzag materials under study that, except for $A_1^z$, all TLL quantities in expression of 
$1/T_1^{\parallel}$, Eq. (\ref{T1-1limit}) in units of $\gamma^2\,\vert A_{\parallel} (2k_{F\downarrow})=1$,
refer to those predicted by the 1D physics. 

Fits of the magnetization measurements \cite{Okunishi_07,Kimura_08,Canevet_13} 
lead to $J'/J = 0.00138$ for BaCo$_2$V$_2$O$_8$. Consistently, 
it was found in Ref. \onlinecite{Klanjsek_15} by the use of the
expression for $T_c^{z} (h)$ given in Eq. (\ref{TcTc}) with $\tilde{A}_1^z$ replaced by 
$A^z_1$ that for fields $h>h_{c1}$ up to $8.5$\,T the effective interchain coupling
in that expression reads $J'/K_B = 0.042$\,K and thus $J' = 0.0036$\,meV.
For $J = 2.60$\,meV this gives $J'/J = 0.00139$, consistently with the
magnetization measurements value $J'/J = 0.00138$. 
Nonetheless, a giant variation of the effective interchain coupling $J'(h)$ by a factor up to 24 was found
upon increasing the magnetic field $h$ from $h=8.5$\,T towards $h=h_{c2}$ \cite{Klanjsek_15}.

The pre-factors $A_1^z$ and $A_0^x$ in the expressions of $1/T_1^{\parallel}$ and $1/T_1^{\perp}$
given Eq. (\ref{T1-1limit-1D}) are controlled by matrix-element's overlaps within the
dynamic structure factor's components $S^{zz} (k,\omega_0)$ and $S^{+-} (k,\omega_0)+S^{-+} (k,\omega_0)$,
respectively, in the NMR relaxation expression, Eq. (\ref{T1-1}). According to the selection rule
associated with selective interchain couplings, $A_0^x$ remains insensitive to the latter.
Such selective interchain couplings though affect the spin-states quantum overlaps that control 
the pre-factor $A_1^z$ associated with $S^{zz} (k,\omega)$, which are sensitive to the variation of $J'(h)$. 

Hence we propose that beyond interchain mean-field theory \cite{Okunishi_07}  in the expression for $T_c^{z} (h)$
(Eq. (\ref{TcTc})), the giant enhancement of $J'(h)$ for $h>h_*=8.5$\,T detected in Ref. \onlinecite{Klanjsek_15}
is actually distributed between $J'$ and $\tilde{A}_1^z$. This implies that
such a giant variation refers to the {\it product} $J' \times \tilde{A}_1^z$
rather than to $J'$ alone. It then follows that the effective interchain coupling of Ref. \onlinecite{Klanjsek_15},
which we denote by $J'_{\rm Ref. 10} (h)$, is replaced by the quantity,
\begin{equation}
C_1^z\,C'\,J'_{\rm min}\hspace{0.20cm}{\rm where}\hspace{0.20cm}
C_1^z = {\tilde{A}_1^z\over A_1^z}
\hspace{0.20cm}{\rm and}\hspace{0.20cm}
C' = {J'\over J'_{\rm min}} \, ,
\label{subs}
\end{equation}
such that $C_1^z\,C'\,J'_{\rm min}=J'_{\rm Ref. 10} (h)$.
Here  $J'_{\rm min} = 0.00139 J$, $J'= J'(h)<J'_{\rm Ref. 10} (h)$ is the enhanced effective coupling,
and $A_1^z$ is the non-universal TLL longitudinal pre-factor of the static spin correlation functions
plotted in Fig. \ref{figure10PR} (b). While both $\tilde{A}_1^z$ and $J'$ are enhanced, we cannot 
access the precise values of their separate enhancement factors $C_1^z=\tilde{A}_1^z/A_1^z$ and $C'=J'/J'_{\rm min}$, respectively, 
although we know that their product gives $C_1^z (h)\times C'(h)\in [1,24]$ for $h\in [h_*,h_{c2}]$.

The field interval $h\in [h_*,h_{c2}]$ for which the enhancement of $C_1^z\,C'\,J'_{\rm min}=J'_{\rm Ref. 10} (h)$ 
was found in Ref. \onlinecite{Klanjsek_15} is precisely that for which {\it in contrast to the 1D physics} there is 
unexpected experimental dominance of the relaxation rate longitudinal component $1/T_1^{\parallel}\propto T^{\zeta_{\parallel}}$ 
relative to $1/T_1^{\perp}\propto T^{\zeta_{\perp}}$, in spite of $\zeta_{\parallel} >0$ and $\zeta_{\perp} < 0$. 
This is thus consistent with the corresponding enhancement by $C_1^z = \tilde{A}_1^z/A_1^z$ of the pre-factor 
$\tilde{A}_1^z$ in the $1/T_1^{\parallel}$'s expression, Eq. (\ref{T1-1limit}).
Indeed, due to selective interchain couplings that act on $S^{zz} (k,\omega)$,
also the ratio $\tilde{A}_1^z/A_0^x$ of the pre-factors $\tilde{A}_1^z$ and $A_0^x$ of
the expressions of $T_1^{\parallel}$ and $1/T_1^{\perp}$ in Eq. (\ref{T1-1limit}), respectively,
is enhanced relative to the corresponding ratio of the 1D physics, $A_1^z/A_0^x$.

The unexpected experimental low-temperature dominance of the longitudinal NMR relaxation rate term 
$1/T_1 = T_1^{\parallel}\propto T^{\zeta_{\parallel}}$ for magnetic fields $h\in [h_*,h_{c2}]$ found both
in BaCo$_2$V$_2$O$_8$ \cite{Klanjsek_15} and in SrCo$_2$V$_2$O$_8$ \cite{Cui_22} is thus here
associated with the enhancement of $\tilde{A}_1^z$ by $C_1^z = \tilde{A}_1^z/A_1^z$ in both such zigzag materials.
That dominance is not mainly due to the relative values of the hyperfine form 
factors $A_{\parallel} (k)$ and $A_{\perp} (k)$ in Eqs. (\ref{T1-1}) and (\ref{T1-1limit}): It rather 
mainly follows from the effects of selective interchain couplings
on the quantum overlaps within the matrix elements of $S^{zz} (k,\omega)$.

Note though that the weaker effects of transverse staggered fluctuations 
are behind the experimental studies of SrCo$_2$V$_2$O$_8$ showing a NMR line splitting that indicates 
the onset of transverse fluctuations \cite{Cui_22} at $h=h_*\approx 7.0$\,T. This confirms that the transverse NMR form factor $A_{\perp} (k)$ 
does not vanish. Consistently, transverse fluctuations and corresponding peaks have been
observed by neutron scattering for magnetic fields $h\in [h_*,h_{c2}]$ both in BaCo$_2$V$_2$O$_8$ 
\cite{Grenier_15} and in SrCo$_2$V$_2$O$_8$ \cite{Shen_19}. This suggests some degree of coexistence of 
both longitudinal and transverse orders \cite{Grenier_15}, in spite of the experimental dominance of the longitudinal
NMR relaxation rate term $1/T_1 = T_1^{\parallel}\propto T^{2\xi^2 - 1}$. 
  
Importantly, the additional $S^{zz} (k,\omega)$'s spectral-weight intensity brought about by selective interchain couplings
also applies to higher energy scales. Indeed, it is also clearly visible by neutron scattering in 
$S^{zz} (k,\omega)$ for larger $\omega$ values, as shown in Fig. 5-b of Ref. \onlinecite{Bera_20} 
for the magnetic field interval $h\in [3.8\,{\rm T},15\,{\rm T}]$, in what the longitudinal 
sharp peak $R^{\rm PAP(zz)}_{\pi}$ (called in this paper $R_{\pi}^{zz}$) is concerned. 
The intensity of such a sharp peak's spectral weight and that of the longitudinal sharp peak $R^{\rm PAP(zz)}_{\pi/2}$ 
(called here $R_{\pi/2}^{zz}$) shown in Fig. 5-a of that reference for fields larger than $h_{c1}$, called $B_c$ in these figures, is 
larger than that of the transverse sharp peaks. 
Note that for higher energies the enhancement occurs for a larger field interval than reported above
for low energy.

\section{Concluding remarks}
\label{SECV}

In this paper we have explained the coexistence in the low-temperature spin-conducting phases 
of the zigzag materials BaCo$_2$V$_2$O$_8$ and SrCo$_2$V$_2$O$_8$ of 1D physics with important 
deviations from it as a result of selective interchain couplings. Those involve a selection rule that follows from interchain spin 
states being coupled more strongly within the spin dynamical structure factor 
whenever they are connected by a symmetry operation of the underlying lattice reported in Sec. \ref{SECIV}.
In the case of excited states, this symmetry operation 
is only a symmetry in spin-space if {\it no} electronic spin flip is performed within the generation of such states.

Deviations from 1D physics due to selective interchain couplings 
are behind the enhancement of the spectral-weight intensity of the longitudinal component $S^{zz} (k,\omega)$ 
and the corresponding dominance at low energy $\omega/(k_B T)\ll 1$ and for fields $h\in [h_*,h_{c2}]$ of the longitudinal NMR relaxation rate term
$1/T_1 = 1/T_1^{\parallel}\propto T^{2\xi^2 - 1}$ of both BaCo$_2$V$_2$O$_8$ \cite{Klanjsek_15} and
SrCo$_2$V$_2$O$_8$ \cite{Cui_22}, in contrast to the 1D physics.  

Concerning the 1D physics protected by such selective interchain couplings, the excellent 
quantitative agreement between theoretical results and the experimentally observed $(k,\omega)$-plane
and $(h,\omega)$-plane locations of the sharp peaks 
confirmed by our study is consistent with the physical-spins $1$-pair - $1$-pair and $1$-pair - $n$-pair scattering 
controlling the $(k,\omega)$-plane line shape at and in the vicinity of the sharp peaks in 
$S^{+-} (k,\omega)$, $S^{-+} (k,\omega)$, and $S^{zz} (k,\omega)$
experimentally observed in SrCo$_2$V$_2$O$_8$ and BaCo$_2$V$_2$O$_8$ \cite{Wang_18,Wang_19,Bera_20}. 
In Appendix \ref{B} we have also identified the spin carriers behind
the spin transport properties of the spin conducting phases.

In the case of $S^{zz} (k,\omega)$ it is found that selective interchain couplings
enhance its spectral-weight intensity without changing its sharp-peaks's energies
and the $1$-pair scattering phase shifts in the  expressions of the exponents 
that control the line shape at and near the sharp peaks. 
We suggest neutron scattering experiments in 
$S^{zz} (k,\omega)$ for magnetic fields above $15$\,T to search for further effects 
of the selective interchain couplings in what the enhancement of its spectral-weight intensity
is concerned. 

The main results of this paper are: 1) The physical origin of the coexistence of 1D physics with deviations from it
results in the low-temperature spin-conducting phases of BaCo$_2$V$_2$O$_8$ and SrCo$_2$V$_2$O$_8$ from 
selective interchain couplings, which are behind the enhancement of the spectral-weight 
intensity of $S^{zz} (k,\omega)$ and of the resulting dominance at low energy $\omega/(k_B T)\ll 1$ 
of the longitudinal NMR relaxation rate term for fields $h\in [h_*,h_{c2}]$; and 2) The scattering of the physical-spins
$1$-pair - $1$-pair, $1$-pair - $2$-pair, and $1$-pair - $3$-pair directly controls 
the line shape at and near the sharp peaks in $S^{+-} (k,\omega)$, $S^{-+} (k,\omega)$, and $S^{zz} (k,\omega)$ 
of these zigzag materials. These insights have opened the door to a key 
advance in the understanding of the physics of the spin chains in BaCo$_2$V$_2$O$_8$ and SrCo$_2$V$_2$O$_8$. 

\acknowledgements
We thank Francisco (Paco) Guinea, Masanori Kohno, Toma\v{z} Prosen, and Zhe Wang for fruitful discussions.
J. M. P. C. would like to thank the Boston University's Condensed Matter Theory Visitors Program for support
Boston University for hospitality during the initial period of this research, and the Inst. Madrileno Estudios Avanzados Nanociencia 
IMD for hospitality during its last period. J.M.P.C. also acknowledges support from
FCT through the Grants  Grant UID/FIS/04650/2013, PTDC/FIS-MAC/29291/2017, and SFRH/BSAB/142925/2018.
P. D. S. acknowledges the support from FCT through the Grant UID/CTM/04540/2019.
T. S. was supported by grant PID2020-113164GBI00 funded by MCIN/AEI/10.13039/501100011033.\\ \\ 
\appendix

\section{$n$-pairs quantities}
\label{A}

For the spin-$1/2$ $XXZ$ chain with anisotropy $\Delta\geq 1$,
the $n$-pairs energy dispersions that appear in the expressions of the spin dynamic structure factor
spectra have the following general form for $n\geq 1$ \cite{Carmelo_22},
\begin{eqnarray}
\varepsilon_{n} (q) & = & {\bar{\varepsilon}_{n}} (\varphi_n (q)) \hspace{0.20cm}{\rm and}\hspace{0.20cm}
\varepsilon_{n}^0 (q)={\bar{\varepsilon}_{n}^0} (\varphi_n (q))\hspace{0.20cm}{\rm where}
\nonumber \\
{\bar{\varepsilon}_{n}} (\varphi) & = & {\bar{\varepsilon}_{n}^0} (\varphi) + \Bigl(n - \delta_{n,1}{1\over 2}\Bigr)\, g\mu_B\,h
\hspace{0.20cm}{\rm for}\hspace{0.20cm}h\in [0,h_{c1}]
\nonumber \\
{\bar{\varepsilon}_{n}} (\varphi) & = &
{\bar{\varepsilon}_{n}^0} (\varphi) + n\,g\mu_B\,h
\hspace{0.20cm}{\rm for}\hspace{0.20cm}h\in ]h_{c1},h_{c2}] \, .
\label{equA4n}
\end{eqnarray}
Here the $n$-band momenta read $q \in [-k_{F\uparrow},k_{F\uparrow}]$ for $n=1$ and
$q \in [-(k_{F\uparrow} - k_{F\downarrow}),(k_{F\uparrow} - k_{F\downarrow})]$ for $n>1$,
$\varphi = \varphi_{n}(q)\in [-\pi,\pi]$ are for $n\geq 1$ the ground-state rapidity functions 
that are solutions of Bethe-ansatz equations \cite{Carmelo_22,Gaudin_71}, 
$B = \varphi_{1}(k_{F\downarrow})$, and the rapidity-dependent dispersions ${\bar{\varepsilon}_{n}^0} (\varphi)$
are defined below.

\begin{figure}
\begin{center}
\centerline{\includegraphics[width=8.5cm]{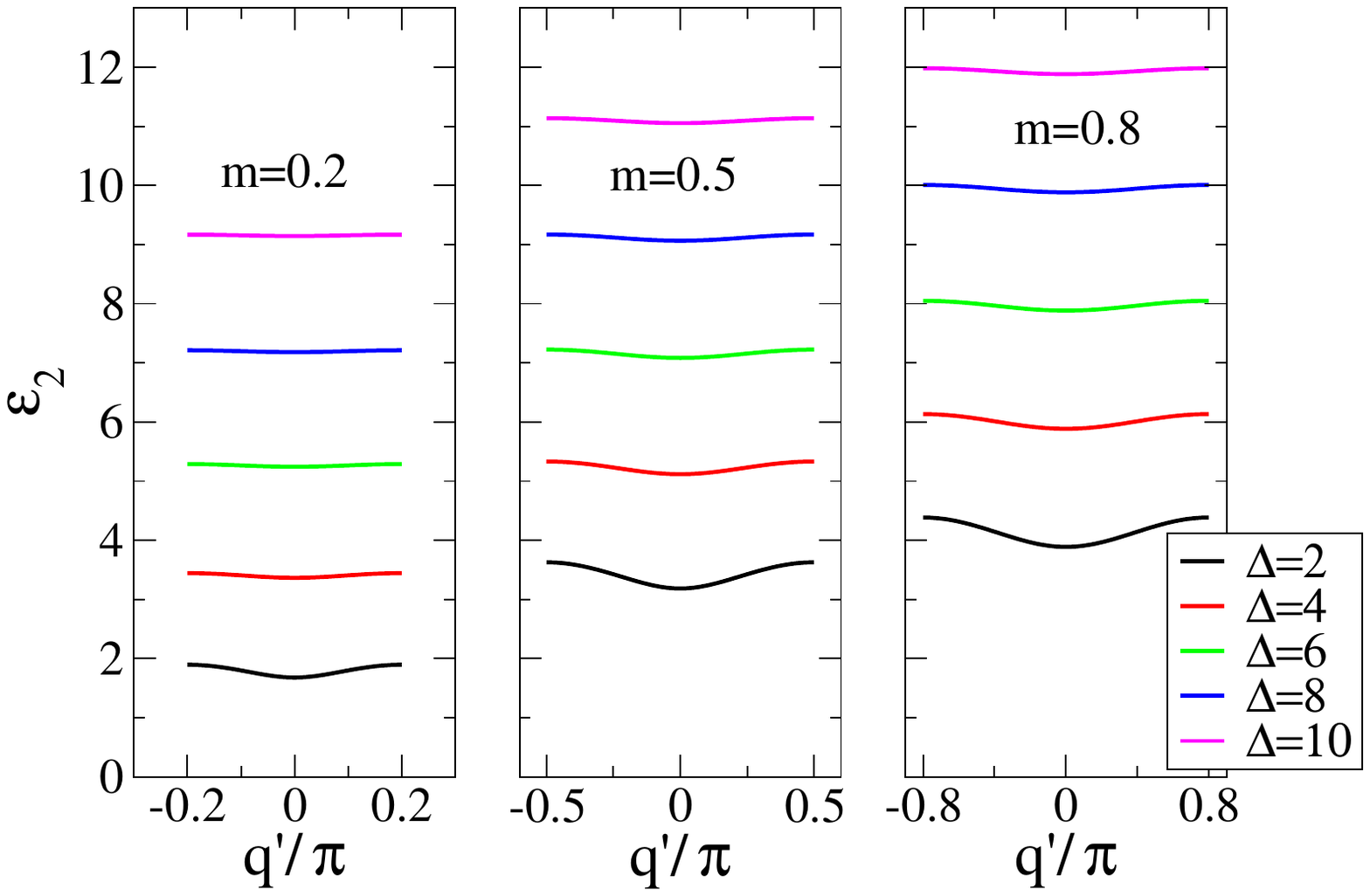}}
\caption{The $2$-string-pair energy dispersion $\varepsilon_{2} (q')$ 
in units of $J$ is plotted as a function of $q'/\pi$ for $2$-band momentum $q'\in [-(k_{F\uparrow} - k_{F\downarrow}),(k_{F\uparrow} - k_{F\downarrow})]$,
spin densities $m=0.2$, $m=0.5$, $m=0.8$, and anisotropies $\Delta =2,4,6,8,10$.
It is associated with a Bethe string of length two.}
\label{figure12PR}
\end{center}
\end{figure}

\begin{figure}
\begin{center}
\centerline{\includegraphics[width=8.5cm]{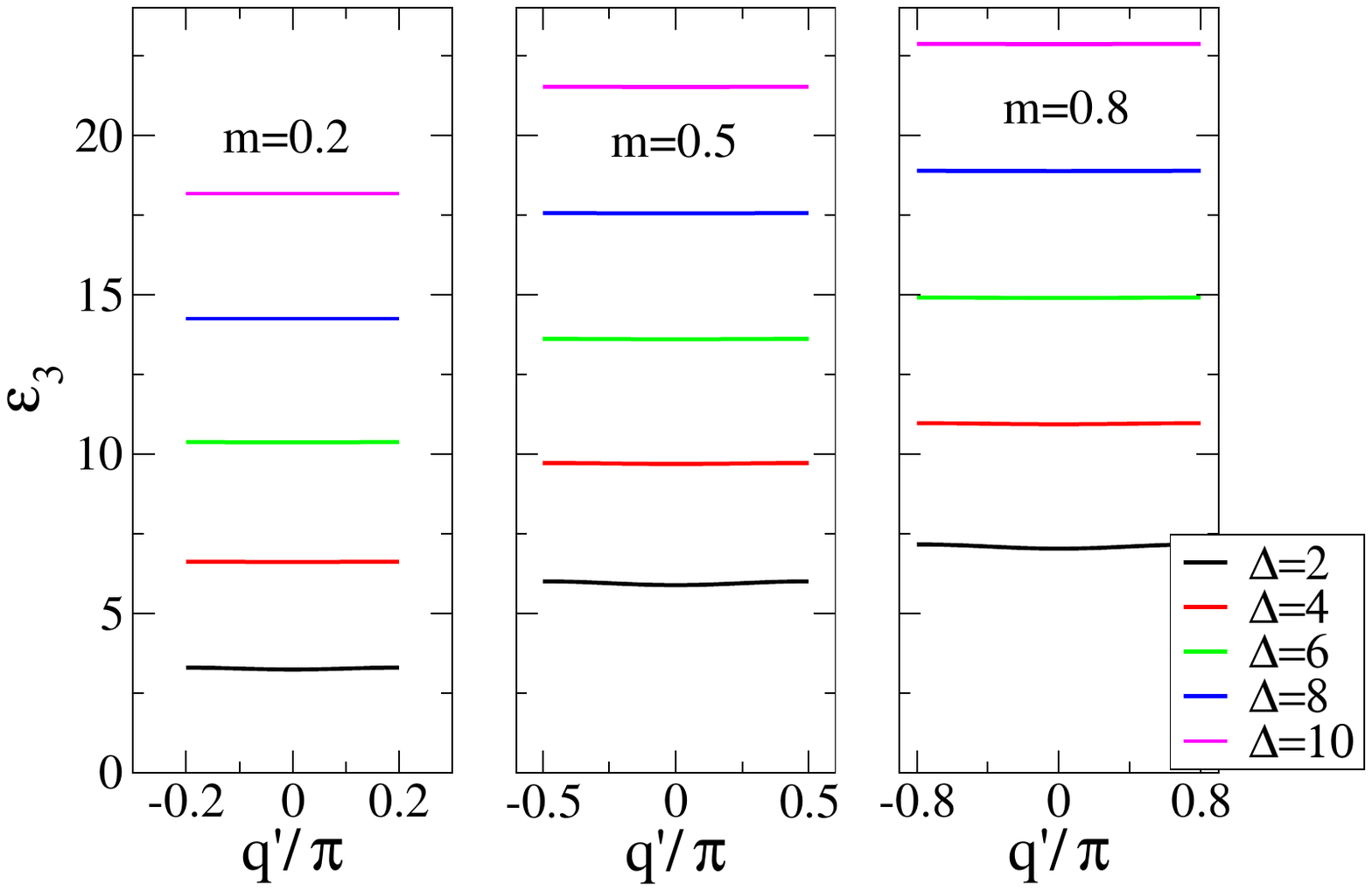}}
\caption{The same as in Fig. \ref{figure12PR} for the
$3$-string-pair energy dispersion $\varepsilon_{3} (q')$ 
associated with a Bethe string of length three.}
\label{figure13PR}
\end{center}
\end{figure}

The $n$-string-pair energy dispersion $\varepsilon_{n} (q')$, Eq. (\ref{equA4n}), in units of $J$ is plotted in 
Figs. \ref{figure12PR} and \ref{figure13PR} for $n=2$ and $n=3$, respectively, as a function of $q'/\pi$ for
$n$-band momentum $q'\in [-(k_{F\uparrow} - k_{F\downarrow}),(k_{F\uparrow} - k_{F\downarrow})]$, 
spin densities $m=0.2$, $m=0.5$, $m=0.8$, and several anisotropy values.
The $n=2$ and $n=3$ $n$-string-pairs are associated with Bethe strings of length two 
and three, respectively.

For simplicity, we provide here expressions of the rapidity functions $\varphi_{n}(q)$
for the limiting cases of spin density values $m=0$ and $m=1$.
In the spin-insulating quantum phase for fields $h\in [0,h_{c1}]$ and $m=0$,
the interval $q' \in [-(k_{F\uparrow} - k_{F\downarrow}),(k_{F\uparrow} - k_{F\downarrow})]$ 
of the $n>1$ rapidity functions $\varphi_{n}(q')\in [-\pi,\pi]$ argument collapses to
$q' =0$. On the other hand, the expression of the $n=1$ function $\varphi_1 (q)$ 
where $q\in [-\pi/2,\pi/2]$ simplifies to,
\begin{equation}
\varphi_1 (q) = \pi {F (q,u_{\eta})\over K (u_{\eta})} \, .
\label{varphi1}
\end{equation}
Here $F (q,u_{\eta})$ and $K (u_{\eta})=F (\pi/2,u_{\eta})$ are the elliptic integral of the first kind
and the complete elliptic integral of the first kind given by
\begin{equation}
F (q,u_{\eta}) = \int_0^{q}d\theta {1\over \sqrt{1 - u_{\eta}^2\sin^2\theta}} \, ,
\label{Felliptic}
\end{equation}  
and
\begin{equation}
K (u_{\eta}) = F (\pi/2,u_{\eta}) = \int_0^{\pi\over 2}d\theta {1\over \sqrt{1 - u_{\eta}^2\sin^2\theta}} \, ,
\label{elliptic}
\end{equation}  
respectively. The dependence of the function $u_{\eta}$ in them on the parameter
$\eta$ associated with anisotropy $\Delta = \cosh\eta$ is defined by its inverse function as,
\begin{equation}
\eta = \pi {K (u_{\eta}') \over K (u_{\eta})}\hspace{0.20cm}{\rm where}\hspace{0.20cm}
u_{\eta}' =\sqrt{1- u_{\eta}^2} \, .
\label{uphi}
\end{equation} 

In the opposite limit of $h=h_{c2}$ and $m=1$, the rapidity function $\varphi_n (q)$
has the following closed-form expression valid for $n\geq 1$,
\begin{equation}
\varphi_n (q) = 2\arctan\left(\tanh\left({n\,\eta\over 2}\right)\tan\left({q\over 2}\right)\right)
\hspace{0.10cm}{\rm for}\hspace{0.20cm} q \in [-\pi,\pi] \, .
\label{varphinqm1}
\end{equation}

The rapidity-dependent dispersions ${\bar{\varepsilon}_{n}^0} (\varphi)$ in Eq. (\ref{equA4n}) are defined by the equations,
\begin{eqnarray}
{\bar{\varepsilon}_{n}^0} (\varphi) & = & \int_{0}^{\varphi }d\varphi^{\prime}2J\gamma_{n} (\varphi^{\prime}) + A_n^{0}
\hspace{0.20cm}{\rm where}
\nonumber \\
A_1^{0} & = & - J(1 + \cosh \eta) 
\nonumber \\
& + & {1\over\pi}\int_{-B}^{B}d\varphi^{\prime}\,2J\gamma_{1} (\varphi^{\prime})
\arctan\left(\coth \eta\tan\left({\varphi^{\prime}\over 2}\right)\right)
\nonumber \\
&& {\rm and}
\nonumber \\
A_n^{0} & = & -J {\sinh \eta\over\sinh (n\,\eta)}\left(1 + \cosh (n\,\eta)\right) 
\nonumber \\
& + & {1\over\pi}\sum_{\iota=\pm 1}\int_{-B}^{B}d\varphi^{\prime}\,2J\gamma_{1} (\varphi^{\prime})
\nonumber \\
& \times & \arctan\left(\coth \left({(n + \iota)\,\eta\over 2}\right)\tan\left({\varphi^{\prime}\over 2}\right)\right) \, ,
\label{equA4n10}
\end{eqnarray}
for $n> 1$. The distribution $2J\gamma_{n} (\varphi)$ obeys the following equation for $n\geq 1$,
\begin{eqnarray}
2J\gamma_{n} (\varphi) & = & J\,{\sinh \eta\,\sinh (n\,\eta)\sin (\varphi)\over (\cosh (n\,\eta) - \cos (\varphi))^2} 
\nonumber \\
& + & \int_{-B}^{B}d\varphi^{\prime}\,G_n (\varphi - \varphi^{\prime})\,2J\gamma_{1} (\varphi^{\prime})  \, ,
\label{equA6}
\end{eqnarray}
where $G_n (\varphi) = - {1\over{2\pi}}\sum_{\iota=\pm 1}{\sinh ((n+\iota)\,\eta)\over \cosh ((n+\iota)\,\eta) - \cos (\varphi)}$.

For $h \in [0,h_{c1}]$ and $m=0$ and for $h=h_{c2}$ and $m=1$, the energy dispersions
$\varepsilon_{n} (q)$ and $\varepsilon_{n}^0 (q)$ have the following simple analytical expressions,
\begin{eqnarray}
\varepsilon_{1} (q) & = & \varepsilon_{1}^0 (q) + {1\over 2}\, g\mu_B\,h
\nonumber \\
\varepsilon_{1}^0 (q) & = & - {J\over\pi}\sinh\eta\,K (u_{\eta}) \sqrt{1 - u_{\eta}^2\sin^2 q} 
\nonumber  \\
&& {\rm for}\hspace{0.20cm}q \in [-\pi/2,\pi/2]\hspace{0.20cm}{\rm and}\hspace{0.20cm}h \in [0,h_{c1}] 
\nonumber \\
&& {\rm and}
\nonumber  \\
\varepsilon_{1} (q) & = & \varepsilon_{n}^0 (q) + J (1+ \Delta) = J (1 - \cos q)
\nonumber \\
\varepsilon_{1}^0 (q) & = & - J (\Delta + \cos q)
\nonumber  \\
&& {\rm for}\hspace{0.20cm}q \in [-\pi,\pi]\hspace{0.20cm}{\rm and}\hspace{0.20cm}h = h_{c2} \, ,
\label{vareband1m0m1}
\end{eqnarray}
at $n=1$ and,
\begin{eqnarray}
\varepsilon_{n} (q') & = & \varepsilon_{n}^0 (q') + n\,g\mu_B\,h
\nonumber  \\
\varepsilon_{n}^0 (q') & = & - g\mu_B\,h 
\nonumber \\
&& {\rm for}\hspace{0.20cm} q' = 0 \hspace{0.20cm}{\rm and}\hspace{0.20cm}h \in [0,h_{c1}] 
\nonumber \\
&& {\rm and}
\nonumber \\
\varepsilon_{n} (q') & = & \varepsilon_{n}^0 (q') + n\,J (1+ \Delta) 
\nonumber \\
\varepsilon_{n}^0 (q') & = & - J\,{\sinh \eta\over\sinh (n\,\eta)}(\cosh (n\,\eta) + \cos q') 
\nonumber  \\
&& {\rm for}\hspace{0.20cm} q' \in [-\pi,\pi] \hspace{0.20cm}{\rm and}\hspace{0.20cm}h = h_{c2} \, ,
\label{qdepvarepsilonm0m1}
\end{eqnarray}
for $n>1$. For the same magnetic field $h$ and $m$ values, the $n>1$ binding energy $E_{{\rm bind},n}$
and energy $T_{n} (q')$, Eq. (\ref{EbnTn}), read,
\begin{eqnarray}
E_{{\rm bind},n} & = & - g\mu_B\,h \hspace{1.60cm}T_{n} (q') = 0 
\nonumber \\
&& {\rm for}\hspace{0.20cm} q' = 0 \hspace{0.20cm}{\rm and}\hspace{0.20cm}h \in [0,h_{c1}] \hspace{0.20cm}{\rm and}
\nonumber \\
E_{{\rm bind},n} & = & - J\,{\sinh \eta\sinh (n\,\eta)\over \cosh (n\,\eta) - 1} 
\nonumber \\
T_{n} (q') & = & J{\sinh\eta\over\sinh (n\,\eta)}(1 - \cos q')
\nonumber  \\
&& {\rm for}\hspace{0.20cm} q' \in [-\pi,\pi] \hspace{0.20cm}{\rm and}\hspace{0.20cm}h = h_{c2} \, .
\label{BnTnexpresshc2}
\end{eqnarray}

The use of the expression of $\varepsilon_1^0 (q)$ in 
$E_{1}^{\uparrow\downarrow} = \varepsilon_{1}^0 (k_{F\downarrow})$, Eq. (\ref{Eminm}),
confirms that $\vert E_{1}^{\uparrow\downarrow}\vert/(g\mu_B)$
gives in the $m\rightarrow 0$ and $m\rightarrow 1$ limits the known Bethe ansatz 
expressions \cite{Takahashi_99} of the critical magnetic fields,
\begin{eqnarray}
h_{c1} & = & \lim_{m\rightarrow 0}\vert E_{1}^{\uparrow\downarrow}\vert /(g\mu_B)
\nonumber \\
& = & {2J\over\pi\,g\mu_B}\,K (u_{\eta})\sqrt{(\Delta^2 - 1)(1 - u_{\eta}^2)} \hspace{0.20cm}{\rm and}
\nonumber \\
h_{c2} & = & \lim_{m\rightarrow 1}\vert E_{1}^{\uparrow\downarrow}\vert /(g\mu_B)
= {J\over g\mu_B} (\Delta +1) \, ,
\label{hc1c2}
\end{eqnarray}
respectively, where $\sqrt{\Delta^2 - 1} = \sinh \eta$.

The momentum dependent exponents that control the line shape of the
dynamic structure factor components at and near
their sharp peaks involve the $1$-pair phase shifts. They are given by,
\begin{equation}
2\pi\,\Phi_{1,n}(q,q') = 2\pi\,\bar{\Phi }_{1,n} \left(\varphi_1 (q),\varphi_n(q')\right) 
\hspace{0.20cm}{\rm for}\hspace{0.20cm}n\geq 1 \, ,
\label{Phi-barPhi}
\end{equation}
where the rapidity-dependent phase shifts 
$2\pi\,\bar{\Phi }_{1,n} \left(\varphi,\varphi'\right)$ are in units
of $2\pi$ defined by the following integral equations,
\begin{eqnarray}
\bar{\Phi }_{1,1} \left(\varphi,\varphi'\right) & = & 
{1\over \pi}\arctan\left(\coth \eta\tan\left({\varphi - \varphi'\over 2}\right)\right)
\nonumber  \\
& + & \int_{-B}^{B} d\varphi''\,G_1 (\varphi - \varphi'')\,\bar{\Phi }_{1,1} \left(\varphi'',\varphi'\right) \, ,
\nonumber  \\
{\rm and} &&
\nonumber  \\
\bar{\Phi }_{1,n} \left(\varphi,\varphi'\right) & = & 
{1\over \pi}\sum_{\iota=\pm 1}
\nonumber  \\
&& \arctan\left(\coth\left({(n + \iota)\,\eta\over 2}\right)\tan\left({\varphi - \varphi'\over 2}\right)\right)
\nonumber  \\
& + & \int_{-B}^{B} d\varphi''\,G_1 (\varphi - \varphi'')\,\bar{\Phi }_{1,n} \left(\varphi'',\varphi'\right) \, ,
\label{Phis1n}
\end{eqnarray}
for $n>1$. The kernel reads
$G_1 (\varphi) = - {1\over{2\pi}}{\sinh (2\eta)\over \cosh (2\eta) - \cos (\varphi)}$.

Specifically, the following phase shifts in units of $2\pi$ and phase-shift parameters given by,
\begin{eqnarray}
\Phi_{1,n}(\iota\,k_{F\downarrow},q) & = & \bar{\Phi }_{1,n} \left(\iota\,B,\varphi_n (q)\right)
\nonumber \\
\xi & = & 1
+ \sum_{\iota=\pm 1} (\iota)\,\Phi_{1,1}\left(k_{F\downarrow},\iota k_{F\downarrow}\right) 
\nonumber \\
\xi_{1\,n}^0 & = & 2\Phi_{1,n}\left(k_{F\downarrow},0\right)\hspace{0.20cm}{\rm for}\hspace{0.20cm}n=2,3 \, ,
\label{x-aa}
\end{eqnarray}
where $\iota =\pm 1$ appear in the expressions given below in  Appendix \ref{C}
of the exponents that control the power-law behaviors of
the spin dynamic structure factor components at and near their sharp peaks.

\section{The physical-spins representation applies to the whole Hilbert space}
\label{B}

On the one hand, the translational degrees of freedom of the $M=2S_q$ unpaired physical spins $1/2$
are described within the Bethe ansatz: They are described by a number $M=2S_q$ of $n$-band momentum 
values, $q_j = {2\pi\over L}I_j^n$, out of the $N^h_{n} = 2S_q +\sum_{n'=n+1}^{\infty}2(n'-n)N_{n'}$ unoccupied 
such values, {\it i.e.} $n$-holes, of each $n$-band with finite $N_n>0$ occupancy. Note that for states without 
$n$-string-pairs one has that $N^h_{1} = 2S_q =M$.

On the other hand, the spin internal degrees of freedom of such $M=2S_q$ 
unpaired physical spins $1/2$ is an issue beyond the Bethe ansatz. We confirm
in the following that the physical-spins representation applies to the whole Hilbert space
because it accounts for their spin internal degrees of freedom.

Let $\left\vert l_{\rm r},S_q,S^z,\Delta\right\rangle$ be an energy eigenstate 
of the Hamiltonian $\hat{H}$, Eq. (\ref{Hphi}), whose quantum
numbers beyond $S_q$, $S^z$, and $\Delta = \cosh \eta >1$ needed to specify it are here denoted by $l_{\rm r}$.
Consider a HWS $\left\vert l_{\rm r},S_q,S_q,\Delta\right\rangle$. 
A number $2S_q$ of $SU_q(2)$ symmetry non-HWSs outside the Bethe-ansatz 
solution referring to different multiplet configurations of the
$M=2S_q$ unpaired physical spins $1/2$ are generated from that HWS as,
\begin{equation} 
\left\vert l_{\rm r},S_q,S_q-n_z,\eta\right\rangle = 
{1\over \sqrt{{\cal{C}}_{\eta}}}({\hat{S}}^{-}_{\eta})^{n_z}\left\vert l_{\rm r},S_q,S_q,\eta\right\rangle \, .
\label{state}
\end{equation} 
Here $n_z\equiv S_q - S^z = 1,...,2S_q$ so that $S^z = S_q - n_z$ and,
\begin{equation}
{\cal{C}}_{\eta} = 
\prod_{l=1}^{n_z}{\sinh^2 (\eta\,(S_q+1/2)) - \sinh^2 (\eta\,(l - S_q - 1/2))\over\sinh^2 \eta} \, ,
\label{nonBAstatesDelta1}
\end{equation}
for $n_z= 1,...,2S_q$. Similarly to the $\Delta =1$ bare ladder spin operators $\hat{S}^{\pm}$, the 
action of the $\Delta = \cosh \eta >1$ $q$-spin ladder operators $\hat{S}^{\pm}_{\eta}$
on $S_q > 0$ energy eigenstates flips an {\it unpaired} physical spin $1/2$ projection. 
(The expression of the operators $\hat{S}^{\pm}_{\eta}$ is given in Ref. \onlinecite{Carmelo_22}.)

For the non-HWSs, Eq. (\ref{state}), the two sets of $n_z\equiv S_q - S^z = 1,...,2S_q$ 
and $2S_q-n_z = S_q + S^z$ unpaired physical spins $1/2$ have opposite $\downarrow$
and $\uparrow$ spin projections, respectively. Hence, the multiplet configurations that
involve the internal degrees of freedom of the $M=2S_q$ unpaired physical spins $1/2$
are generated as given in Eq. (\ref{state}). An important property that follows
from the $SU_q(2)$ symmetry is that all $2S_q+1$ states of the same 
$q$-spin tower have exactly the same $n$-pairs occupancy configurations
and thus the same values for the set of $n=1,...,\infty$ distributions $\{N_n (q_j)\}$ and 
rapidity functions $\{\varphi_{n} (q_{j})\}$.

Let $E_{{\rm r},S_q,\Delta}$ be the energy eigenvalue of a HWS $\left\vert l_{\rm r},S_q,S_q,\Delta\right\rangle$
relative the Hamiltonian $\hat{H}$, Eq. (\ref{Hphi}). Then the energy eigenvalue 
$E_{l_{\rm r},S_q,S^z,\Delta}$ of a corresponding non-HWS, Eq. (\ref{state}), reads,
$E_{l_{\rm r},S_q,S^z,\Delta} = E_{l_{\rm r},S_q,\Delta} + n_z\,g\mu_B h$.
This reveals that a $\uparrow\rightarrow\downarrow$ spin flip requires an excitation energy $g\mu_B h$.
The excitation energy for a $\downarrow\rightarrow\uparrow$ spin flip is actually $-g\mu_B h$, {\it i.e.}
it is an energy release process.

A ground state of energy $E_{l_{\rm r},S_q,\Delta}^{\rm GS}$  is for $0<m<1$ and $h_{c1}<h<h_{c2}$ a HWS. Hence the excitation
energy of non-HWSs generated from it as given in Eq. (\ref{state}) reads,
\begin{equation}
E_{l_{\rm r},S_q,S^z,\Delta} - E_{l_{\rm r},S_q,\Delta}^{\rm GS} = n_z\,g\mu_B h \, .
\label{EEGS}
\end{equation}

The $M = 2S_q$ unpaired physical spins $1/2$ whose translational and internal degrees of freedom
we have just identified play an important role for spin transport \cite{Jepsen_20}: As shown in the following,
they are the spin transport carriers whereas $n$-pairs do not couple to a vector potential and thus 
do not carry spin current. This results from their singlet nature.

To show this one considers the Hamiltonian, Eq. (\ref{Hphi}), in the 
presence of a uniform vector potential \cite{Shastry_90}, $\hat{H}=\hat{H} (\Phi/L)$ where $\Phi = \Phi_{\uparrow} = -\Phi_{\downarrow}$.
It remains solvable by the Bethe ansatz \cite{Zotos_99,Herbrych_11}.
After some straightforward algebra using the corresponding $\Phi\neq 0$ Bethe-ansatz equations \cite{Carmelo_18}, 
one finds that the momentum eigenvalues for HWSs in the thermodynamic limit read,
\begin{equation}
P = \pi\sum_{j=1}^{L_n}N_{n} + \sum_{n=1}^{\infty}\sum_{j=1}^{L_n}N_{n} (q_j)\,q_j + {\Phi\over L}\,(N - \sum_{n=1}^{\infty}2n\,N_{n}) \, .
\label{PPhi}
\end{equation}
The number of physical spins $1/2$ that couple to the vector potential is 
given by the factor that multiplies ${\Phi\over L}$ in Eq. (\ref{PPhi}).
From the use of the thermodynamic-limit exact sum rule, 
$2\Pi = N-2S_q = \sum_{n=1}^{\infty}2n\,N_n$, one finds that such a
number actually reads $2S_q = N - 2\Pi = N - \sum_{n=1}^{\infty}2n\,N_{n}$.

The term ${\Phi\over L}\,N$ in ${\Phi\over L}\,2S_q = {\Phi\over L}\,(N - \sum_{n=1}^{\infty}2n\,N_{n})$
refers to {\it all} $N$ physical spins $1/2$ coupling to the vector potential
in the absence of physical spins pairing. Indeed, the negative coupling counter terms $-\sum_{n=1}^{\infty}2n\,N_n$ 
refer to the number $2n$ of paired physical spins $1/2$ in each $n$-pair both
for $n=1$ and $n>1$. They {\it exactly cancel} the positive coupling of the 
corresponding $2n$ paired physical spins $1/2$ in each $n$-pair. 
As a result of such counter terms, only the $M = 2S_q = N - \sum_{n=1}^{\infty}2n\,N_n$ 
unpaired physical spins $1/2$ couple to the vector potential and thus carry
spin current.

A similar analysis for non-HWSs, Eq. (\ref{state}), gives Eq. (\ref{PPhi}) with ${\Phi\over L}\,2S_q={\Phi\over L}\,M$
replaced by ${\Phi\over L}(M_{+1/2} - M_{-1/2})$. Here $M_{\pm 1/2} =  N/2 - \sum_{n=1}^{\infty}n\,N_{n} \mp S^z$ where
$M_{\pm 1/2} = S_q \mp S^z$ is the number of unpaired physical spins of projection $\pm 1/2$
that couple to the vector potential.
This again confirms that only the $M = 2S_q = N - \sum_{n=1}^{\infty}2n\,N_n$ unpaired 
physical spins $1/2$ couple to a uniform vector potential and thus carry spin current,
so that they are indeed the spin transport carriers.

\section{Dynamical theory for the $1$-pair - $1$-pair and $1$-pair - $n$-pair scattering}
\label{C}

Here we provide some basic information on the dynamical theory for the 
$1$-pair - $1$-pair and $1$-pair - $n$-pair scattering that involves the $2n$-physical-spins $n$-pairs 
\cite{Carmelo_22} used in the studies of this paper. In addition, the expressions of 
spectra and exponents associated with the set of sharp peaks studied in this paper are given.
The theory is valid in the thermodynamic limit
and provides the line shape of the spin dynamic structure factor $ab = +-,-+,zz$ components $S^{ab} (k,\omega)$ 
at and just above the $(k,\omega)$-plane $n=1,2,3$ $n$-continua lower thresholds 
where there are sharp peaks. Such continua are shown in Figs. \ref{figure2PR} (a)-(c), 
\ref{figure3PR} (a)-(c), and \ref{figure4PR} (a)-(c) for $S^{+-} (k,\omega)$, $S^{-+} (k,\omega)$, 
and $S^{zz} (k,\omega)$, respectively.

At fixed excitation momentum $k$ and small values of the energy deviation
$(\omega - E^{ab}_{n} (k))\geq 0$, the spin dynamic structure factor $ab = +-,-+,zz$ components 
have the power-law form,
\begin{equation}
S^{ab} (k,\omega) = C_{ab}^n (k)
\left({\omega - E^{ab}_{n} (k)\over 4\pi\,B_1^{ab}\,v_1 (k_{F\downarrow})}\right)^{\zeta_{n}^{ab} (k)} \, .
\label{MPSs}
\end{equation}
Here $E^{ab}_{n} (k)$ denotes the $n$-continua lower-threshold spectra of the excited states. Their expressions 
for the experimentally observed sharp peaks at fixed momenta $k=0,\pi/2,\pi$ are
given below in Eqs. (\ref{EPM0})-(\ref{EPM3PI2}). They involve simple combinations of the $n$-band 
energy dispersions $\varepsilon_{n} (q)$, Eq. (\ref{equA4n}) of Appendix \ref{A}. In Eq. (\ref{MPSs}),
$v_1 (k_{F\downarrow})$ denotes the $1$-pair group velocity
$v_1 (q) = \partial\varepsilon_1 (q)/ \partial q$ at $q=k_{F\downarrow}$,  $0<B_1^{ab}\leq 1$ is a $\eta$ and $m$ dependent 
constant, and expressions for the exponents $\zeta_{n}^{ab} (k)$ and factor functions $C_{ab}^n (k)$ are given below.
Such exponents are fully controlled by the $1$-pair - $1$-pair, $1$-pair - $2$-pair, and
$1$-pair - $3$-pair scattering involving $2$-physical-spins $1$-pairs and $n$-string-pairs with $n=2$ and $n=3$ pairs of physical 
spins $1/2$ bound within them. 

The $(k,\omega)$-plane $n=1,2,3$ $n$-continua of $S^{+-}  (k,\omega)$, $1$-continuum of
$S^{-+} (k,\omega)$, and the $n=1,2$ $n$-continua of $S^{zz} (k,\omega)$ shown in 
Figs. \ref{figure2PR} (a)-(c), \ref{figure3PR} (a)-(c), and \ref{figure4PR} (a)-(c), respectively, 
are those where in the thermodynamic limit there is significant spectral weight. 
Such figures refer to the anisotropy $\Delta =2$ suitable to the spin chains in SrCo$_2$V$_2$O$_8$
and spin densities $m=0.209\approx 0.2$, $m=0.514\approx 0.5$, and $m=0.793\approx 0.8$.
The $k$ intervals of the lines marked in these figures refer to the location of sharp peaks
of form, Eq. (\ref{MPSs}), for which $\zeta_{n}^{ab} (k)<0$. Corresponding figures for anisotropy 
$\Delta =2.17$ suitable to the spin chains in BaCo$_2$V$_2$O$_8$ are very similar. 

The singlet nature of the pairs of physical spins $1/2$ 
contained in the $1$-pairs and $n>1$ $n$-string-pairs
determines the form of the $\cal{S}$ matrices associated with the general 
physical-spins $n$-pair - $n'$-pair scattering 
where $n,n'\geq 1$. They are dimension-one scalars of the form,
\begin{eqnarray}
{\cal{S}}_{n} (q_j) & = & \prod_{n'=1}^{\infty}\,\prod_{j'=1}^{L_{n'}}\,{\cal{S}}_{n ,n'} (q_j, q_{j'})
\hspace{0.20cm}{\rm where}
\nonumber \\
{\cal{S}}_{n,n'} (q_j, q_{j'}) & = &
e^{i\,\delta N_{n'}(q_{j'})\,2\pi\Phi_{n,n'}(q_j,q_{j'})} \, .
\label{Smatrix}
\end{eqnarray}
Here $\delta N_{n'}(q_{j'})$ are deviations from the ground-state 
$n'$-band momentum distributions $N_{n'}(q_{j'})$ suitable to specific excited states.
The quantities $2\pi\Phi_{n,n'}(q_j,q_{j'})$ in Eq. (\ref{Smatrix}) are $n$-pair 
phase shifts and $n'$ refers to the corresponding $n'$-pair scattering centers.

For the line shape at and near the sharp peaks in $S^{ab} (k,\omega)$ 
only the phase shifts $2\pi\Phi_{1,1}(q,q')$ and $2\pi\Phi_{1,n}(q,q')$ where $n=1,2,3$
play an active role. They are defined by Eqs. (\ref{Phi-barPhi})-(\ref{Phis1n}) of Appendix \ref{A}.
Indeed, ground states are not populated by $n$-string-pairs. Hence only the
ground-state preexisting $1$-pairs play the role of scatterers. $1$-pairs, $1$-holes, 
and $n$-string-pairs created under transitions to excited states play the role of scattering centers. 

The corresponding $1$-pair $\cal{S}$ matrix then determines the momentum $k$ dependence 
of the exponents $\zeta_{n}^{ab} (k)$ and pre-factor functions $C_{ab}^n (k)$ in Eq. (\ref{MPSs}). They read, 
\begin{equation}
\zeta_{n}^{ab} (k) = - 1  + \sum_{\iota = \pm 1}\Phi_{\iota}^2 (k) \, ,
\label{zetaabk}
\end{equation}
and
\begin{eqnarray}
C_{ab}^n (k) & = & {1\over \vert\zeta_{n}^{ab} (k)\vert}
\nonumber \\
& \times & \prod_{\iota =\pm 1}
{e^{-f_0^{ab} + f_2^{ab}\left(2{\tilde{\Phi}}_{\iota}\right)^2 - f_4^{ab}\left(2{\tilde{\Phi}}_{\iota}\right)^4} \over \Gamma (\Phi_{\iota}^2 (k))} \, ,
 \label{Cabn}
\end{eqnarray}
respectively. Here $ab = +-,-+,zz$, the index $n=1,2,3$ refers the $(k,\omega)$-plane $n$-continua shown in 
Figs. \ref{figure2PR} (a)-(c), \ref{figure3PR} (a)-(c), and \ref{figure4PR} (a)-(c), the $l=0,2,4$ coefficients $0<f_l^{ab}<1$ 
depend on $\eta$ and are different for each spin dynamic structure factor component,
and ${\tilde{\Phi}}_{\iota} = - {i\over 2\pi}\ln {\cal{S}}_{1} (\iota k_{F\downarrow})$ is the scattering part of the general functional,
\begin{eqnarray}
\Phi_{\iota} & = & \iota\,\delta N_{1,\iota}^F  - {i\over 2\pi}\ln {\cal{S}}_{1} (\iota k_{F\downarrow})
\nonumber \\
& = & \iota{\delta N_1^F\over 2} + \delta J_1^F  - {i\over 2\pi}\ln {\cal{S}}_{1} (\iota k_{F\downarrow})
\hspace{0.20cm}{\rm where}
\nonumber \\
{\cal{S}}_{1} (\iota k_{F\downarrow}) & = & 
\prod_{n=1}^{3}\,\prod_{j=1}^{L_{n}}\,e^{i\,\delta N_{n}(q_{j})\,2\pi\Phi_{1,n}(\iota k_{F\downarrow},q_{j})} \, .
\label{functional}
\end{eqnarray}
It involves the $\cal{S}$ matrix ${\cal{S}}_{1} (q)$ at the $1$-band Fermi points $q=\iota k_{F\downarrow} = \pm k_{F\downarrow}$.
Its dependence on the excitation momentum $k$ occurs through its direct
relation to the $n$-band momenta $q_j$ in the phase shifts $2\pi\Phi_{1,n}(\iota k_{F\downarrow},q_{j})$.
The index $\iota =\pm 1$ in $\iota k_{F\downarrow}$ refers to the left $(\iota = -1)$ and right $(\iota = +1)$ 
$1$-band Fermi points and $\delta N_{1}^F = \sum_{\iota = \pm 1}\delta N_{1,\iota}^F$ and
$\delta J_{1}^F =  {1\over 2}\sum_{\iota = \pm 1}\iota\,\delta N_{1,\iota}^F$ are deviations under the ground-state - excited 
state transitions. Here $\delta N_{1,\iota}^F$ is the deviation in the number of $1$-pairs at and very near such $\iota = \pm 1$
$1$-band Fermi points.

The exponent expressions for specific types of excited states are determined by the corresponding values
of the deviations $\delta N_{1}^F$, $\delta J_{1}^F$, and $\delta N_{n}(q_{j})$ for $n=1,2,3$ in
Eqs. (\ref{zetaabk}) and (\ref{functional}). Specific values for such deviations determine for instance the exponents plotted in
Figs. \ref{figure2PR} (d)-(f), \ref{figure3PR} (d)-(f), and \ref{figure4PR} (d)-(f) for the spin dynamical structure factor
components $S^{-} (k,\omega)$, $S^{-+} (k,\omega)$, and $S^{zz} (k,\omega)$, respectively. The same applies to the exponents 
whose specific expressions are given below in Eqs. (\ref{EPM0})-(\ref{EPM3PI2}). They control the line shape 
at and in the vicinity of the sharp peaks experimentally 
observed in BaCo$_2$V$_2$O$_8$ and SrCo$_2$V$_2$O$_8$
at momentum values $k=0$, $k=\pi/2$, and $k=\pi$.

In (i) the spectra and (ii) the exponents given in the following, (i) $\varepsilon_n (q)$ are for $n=1,2,3$ the
$n$-pair energy dispersions, Eqs. (\ref{equA4n})-(\ref{equA6}) of Appendix \ref{A},
whose limiting behaviors are provided in Eqs. (\ref{vareband1m0m1}) and (\ref{qdepvarepsilonm0m1}) of 
that Appendix for $n=1$ and $n>1$, respectively, and (ii) $\Phi_{1,1}(\iota\,k_{F\downarrow},q)$
where $\iota =\pm 1$ and $\{\xi,\xi_{1\,n}^0\}$ for $n=2,3$ are the phase shifts
in units of $2\pi$ and related phase-shift parameters, respectively, Eq. (\ref{x-aa}) of
Appendix \ref{A}. In the cases of sharp peaks in (a) $S^{+-} (k,\omega)$ and $S^{zz} (k,\omega)$ and (b) $S^{-+} (k,\omega)$
located in the lower thresholds of the corresponding $n$-continua, the smallest and largest 
values given in the following for the energy intervals refer to (a) the smallest and largest magnetic field 
and to (b) the largest and smallest magnetic field, respectively. 

The $n=1,2,3$ lower threshold energies $E^{+-}_{n} (k,h)$, $n=1$ lower threshold energy $E^{-+}_{1} (k,h)$,
$n=1$ lower threshold energy $E^{zz}_{1} (k,h)$, and exponents $\zeta_{n}^{ab} (k,h)$, Eqs. (\ref{zetaabk}) 
and (\ref{functional}), appearing in the expressions, Eq. (\ref{Rkab}), of the line shape 
at and near the sharp peaks at anisotropies $\Delta =2$ and $\Delta =2.17$ representative
of SrCo$_2$V$_2$O$_8$ and BaCo$_2$V$_2$O$_8$, respectively, are given by,
\begin{eqnarray}
E^{+-}_1 (0,h) & = & \varepsilon_1 (k_{F\uparrow})\in [0,2J]
\hspace{0.20cm}{\rm at}\hspace{0.20cm}\Delta = 2\hspace{0.20cm}{\rm and}\hspace{0.20cm}\Delta = 2.17
\nonumber \\
\zeta^{+-}_1 (0,h) & = & -1 + \sum_{\iota =\pm 1}\Bigl(- {\xi\over 2} 
+ \Phi_{1,1}(\iota k_{F\downarrow},-k_{F\uparrow})\Bigr)^2
\nonumber \\
&& {\rm for}\hspace{0.20cm}h \in [h_{c1},h_{c2}] \, ,
\label{EPM0}
\end{eqnarray}
\begin{eqnarray}
E^{-+}_1 (\pi/2,h) & = & - \varepsilon_1 \Bigl({(k_{F\uparrow}-k_{F\downarrow})\over 2}\Bigr) \in [0,1.876 J]
\hspace{0.20cm}{\rm at}\hspace{0.20cm}\Delta = 2
\nonumber \\
&& \hspace{2.6cm} \in [0,2.153 J]\hspace{0.20cm}{\rm at}\hspace{0.20cm}\Delta = 2.17
\nonumber \\
\zeta^{-+}_1 (\pi/2,h) & = & -1 
\nonumber \\
& + & \sum_{\iota =\pm 1}\Bigl(- {\xi\over 2} 
- \Phi_{1,1}\Bigl(\iota k_{F\downarrow},{(k_{F\uparrow}-k_{F\downarrow})\over 2}\Bigr)\Bigr)^2 
\nonumber \\
&& {\rm for}\hspace{0.20cm}h \in [h_{c1},h_{1/2}] \, ,
\label{EMPPI2}
\end{eqnarray}
\begin{eqnarray}
E^{+-}_1 (\pi/2,h) & = & \varepsilon_1 \Bigl({(k_{F\uparrow}-k_{F\downarrow})\over 2}\Bigr)\in [0,J]
\nonumber \\
&& {\rm at}\hspace{0.20cm}\Delta = 2\hspace{0.20cm}{\rm and}\hspace{0.20cm}\Delta = 2.17
\nonumber \\
\zeta^{+-}_1 (\pi/2,h) & = & -1\nonumber \\
& + & \sum_{\iota =\pm 1} \Bigl(-{\xi\over 2} 
+ \Phi_{1,1}\Bigl(\iota k_{F\downarrow},- {(k_{F\uparrow}-k_{F\downarrow})\over 2}\Bigr)\Bigr)^2 
\nonumber \\
&& {\rm for}\hspace{0.20cm}h \in [h_{1/2},h_{c2}] \, ,
\label{EPMPI2}
\end{eqnarray}
\begin{eqnarray}
E^{zz}_1 (\pi,h) & = & \varepsilon_1 (k_{F\uparrow})\in [0,2J]
\hspace{0.20cm}{\rm at}\hspace{0.20cm}\Delta = 2
\nonumber \\
\zeta_{1}^{zz} (\pi,h) & = & -1 
\nonumber \\
& + & \sum_{\iota =\pm 1}\left(- {\iota\over 2\xi_{1\,1}} + {\xi_{1\,1}\over 2} 
+ \Phi_{1,1}(\iota k_{F\downarrow},k_{F\uparrow})\right)^2 
\nonumber \\
&& {\rm for}\hspace{0.20cm}h \in [h_{c1},h_{c2}] \, ,
\label{EZZPI}
\end{eqnarray}
\begin{eqnarray}
E^{+-}_{2} (0,h) & = & \varepsilon_2 (0)\in [0.389 J,4J] \hspace{0.20cm}{\rm at}\hspace{0.20cm}\Delta = 2
\nonumber \\
&& \hspace{0.85cm}\in [0.518 J,4.170 J]\hspace{0.20cm}{\rm at}\hspace{0.20cm}\Delta = 2.17
\nonumber \\
\zeta^{+-}_{2} (0,h) & = & -1 + {1\over 2}\left({1\over 2\xi} - \xi_{1\,2}^0\right)^2 
\nonumber \\
&& {\rm for}\hspace{0.20cm}h \in [h_{c1},h_{c2}] \, ,
\label{EPM20}
\end{eqnarray}
\begin{eqnarray}
&& E^{+-}_{2} (\pi/2,h) = \varepsilon_ 2 (0) - \varepsilon_1 \Bigl({(k_{F\uparrow}-k_{F\downarrow})\over 2}\Bigr)
\nonumber \\
&& \hspace{2cm}\in [2.265 J,3.190 J]\hspace{0.20cm}{\rm at}\hspace{0.20cm}\Delta = 2
\nonumber \\
&& \zeta^{+-}_{2} (\pi/2,h) = -1 
\nonumber \\
&& + \sum_{\iota=\pm 1} \Bigl(\iota {\xi_{1\,2}^0\over 2} + {\xi\over 2}
- \Phi_{1,1}\Bigl(\iota k_{F\downarrow},- {(k_{F\uparrow}-k_{F\downarrow})\over 2}\Bigr)\Bigr)^2 
\nonumber \\
&& \hspace{2cm}{\rm for}\hspace{0.20cm}h \in [h_{c1},h_{1/2}] \, ,
\label{EPM2PI2}
\end{eqnarray}
\begin{eqnarray}
E^{+-}_{2} (\pi,h) & = & \varepsilon_ 2 (k_{F\uparrow}-k_{F\downarrow})\in [0.389 J,4.5J]
\hspace{0.20cm}{\rm at}\hspace{0.20cm}\Delta = 2
\nonumber \\
&& \hspace{1.3cm} \in [0.518 J,4.631 J]\hspace{0.20cm}{\rm for}\hspace{0.20cm}\Delta = 2.17
\nonumber \\
\zeta_{2}^{+-} (\pi) & = & -1 
\nonumber \\
& + & \sum_{\iota=\pm 1}
\Bigl(- {\iota\over 2\xi} + \xi 
+ \Phi_{1,2}(\iota k_{F\downarrow},k_{F\uparrow}-k_{F\downarrow})\Bigr)^2 
\nonumber \\
&& {\rm for}\hspace{0.20cm}h \in [h_{c1},h_{c2}] \, ,
\label{EPM2pi}
\end{eqnarray}
and
\begin{eqnarray}
&& E^{+-}_{3} (\pi/2,h) = \varepsilon_ 3 (0) - \varepsilon_1 \Bigl({(k_{F\uparrow}-k_{F\downarrow})\over 2}\Bigr)
\nonumber \\
&& \hspace{2cm}\in [2.654 J,5.891 J]\hspace{0.20cm}{\rm at}\hspace{0.20cm}\Delta = 2
\nonumber \\
&& \hspace{2cm}\in [2.912 J,6.165 J]\hspace{0.20cm}{\rm at}\hspace{0.20cm}\Delta = 2.17
\nonumber \\
&& \zeta^{+-}_{3} (\pi/2,h) = -1 
\nonumber \\
&& + \sum_{\iota=\pm 1}\Bigl(- {\iota\over 2\xi} + \iota {\xi_{1\,3}^0\over 2} + {\xi\over 2}
- \Phi_{1,1}\Bigl(\iota k_{F\downarrow},{(k_{F\uparrow}-k_{F\downarrow})\over 2}\Bigr)\Bigr)^2 
\nonumber \\
&& \hspace{2cm}{\rm for}\hspace{0.20cm}h \in [h_{c1},h_{1/2}] \, .
\label{EPM3PI2}
\end{eqnarray}

Finally, the line shape at and near the momentum $k=\pi/2$ sharp peak $R_{\pi/2}^{zz}$ called $R^{\rm PAP(zz)}_{\pi/2}$ in Fig. 5-a 
of Ref. \onlinecite{Bera_20} is for small values of the energy deviation $(\omega - E^{zz}_1 (\pi/2,h))\geq 0$ of the form, 
\begin{eqnarray}
S^{zz} (\pi/2,\omega) & \propto & \Bigl(\omega - E^{zz}_1 (\pi/2,h)\Bigr)^{-1/2}\hspace{0.20cm}{\rm where}
\nonumber \\
E^{zz}_1 (\pi/2,h) & = & (\varepsilon_1 (q+\pi/2) - \varepsilon_1 (q))\delta_{v_1 (q+\pi/2),v_1 (q)} 
\nonumber \\
& \in & [1.632\,J,1.876\,J]\hspace{0.20cm}
\nonumber \\
&& {\rm for}\hspace{0.20cm}h \in [h_{c1},h_{\diamond}] \, .
\label{EZZPI2}
\end{eqnarray}
Here the limiting energies $1.632\,J$ and $1.876\,J$ refer to magnetic fields $h_{\diamond}$
and $h=h_{c1}$, respectively, at anisotropy $\Delta =2$, the field $h_{\diamond}$ is given below, 
$E^{zz}_1 (\pi/2,h)$ is the $1$-continuum upper-threshold energy of $S^{zz} (k,\omega)$ at $k=\pi/2$,
and $v_1 (q)$ is the $1$-band group velocity, $v_1 (q) = \partial\varepsilon_1 (q)/ \partial q$.

The line shape at and in the vicinity of the sharp peak $R_{\pi/2}^{zz}$ is controlled
by a field-independent classical exponent $-1/2$. Indeed, the origin of this sharp peak is a density of states
singularity of a spectrum associated with the creation of one $1$-hole and one $1$-pair with {\it the same} group
velocity in the intervals $q\in [0,k_{F\downarrow}]$ and $q+\pi/2\in [\pi/2,k_{F\downarrow}+\pi/2]$, respectively. Here 
$q$ continuously increases upon increasing $m$ from $q=0$ for $m\rightarrow 0$ and $h\rightarrow h_{c1}$, reaching
$q=k_{F\downarrow}$ at a maximum spin density that for anisotropy $\Delta =2$ reads $m_{\diamond} = 0.627$ and 
a magnetic field $h_{\diamond} = 2.76$ in units of $J/(g\mu_B)$. For $J=3.55$\,meV it corresponds to $h_{\diamond} = 27.30$\,T. Indeed,
the sharp peak $R_{\pi/2}^{zz}$ exists only for fields $h\in [h_{c1},h_{\diamond}]$ for which the relation $v_1 (q)=v_1 (q+\pi/2)$ is satisfied.


\begin{references}
 \bibitem{He_05} 
	Z. He, D. Fu, T. Ky\^omen, T. Taniyama, and M. Itoh, 
	{\it Crystal growth and magnetic properties of BaCo$_2$V$_2$O$_8$},
	Chem. Mater. {\bf 17}, 2924 (2005).    
\bibitem{He_06}
	Z. He, T. Taniyama, and M. Itoh,
	{\it Antiferromagnetic-paramagnetic transitions in longitudinal and transverse magnetic fields
	in a SrCo$_2$V$_2$O$_8$ crystal},
	Phys. Rev. B {\bf 73}, 212406 (2006).
\bibitem{Kimura_07}
	S. Kimura, H. Yashiro, K. Okunishi, M. Hagiwara, Z. He, K. Kindo, T. Taniyama, and M. Itoh,
	{\it Field-induced order-disorder transition in antiferromagnetic BaCo$_2$V$_2$O$_8$ driven by a softening of spinon excitation},
	Phys. Rev. Lett. {\bf 99}, 087602 (2007).	
\bibitem{Okunishi_07}
	Okunishi, K. and Suzuki, T.
	{\it Field-induced incommensurate order for the quasi-one-dimensional $XXZ$ model in a magnetic field},
	Phys. Rev. B {\bf 76}, 224411 (2007),
\bibitem{Kimura_08}
	S. Kimura, T. Takeuchi, K. Okunishi, M. Hagiwara, Z. He, K. Kindo, T. Taniyama, and M. Itoh,
	{\it Novel ordering of an $S=1/2$ quasi-1d Ising-like antiferromagnet in magnetic field},
	Phys. Rev. Lett. {\bf 100}, 057202 (2008).	
\bibitem{Suga_08}
	S. Suga, 
	{\it Tomonaga-Luttinger liquid in quasi-one-dimensional sntiferromagnet BaCo$_2$V$_2$O$_8$ in magnetic fields},
	J. Phys. Soc. Jpn. {\bf 77}, 074717 (2008).	
\bibitem{Canevet_13}
	E. Can\'evet, B. Grenier, M. Klanj\v{s}ek, C. Berthier, M. Horvati\'c, V. Simonet, and P. Lejay,
	{\it Field-induced magnetic behavior in quasi-one-dimensional Ising-like antiferromagnet BaCo$_2$V$_2$O$_8$:
	A single-crystal neutron diffraction study},
	Phys. Rev. B {\bf 87}, 054408 (2013).		
\bibitem{Kimura_13}
	T. Kimura, K. Okunishi, M. Hagiwara, K. Kindo, Z. He, T. Taniyama, M. Itoh, K. Koyama, and K. Watanabe,
	{\it Collapse of magnetic order of the quasi one-dimensional Ising-like antiferromagnet BaCo$_2$V$_2$O$_8$ in transverse fields},
	J. Phys. Soc. Jpn. {\bf 82}, 033706 (2013).
\bibitem{Okutani_15}
	A. Okutani, T. Kida, T. Usui, T. Kimura, K. Okunishi, and M. Hagiwara, 
	{\it High field magnetization of single crystals of the $S=1/2$ quasi-1D Ising-like Antiferromagnet SrCo$_2$V$_2$O$_8$},
	Phys. Procedia {\bf 75}, 779 (2015).	 
\bibitem{Klanjsek_15}
	M. Klanj\v{s}ek, M. Horvati\'c, S. Kr\"amer, S. Mukhopadhyay, H. Mayaffre, C. Berthier, E. Can\'evet,
	B. Grenier, P. Lejay, and E. Orignac,
	{\it Giant magnetic field dependence of the coupling between spin chains in BaCo$_2$V$_2$O$_8$},
	Phys. Rev. B {\bf 92}, 060408(R) (2015).	
\bibitem{Grenier_15A}	
	B. Grenier, S. Petit, V. Simonet, E. Can\'evet, L.-P. Regnault, S. Raymond, B. Canals, C. Berthier, and P. Lejay, 
	{\it Longitudinal and transverse Zeeman ladders in the Ising-like Chain antiferromagnet BaCo$_2$V$_2$O$_8$},
	Phys. Rev. Lett. {\bf 114}, 017201 (2015).
\bibitem{Grenier_15}
	B. Grenier, V. Simonet, B. Canals, P. Lejay, M. Klanj\v{s}ek, M. Horvati\'c, and C. Berthier,
	{\it Neutron diffraction investigation of the $H-T$ phase diagram above the longitudinal incommensurate phase of BaCo$_2$V$_2$O$_8$},
	Phys. Rev. B {\bf 92}, 134416 (2015).	
\bibitem{Wang_18}
	Z. Wang, J. Wu, W. Yang, A. K. Bera, D. Kamenskyi, A. T. M. N. Islam, S. Xu, J. M. Law, B. Lake, C. Wu, and A. Loidl,
	{\it Experimental observation of Bethe strings},
	Nature {\bf 554}, 219 (2018).
\bibitem{Shen_19}
	L. Shen, O. Zaharko, J. O. Birk, E. Jellyman, Z. He, and E. Blackburn, 
	{\it Magnetic phase diagram of the quantumspin chains compound SrCo$_2$V$_2$O$_8$: 
	a single-crystal neutron diffraction study},
	New J. Phys. {\bf 21}, 073014 (2019).	
\bibitem{Wang_19}
	Z. Wang, M. Schmidt, A. Loidl, J. Wu,3, H. Zou, W. Yang, C. Dong, Y. Kohama, 
	K. Kindo, D. I. Gorbunov, S. Niesen, O. Breunig, J. Engelmayer, and T. Lorenz,
	{\it Quantum critical dynamics of a Heisenberg-Ising chain in a longitudinal field: 
	many-body strings versus fractional excitations},
	Phys. Rev. Lett. {\bf 123}, 067202 (2019).
\bibitem{Horvatic_20}
	M. Horvati\'c, M. Klanj\v{s}ek, and E. Orignac, 
	{\it Direct determination of the Tomonaga-Luttinger parameter $K$ in quasi-one-dimensional spin systems},
	Phys. Rev. B {\bf 101}, 220406(R) (2020).
\bibitem{Bera_20}
	A. K. Bera, J. Wu, W. Yang, R. Bewley, M. Boehm, J. Xu, M. Bartkowiak, O. Prokhnenko, B. Klemke, A. T. M. N. Islam, 
	J. M. Law, Z. Wang, and B. Lake, 
	{\it Dispersions of many-body Bethe strings},
	Nature Phys. {\bf 16}, 625 (2020).	
\bibitem{Han_21}
	Y. Han, S. Kimura, K. Okunishi, and M. Hagiwara,	
	{\it Unconventional magnetic excitations and spin dynamics of exotic quantum spin systems 
	BaCo$_2$V$_2$O$_8$ and Ba$_3$CuSb$_2$O$_9$},
	App. Magn. Reson. {\bf 52}, 349 (2021).
\bibitem{Cui_22}
	Y. Cui, Y. Fan, Z. Hu, Z. He, W. Yu, and R. Yu,  
	{\it Field-induced antiferromagnetism and Tomonaga-Luttinger liquid behavior in the quasi-one-dimensional Ising 
	antiferromagnet SrCo$_2$V$_2$O$_8$},
	Phys. Rev. B {\bf 105}, 174428 (2022).
\bibitem{Gaudin_71}	
	M. Gaudin, 
	{\it Thermodynamics of the Heisenberg-Ising ring for $\Delta\geq 1$},
	Phys. Rev. Lett. {\bf 26}, 1301 (1971).
\bibitem{Gaudin_14}	
	M. Gaudin, 
	{\it The Bethe wavefunction}
	(Cambridge University Press, 2014).
\bibitem{Takahashi_99}	
	M. Takahashi,
	{\it Thermodynamics of one-dimensional solvable models}
	(Cambridge University Press, 1999).   
\bibitem{Yang_19}
	W. Yang, J. Wu, S. Xu, Z. Wang, and C. Wu,
	{\it One-dimensional quantum spin dynamics of Bethe string states},
	Phys. Rev. B {\bf 100}, 184406 (2019). 
\bibitem{Carmelo_22} 
	J. M. P. Carmelo and P. D. Sacramento, 
        {\it The role of $q$-spin singlet pairs of physical spins in the dynamical properties of the spin-$1/2$ Heisenberg-Ising $XXZ$ chain},
        Nucl. Phys. B {\bf 974}, 115610 (2022).
\bibitem{Dupont_16} 
	M. Dupont, S. Capponi, and N. Laflorencie,
	{\it Temperature dependence of the NMR relaxation rate $1/T_1$ for quantum spin chains},
	Phys. Rev. B {\bf 94}, 144409 (2016).		
\bibitem{Carmelo_15}
	J. M. P. Carmelo, T. Prosen, and D. K. Campbell,
	{\it Vanishing spin stiffness in the spin-${1\over 2}$ Heisenberg chain for any nonzero temperature},
	Phys. Rev. B {\bf 92}, 165133 (2015).
\bibitem{Carmelo_17}
	J. M. P. Carmelo and T. Prosen, 
	{\it Absence of high-temperature ballistic transport in the spin-$1/2$ $XXX$ chain within the grand-canonical ensemble},
	Nucl. Phys. B {\bf 914}, 62 (2017).	
\bibitem{Pasquier_90}
	V. Pasquier and H. Saleur,
	{\it Common structures between finite systems and conformal field theories through quantum groups},
	Nucl. Phys. B {\bf 330}, 523 (1990).
\bibitem{Jepsen_20}
	P. N. Jepsen, J. Amato-Grill, I. Dimitrova, W. W. Ho, E. Demler, and W. Ketterle,
	{\it Spin transport in a tunable Heisenberg model realized with ultracold atoms},
	Nature {\bf 588}, 403 (2020).
\bibitem{Carmelo_20} 
	J. M. P. Carmelo, T. \v{C}ade\v{z}, and P. D. Sacramento, 
	{\it Bethe strings in the dynamical structure factor of the spin-$1/2$ Heisenberg $XXX$ chain},
	Nucl. Phys. B {\bf 960}, 115175 (2020).
\bibitem{Carmelo_05}
	J. M. P. Carmelo, K. Penc, and D. Bozi, 
	{\it Finite-energy spectral-weight distributions of a 1D correlated metal},
        Nucl. Phys. B {\bf 725}, 421 (2005); {\bf 737}, 351 (2006) (E).	        
\bibitem{Karlo_97}	
	K. Penc, K. Hallberg, F. Mila, and H. Shiba,
	{\it Spectral functions of the one-dimensional Hubbard model in 
	the $U\rightarrow\infty$ limit: How to use the factorized wave function},
	Phys. Rev. B {\bf 55}, 15 475 (1997).			
\bibitem{Imambekov_09}
	A. Imambekov and L. I. Glazman, 
	{\it Universal theory of nonlinear Luttinger liquids},
	Science {\bf 323}, 228 (2009).		
\bibitem{Imambekov_12}  
	A. Imambekov, T. L. Schmidt, and L. I. Glazman, 
	{\it One-dimensional quantum liquids: Beyond the Luttinger liquid paradigm},
	Rev. Mod. Phys. {\bf 84}, 1253 (2012).	
\bibitem{Carmelo_18}
	J. M. P. Carmelo and P. D. Sacramento, 
	{\it Pseudoparticle approach to 1D integrable quantum models},
	Phys. Reports {\bf 749}, 1 (2018).
\bibitem{Carmelo_94}
	J. M. P. Carmelo, A. H. Castro Neto, and D. K. Campbell,	
	{\it Perturbation theory of low-dimensional quantum liquids. II. Operator description of Virasoro algebras in integrable systems},
	Phys. Rev. B {\bf 50}, 3683 (1994).	
\bibitem{Sorella_96}
	S. Sorella and A. Parola,
	{\it Anomalous diffusion properties of wave packets on quasiperiodic chains},
	Phys. Rev. Lett. {\bf 76}, 4604 (1996).	
\bibitem{Sorella_98}
	S. Sorella and A. Parola,
	{\it Theory of hole propagation in one-dimensional insulators and superconductors},
	Phys. Rev. B {\bf 57}, 6444 (1998). 
\bibitem{Kruis_04}	
	H. V. Kruis, I. P. McCulloch, Z. Nussinov, and J. Zaanen, 
	{\it Geometry and the hidden order of Luttinger liquids:?The universality of squeezed space},
	Phys. Rev. B {\bf 70}, 075109 (2004).	
\bibitem{Shastry_90}
	B. S. Shastry and B. Sutherland, 
	{\it Twisted boundary conditions and effective mass in Heisenberg-Ising and Hubbard rings},
	Phys. Rev. Lett. {\bf 65}, 243 (1990).	
\bibitem{Zotos_99}	
	X. Zotos,  
	{\it Finite temperature Drude weight of the one-dimensional spin-$1/2$ Heisenberg model},
	Phys. Rev. Lett. {\bf 82}, 1764 (1999).
\bibitem{Herbrych_11}	
	J. Herbrych, P. Prelov\v{s}ek, and X. Zotos,
	{\it Finite-temperature Drude weight within the anisotropic Heisenberg chain},
	Phys. Rev. B {\bf 84}, 155125 (2011).
\bibitem{Caux_08}
	J. S. Caux, J. Mossel, and I. P. Castillo, 
	{\it The two-spinon transverse structure factor of the gapped Heisenberg antiferromagnetic chain},
	J. Stat. Mech. P08006 (2008).
\bibitem{Karbach_02}
	M. Karbach, D. Biegel, and G. M\"uller,
	{\it Quasiparticles governing the zero-temperature dynamics of the one-dimensional spin-$1/2$
	Heisenberg antiferromagnet in a magnetic field},
	Phys. Rev. B {\bf 66}, 054405 (2002).
\bibitem{Sacramento_95}
	P. D. Sacramento,
	{\it Thermodynamics of the attractive Hubbard chain},
	J. Phys. Cond. Matter {\bf 7}, 143 (1995).
\bibitem{Hikihara_04}
	T. Hikihara and A. Furusaki,
	{\it Correlation amplitudes for the spin-${1\over 2}$ $XXZ$ chain in a magnetic field},
	Phys. Rev. B {\bf 69}, 064427 (2004).	
\end{references}
\end{document}